%% file: galex_nkv.tex
\documentclass[useAMS,usenatbib,usegraphicx]{mn2e}

\usepackage{lscape}
\usepackage{amsmath}
\usepackage{multirow}
\usepackage[figuresright]{rotating}
\usepackage{float}
\usepackage{natbib}
\usepackage{url}
\usepackage{setspace}
\usepackage{subfigure}
\usepackage{txfonts}
\usepackage{graphicx}
\usepackage{graphics}

\usepackage{color}
\usepackage{soul}

\usepackage[]{placeins} 

\makeatletter
\newcommand{\customlabel}[2]{%
\protected@write \@auxout {}{\string \newlabel {#1}{{#2}{}}}}
\makeatother

\title[Hot subdwarfs in the {\sl GALEX} survey]{A selection of hot subluminous stars in the
{\sl GALEX} survey -- II.\\
Subdwarf atmospheric parameters
\thanks{Based on observations made with ESO telescopes at the La Silla Paranal Observatory
under programmes 82.D-0750, 83.D-0540 and 85.D-0866.}}

\author[P. N\'emeth, A. Kawka and S. Vennes]
{P\'eter N\'emeth\thanks{E-mail:
pnemeth1981@gmail.com (PN); kawka@sunstel.asu.cas.cz (AK); vennes@sunstel.asu.cas.cz
(SV)}\thanks{Visiting Astronomer, Kitt Peak National Observatory, National
Optical Astronomy Observatory,
which is operated by the Association of Universities for Research in
Astronomy (AURA)
under cooperative agreement with the National Science Foundation.},
Ad\'{e}la Kawka\footnotemark[2]\footnotemark[3]
and St\'{e}phane Vennes\footnotemark[2]\footnotemark[3]\\\\
{Astronomick\'{y} \'{u}stav AV \v{C}R, Fri\v{c}ova 298, CZ-251 65
  Ond\v{r}ejov, Czech Republic}}

\begin{document}

\date{Accepted 2012 August 27. Received 2012 August 27; in original 
form 2012 May 15}

\pagerange{\pageref{firstpage}--\pageref{lastpage}} \pubyear{2012}

\maketitle

\label{firstpage}

\begin{abstract}
We present an update of our low-resolution spectroscopic follow-up and model
atmosphere analysis of hot subdwarf stars from the {\it Galaxy Evolution
Explorer} ({\sl GALEX}) survey. 
Targets were selected on the basis 
of colour indices calculated from 
the {\sl GALEX} GR6 $N_{\rm UV}$, Guide Star Catalogue (GSC2.3.2) $V$ 
and the Two Micron All Sky Survey (2MASS) $J$ and $H$ photometry. 
High signal-to-noise ratio spectra were
obtained at the European Southern Observatory (ESO) and the 
Kitt Peak National Observatory (KPNO) over the course of three years. 
Detailed H, He and CNO 
abundance analysis helped us improve 
our $T_{\rm eff}$, $\log g$ and He abundance determination and to constrain CNO
abundances. 
We processed 191 observations of 180 targets and found 124 sdB and 42 sdO stars in this sample
while some blue horizontal branch stars were
also found in this programme. 
With quantitative binary decomposition of 29 composite
spectra we investigated the incidence of A, F and G type companions. 
The incidence of late G and K type companions and their
effects on subdwarf atmospheric parameters were also examined.
\end{abstract}


\section{Introduction}

Hot subdwarfs are core He burning stars located at the blue end of the
horizontal branch (HB), also known as the extreme horizontal branch (EHB). 
These subluminous objects are located roughly 
midway between the early-type
main-sequence (MS) stars and the white dwarf (WD) cooling curve 
in the Hertzsprung-Russell diagram (HRD). 
Being
in a relatively long lasting ($\sim$160 Myr), intermediate evolutionary
stage of $\sim$1${\rm M}_{\odot}$ stars, they are
quite common and likely to overwhelm WDs in blue and ultraviolet (UV) surveys
of old stellar populations.
These stars are the primary sources of the UV excess of elliptical
galaxies (\citealt{brown97}, \citealt{oconnell99}).

To look for bright, thus nearby WD and subdwarf 
candidates in the {\it Galaxy Evolution Explorer} ({\sl GALEX}) data base, 
\citet{vennes11a} (hereafter: Paper I) devised a method
based on UV, optical and infrared (IR) colours.
Between 2008 and 2011, our systematic, low-resolution spectroscopic
follow-up with the European Southern Observatory (ESO)/New Technology
Telescope (NTT) and the Kitt Peak National Observatory (KPNO)/Mayall telescopes 
confirmed 166 stars as
hot subdwarfs. 
Here, we present the results of a model atmosphere
analysis of 180 stars. 
Details of previous surveys of blue or ultraviolet excess objects and of our
source selection can be found in Paper I.
Here we modified the original selection criteria to avoid
de-selecting
bright sources with unreliable {\sl GALEX} $F_{\rm UV}$ photometry. 
For {\sl GALEX} $N_{\rm UV} \le 12.5$, where $N_{\rm UV}$ is
corrected for non-linearity effects 
\citep{morrissey07}, we selected 
targets with $N_{\rm UV}-V \le 0.5$, where $V$ is the Guide Star Catalogue
version 2.3.3 ({\small GSC2.3.2}; \citealt{lasker08}) photographic magnitude (see
Paper I), while ignoring the {\sl GALEX} $F_{\rm UV}$ magnitude. 
13 bright objects were added to our sample by using the modified criteria.
Among the 180 stars we found 124 subdwarf B (sdB) and 42
subdwarf O (sdO) stars. 
Most of the remaining objects are classified as
blue HB (BHB) stars. 
The 52 stars from Paper I are also re-analysed here with
an extended abundance pattern and using a new fitting algorithm that
provided consistent results with Paper I. 
Such large and homogeneously modelled
samples of bright subdwarfs 
are useful in identifying candidates for pulsation (e.g.,
\citealp{ostensen10}), radial velocity studies (e.g., \citealp{geier11a},
\citealt{napiwotzki08}) and
eventually
to set observational constraints for formation and evolution theories.

The present paper deals with optical/IR data exclusively. We list 
$N_{\rm UV}$ magnitudes, but refer the interested reader to the UV work of 
\citet{girven12}.   
A similar UV analysis will be reported elsewhere.

There are two main formation concepts for subdwarf stars, but current 
observational data cannot
satisfyingly distinguish the contribution of their formation channels. 
The canonical subdwarf formation theory \citep{mengel76} invokes
binarity and strong mass loss on the red-giant branch (RGB) for the 
formation of sdB stars, and WD binary mergers for the
formation of He-rich hot sdO (He-sdO) stars \citep{webbink84}. 
For a recent review of the canonical formation channels and binary 
population synthesis see \citet{hu08}, \citet{han03a} and \citet{han03b}. 

The late hot-flasher scenario \citep{dcruz96} suggests 
a different evolutionary path that explains the abundance diversities observed
in hot subdwarfs. 
This model also assumes a strong mass-loss on
the RGB, but the star undergoes a core He-flash on the WD cooling track.
Then, convective mixing would bring core material into the hydrogen
envelope and would account for He, C and N overabundances. 

Both models require an effective process to remove most of the 
hydrogen envelope on the RGB. 
The main difference is that the canonical formation scenario 
assumes a core flash on the tip of the RGB without mixing, while in the 
late hot-flasher scenario the star departs from the RGB before 
the flash happens allowing for shallow or deep convective mixing \citep{lanz04}. 

Stringent observational constraints
are needed to determine the relative contribution of these formation models. 
For this reason, detailed model atmosphere and
binary analyses of large samples carried-out independently 
with different model atmosphere codes 
are necessary to reveal the contribution of various formation channels and the condition
of mass loss on the RGB.

Throughout the paper we use the notation 
$[{\rm He/H}]_\bullet=\log_{10}(n_{\rm He}/n_{\rm H})$ for abundance and 
$[{\rm Fe/H}]=[{\rm Fe/H}]_\bullet-[{\rm Fe/H}]_\odot$ for metallicity.
We omit the {\sl GALEX}
prefix, as well as the seconds and arcseconds of the identifier when referring to
stars in the catalogue. 
Finally, we refer to possible short- ({\it p}-mode) and long-period ({\it g}-mode) pulsators as
rapid and slow pulsators using the nomenclature of \citet{kilkenny10}.

In Section \ref{Sec:data} we briefly describe our data acquisition and reduction. 
In Section \ref{Sec:analysis} we give details of our spectral modelling and 
fitting with {\small TLUSTY} and {\small XTGRID}, and our approach for binary decomposition. 
In Section \ref{Sec:properties} we present the properties of hot subdwarfs 
in the $T_{\rm eff}-{\log g}$ and $T_{\rm eff}-{\rm He}$ diagrams, and 
describe our results on CNO abundance trends and binarity. 
In Section \ref{Sec:luminosity} we investigate the sample
completeness by comparing our observed luminosity distribution to a modelled one. 
Modelling homogeneity is examined in Section \ref{Sec:homogeneity}. 
Subdwarf populations are reviewed in light of our results in Section \ref{Sec:sdBd}. 
The catalogue of our {\sl GALEX} sample is presented in Section
\ref{Sec:catalogue} and Section \ref{Sec:conclusions} summarizes our results. 
Finally, in Section \ref{Sec:future}, 
we propose to initiate a large collaborative study on hot subdwarf stars.

\section{Follow-up spectroscopy}\label{Sec:data}

Low-dispersion optical spectra were obtained at two sites during seven
observing runs between 2008 and 2011. 
At ESO, we used the ESO Faint object Spectrograph and Camera (EFOSC2) 
attached to the 3.6-m NTT 
at the La Silla Observatory. 
On UT 2008 October 19-22, 2009 March
2-4 and 2009 August 23-27, we used grism \#11 (300 lines mm$^{-1}$) with
a dispersion of $\sim$$4.17$ \AA\ per binned pixel (2 x 2). 
With a $1$ arcsec slit width we obtained 
a resolution of $\Delta\lambda\approx13.7$ \AA. 
On UT 2010 March 2-4 and 2010 September 18-21 we employed grism \#7 (600
lines mm$^{-1}$, $\sim$$1.96$ \AA\ per binned pixel) and a $1$ arcsec slit width
resulting in a resolution of $\Delta\lambda\approx6.4$ \AA.

At the KPNO on UT 2010 March 23-26 and 2011 January 28-31 
the Ritchey--Chretien Focus 
Spectrograph (RC Spectrograph) was used with the 4-m Mayall telescope and KPC-10A
(316 lines mm$^{-1}$) grating, delivering a spectral resolution of
$\Delta\lambda\approx5$ \AA\ in first order. 
The slit width was set at
$1.5$ arcsec resulting in a dispersion of $\sim$$2.75$ \AA\ per pixel. 

The KPNO spectra and the lower-resolution ESO data
cover the entire Balmer-series from 3700 to 7400 \AA. 
The ESO
$\Delta\lambda\approx6.4$ \AA\ data covers the upper Balmer line series from 
3600 to 5200 \AA.
All data were reduced using standard
IRAF\footnote{\url{http://iraf.noao.edu/}}
procedures.

Our selection also included three hot subdwarf candidates proposed by 
\citet{jimenez11}.
Two of these objects are {\sl GALEX} sources outside of our selection criteria
(TYC 6017-419-1 and TYC 9327-1311-1), while another lies in a region not
covered by the survey (TYC 9044-1653-1).
We included another two targets (J0716+2319 and J2349+4119) that
are not subdwarfs, but accidentally met the criteria due to their erroneous
$V$ magnitude in our source selection. 
Although these stars are not formally part of the UV selection, we analyze their
spectra in Section \ref{Sec:analysis}.

\section{Spectral analysis}\label{Sec:analysis}
\subsection{Model atmospheres}

We computed H/He/CNO non-LTE model atmospheres with {\small TLUSTY 200} and
synthetic spectra with {\small SYNSPEC 48} (\citealt{hubeny95};
\citealt{lanz95}). Model atoms of H\,{\sc i}, He\,{\sc i--ii}, 
C\,{\sc ii--iv}, N\,{\sc iii--v} and O\,{\sc iv--vi}; 
and
detailed line profiles of H and He were used in {\small SYNSPEC}. 
For all relevant ions, we included the most detailed model atoms from 
the {\small OSTAR2002} \citep{lanz03} and {\small BSTAR2006} \citep{lanz07} data base. 

Model atmospheres for BHB stars 
with effective temperatures below 20\,000 K were calculated with 
H\,{\sc i}, He\,{\sc i--ii}, C\,{\sc ii--iv}, 
N\,{\sc ii--iv} and O\,{\sc ii--iii} model atoms. 
A line list compiled from Kurucz CD-ROM 23 and available at the 
{\small SYNSPEC} web page\footnote{\url{http://nova.astro.umd.edu/}} was used. 
Details of the model atoms used in this work are listed in Table \ref{Tab:modatom}. 

\begin{table}
\begin{center}
\caption[3]{Model atoms used for hot (H) and cool (C) subdwarfs.}
\begin{tabular}{lccc}
\hline
Ion & Number of levels & Frequency of & Model\\
    & and line transitions  & highest level, s$^{-1}$ & \\
\hline
H\,{\sc i}   &  9/10   & 3.273$\times10^{15}$ &HC\\
H\,{\sc ii}  &  1/0    & $\dots$               &HC\\ 
\vspace{-2mm}\\
He\,{\sc i}  & 24/89   & 5.138$\times10^{13}$  &HC\\
He\,{\sc ii} & 20/51   & 3.289$\times10^{13}$  &HC\\
He\,{\sc iii}&  1/0    & $\dots$               &HC\\
\vspace{-2mm}\\
C\,{\sc ii}  & 39/253  & -5.299$\times10^{13}$ &HC\\
C\,{\sc iii} & 46/339  & 2.998$\times10^{14}$  &HC\\ 
C\,{\sc iv}  & 37/222  & 5.272$\times10^{14}$  &HC\\
C\,{\sc v}   &  1/0    & 9.480$\times10^{16}$  &HC\\
\vspace{-2mm}\\
N\,{\sc ii}  & 26/93   & 5.564$\times10^{14}$  &C\\
N\,{\sc iii} & 32/187  & 2.519$\times10^{14}$  &HC\\
N\,{\sc iv}  & 23/95   & 1.471$\times10^{15}$  &HC\\
N\,{\sc v}   & 16/95   & 8.227$\times10^{14}$  &HC\\
N\,{\sc vi}  &  1/0    & 1.335$\times10^{17}$  &H\\
\vspace{-2mm}\\
O\,{\sc ii}  & 29/123  & 6.622$\times10^{14}$  &C\\
O\,{\sc iii} & 41/226  & 2.410$\times10^{14}$  &C\\
O\,{\sc iv}  & 39/283  & 5.678$\times10^{14}$  &HC\\
O\,{\sc v}   & 40/225  & 1.685$\times10^{15}$  &H\\
O\,{\sc vi}  & 20/126  & 1.185$\times10^{15}$  &H\\
O\,{\sc vii} &  1/0    & 1.788$\times10^{17}$  &H\\
\hline
\end{tabular}
\label{Tab:modatom}
\end{center}
\end{table}

\subsection{Spectral fitting with {\small XTGRID}}

A spectral fitting method was employed by combining the steepest-descent
and simplex algorithms, implemented in our new $\chi^2$ minimizing 
fitting program {\small XTGRID}. 
This {\sc Python} program is an adjustable interface for {\small TLUSTY} and
{\small SYNSPEC}, designed to carry out iterative multi-wavelength spectral analysis 
of hot stars from soft X-rays to near infrared wavelengths. 
The procedure
requires a starting model and with successive adjustments approaches the
observed spectrum. 

At the beginning, {\small XTGRID} follows 
the gradient of steepest descent with a maximal
step size that {\small TLUSTY} convergence allows. 
New step directions and sizes are calculated in a
small grid (simplex step) and adjusted independently for each parameter. 
To follow the steepest gradient, relative step sizes are normalized to the
leading parameter. 
Model atmospheres and synthetic spectra 
are calculated for a new temperature and gravity in all iterations,
and only synthetic spectra are calculated for new abundances. 
Directions are calculated
from these fits and the starting model gets updated with the new model. 
This process is
pursued until all relative changes ($T$, $\log g$, abundances and $\chi^2$)
decrease below
0.5 per cent in three consecutive iterations 
and all maximum allowed step sizes decrease below 50 per cent of their initial values.
The latter condition ensures that a minimum is found for parameters with small
gradients.

Temperature and gravity have a higher convergence rate than abundances. 
To take
advantage of this observation and accelerate our procedure, 
parameter relaxation is included in the program. 
When relative changes of temperature or gravity decrease below 0.5 per cent
they can be kept fixed for five iterations. 
This allows for slow converging parameters (like abundances) to approach
their final values faster.
Furthermore, such iterations save two out of 
the three model atmosphere calculations per iteration, the
most time consuming part. 
After 11 iterations all models are
calculated again for three iterations, to examine relative changes and
to test if the final model is achieved. 
If the convergence limit is not reached, this loop
of 14 iterations is repeated. 
This strategy provides about a factor of 2 speed-up
with respect to regular iterations when model atmospheres are always
calculated for a new temperature and gravity. 
Additionally, the fitting procedure accelerates
automatically as by approaching the solution the relative differences between 
models get smaller.

{\small XTGRID} regularly reads the optional {\small TLUSTY} flag file, in
which important control parameters can be stored for the fitting process.
This way it is possible to adjust any fitting parameters or procedures during
calculations. 

A critical point in spectral fitting is the method how models are normalized 
to observations. 
Over a wide spectral range the match between theoretical and 
observed
continua is usually poor in the beginning, while over a 
short wavelength range the continuum might not be visible because of 
broad spectral lines or line blanketing.
To account for both effects, {\small XTGRID} samples
the spectrum in small sections. 
The selection of these fitting
windows is either specified by the user or the program partitions the entire
data in equally sized sections. 
The first setup can be used for selecting certain lines or ranges for fitting, like the
hydrogen Balmer series or certain metal lines. When the whole spectrum is
considered, only the segment size needs to be specified. 
The optimal section
size depends on the
spectral resolution, the spectral features being analysed and must be
selected carefully.
The synthetic spectrum is then normalized in each of these bins to the observed data
by finding the ratio of their median fluxes. 
This simple approach tends to normalize for
the continuum and normalization improves with the goodness of fit. 
Low
order variations of the spectrum are significantly reduced this way,
therefore the effects of interstellar reddening and poor fluxing, often seen at short
optical wavelengths, are treated. 
An additional benefit of this 
method is, when the whole spectrum is used, 
that the $\chi^2$ considers the interconnections between 
all spectral lines and the continuum. 
This can be very helpful in investigating the effects of metal abundances 
on hydrogen lines, in particular, to address the Balmer-line problem.  

Although low order variations can be significantly reduced with a careful
sectioning of the spectrum, our normalisation method considers the
continuum as well, hence it is sensitive for flux calibration. 
Large deviations
from the theoretical continuum due to a poor flux calibration can affect the
final parameters. 
Hence, for consistent modelling the same sampling of the spectrum is necessary, or
to work with continuum normalized spectra. {\small XTGRID} is designed to fit
continuum normalized data as well.

To match the resolution of synthetic spectra to the observations, 
they are convolved with a Gaussian
profile at either constant
resolution (full width at half-maximum) or constant resolving power $R=\lambda/\Delta\lambda$. 

Radial
velocity corrections can be defined by specifying $\Delta\lambda/\lambda$ or
taken as free parameter. 
If fitted, a $\chi^2$ is calculated at small shifts of the
model and the minimum is determined by a low order polynomial fit. This
correction yields absolute radial velocities and 
updated after every third iteration with adjusted shifts. 

After fitting is done, parameter errors for 60, 90
and 99 per cent confidence intervals are estimated by mapping the $\Delta\chi^2$ with
respect to the final abundance ($X$) 
at representative points in the range of $-0.99<{\Delta}X/X
<100$. Temperature and gravity errors are measured similarly in the range
of $-0.26<{\Delta}T/T <0.26$ and $-0.195<{\Delta}\log g/\log g <0.195$. 
Parameters are
changed until the statistical limit for the 60 per cent confidence level
at the given number of free parameters is reached. 
Model atmospheres are calculated during error
calculations for all
synthetic spectra to track small changes of the $\chi^2$. 
This way the size
and symmetry of error intervals can reflect the quality of the parameter
determination. 
Errors at 90 and 99 per cent are extrapolated by a parabolic fit for the upper and lower error intervals
independently. 

\begin{figure*}
 \includegraphics[width=0.41\linewidth,clip=,angle=-90]{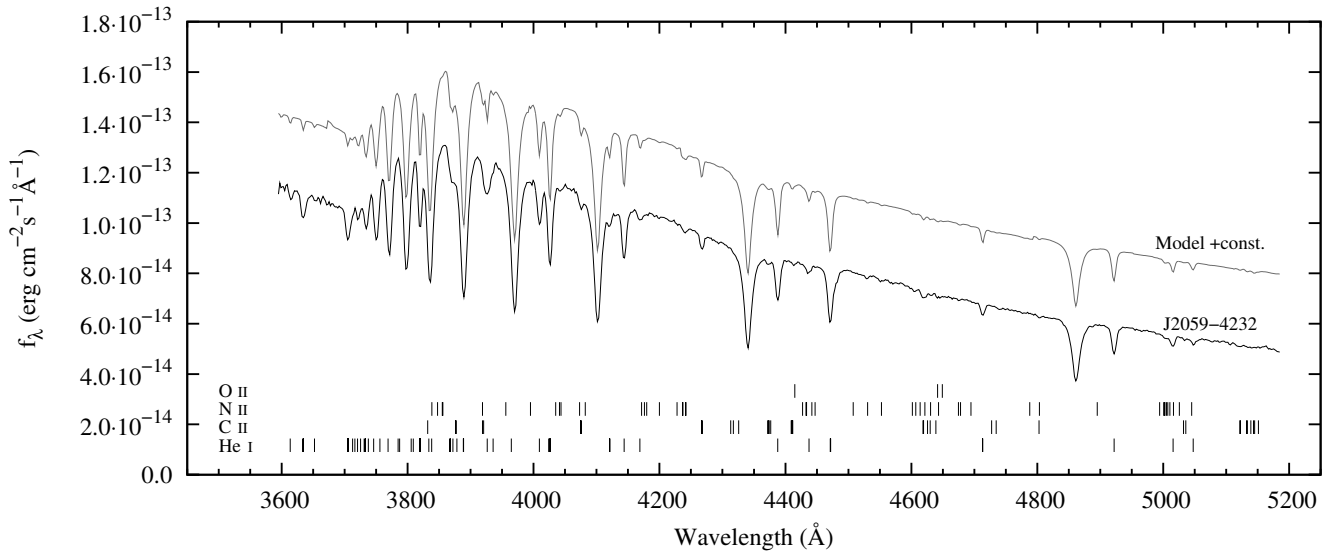}
\caption{{\small
Observed NTT and best-fitting model spectrum for J2059--4232.
 Identifications are given for He and
 CNO lines with theoretical equivalent widths larger than 1 \AA. The 
 presence of He, C and N is statistically confirmed.}
\label{Fig:2059}}
\end{figure*}

{\small XTGRID} is
scalable for cluster calculation and to further 
accelerate the fitting procedure, previously calculated models are
reused until relative changes of temperature and gravity drop below 13 per
cent of
their respective maximum allowed values.
Following this procedure, the program 
builds a model cloud that is more closely spaced near best solutions.
Such accelerations are necessary to cope
with the computing demand of model atmosphere analyses. 
In particular, in batch analysis of multiple spectra, because 
any change in the input
data, physics or fitting procedure requires the recalculation of all 
fits to maintain homogeneity. 
In the final step \LaTeX\ fit summaries and 
{\sc Gnuplot} scripts are produced for graphs shown in this paper.
Figure \ref{Fig:2059} shows the best-fitting of the He-rich BHB star J2059--4232.

\subsection{Binary spectral decomposition}

\begin{figure}
 \includegraphics[width=\linewidth,clip=,angle=0]{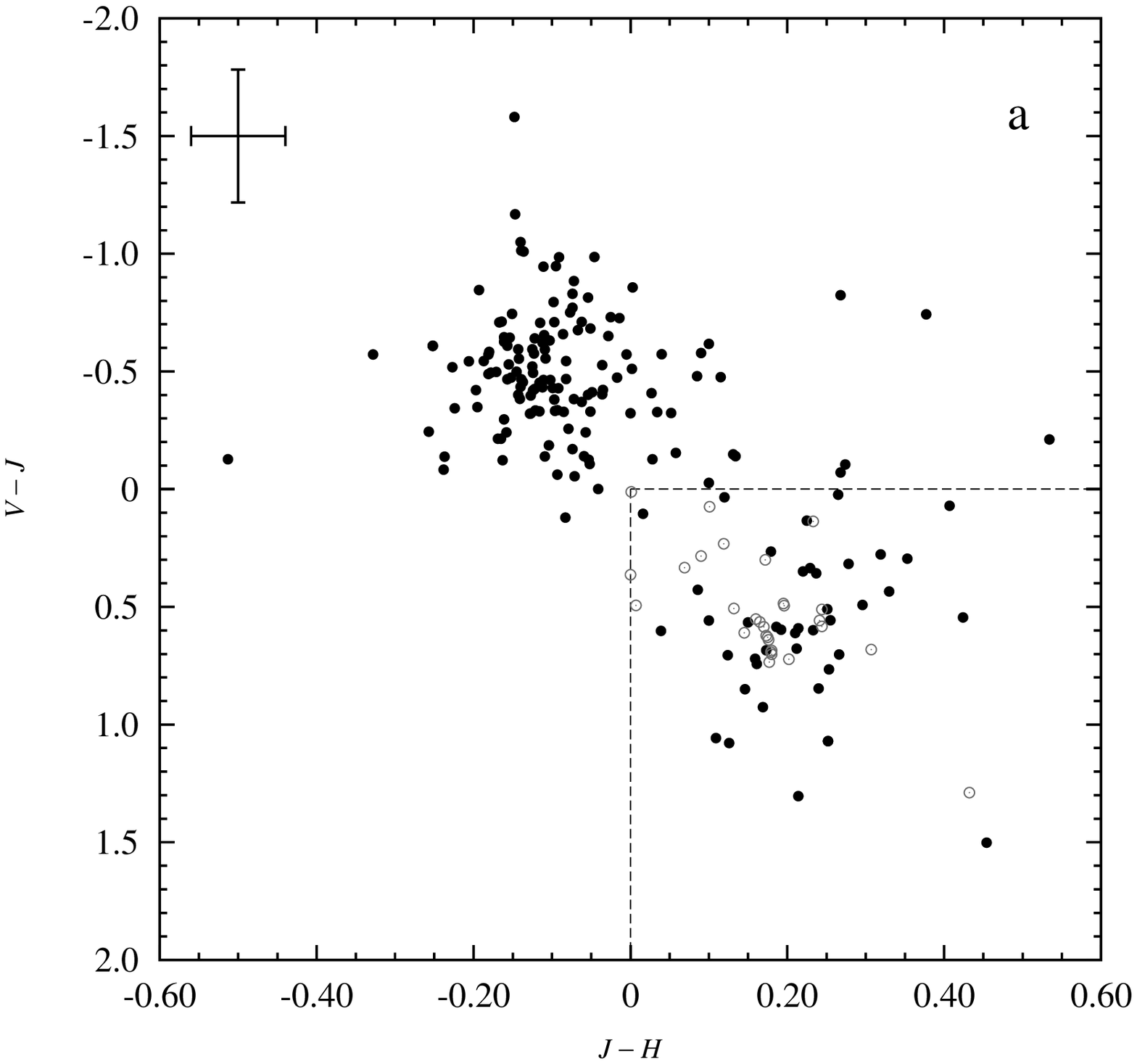}

\centering
\vspace{2pt}
  \includegraphics[width=0.48\linewidth,clip=,angle=0]{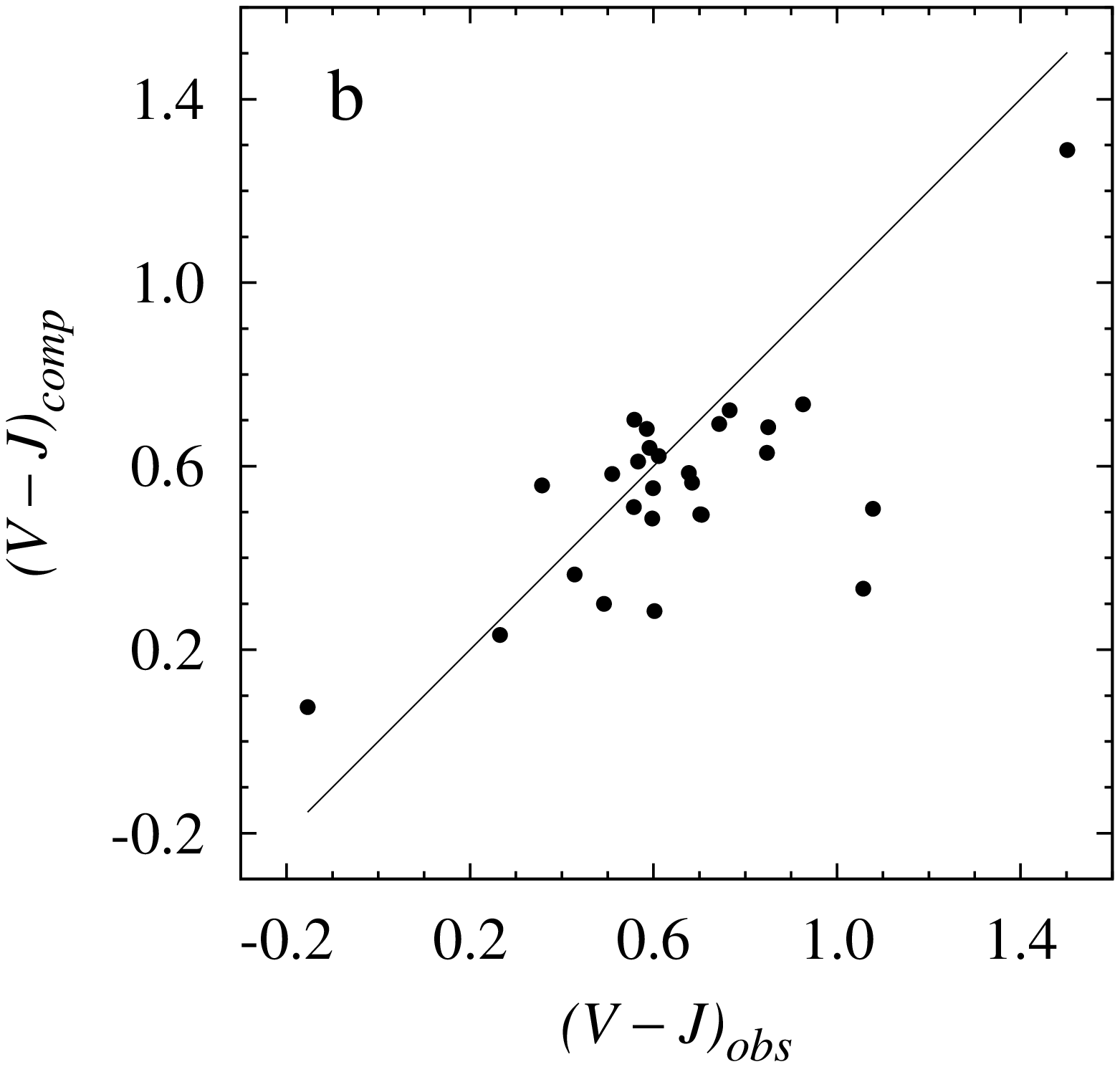}
\hspace{1mm}
 \includegraphics[width=0.48\linewidth,clip=,angle=0]{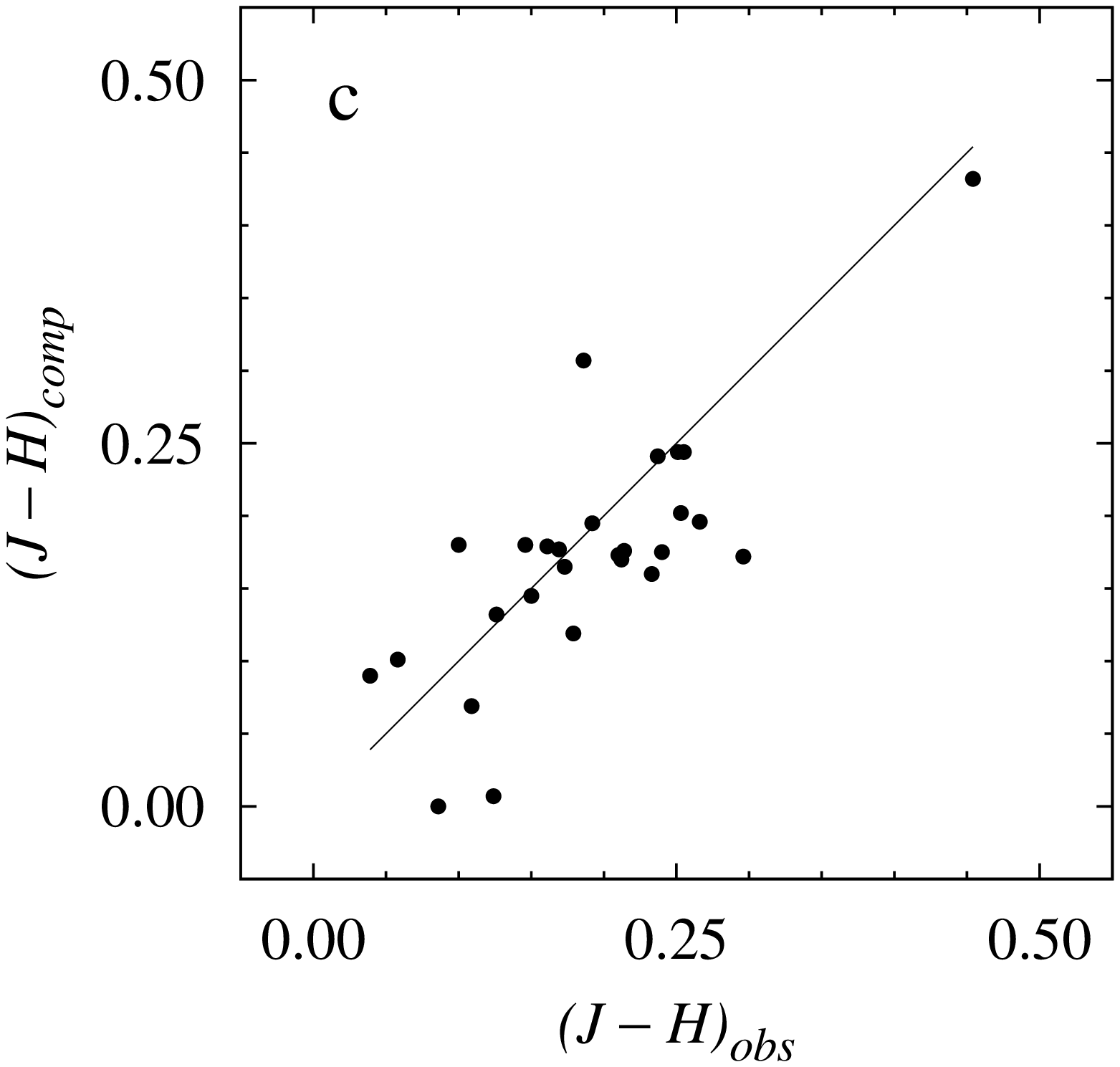}
 \caption{\small Panel {\it a} shows the
 $V$$-$$J$ vs. $J$$-$$H$ 
 colour--colour diagram for our subdwarf
 sample. Optical magnitudes were collected from {\small GSC2.3.2} and infrared from
 the
 {\small 2MASS} data base using VizieR. Apparently single stars aggregate near
 $V$$-$$J$$\approx$$-0.5$ and
 $J$$-$$H$$\approx$$-0.1$, while composite-spectra binaries show
 infrared excess (lower right corner). About 19 per cent of the stars have
 $V$$-$$J>0$
 and $J$$-$$H>0$. Open circles show synthetic colour indices of 
 the 29 binaries resolved in this work. Typical errors of photometric colour
 indices
 are shown in the upper left corner. 
 Panel {\it b} and {\it c} show the correlations of synthetic and observed $V-J$ and $J-H$ colour
 indices, respectively.   
\label{Fig:IRE}}
\end{figure}

Binarity plays an important role in the theory of subdwarf formation and
evolution. In the canonical formation theory \citep{han03a} 
the common-envelope, Roche lobe overflow and binary merger channels 
all depend on binary parameters. 
About 40 per cent of sdBs were found in binary
or multiple systems with MS companions (\citealt{maxted01},
\citealt{napiwotzki04}).
\citet{reed04} derived an sdB--MS (A0 to M2) binary fraction of 
$53\pm6$ per cent by using Two Micron All Sky Survey ({\small 2MASS};
\citealt{skrutskie06}) $J-H$ and
optical $B-V$ colours. 
\citet{thejll95} found $\sim$43 per cent composite spectra in their sample of 27
subdwarfs. 
\citet{stark03} found 40 per cent composite spectra in their magnitude limited sample of
about 600 subdwarfs collected from the Palomar-Green survey (PG,
\citealt{green86}) and the
Catalogue of Spectroscopically Identified Hot Subdwarfs \citep{kilkenny88}.
At least 20 per cent of subdwarfs show significant IR excess in the
$V$$-$$J$ - $J$$-$$H$ diagram \citep{ferguson84}. 
In agreement with this number, about 19 per cent of
our sample has $V$$-$$J>0$ and $J$$-$$H>0$ indicative of an IR excess (Figure
\ref{Fig:IRE}).
The true fraction of composites depends heavily on the quality of the
spectra [signal-to-noise ratio (SNR),
resolution] and selection criteria. 
Combining the frequency
of composite spectra, radial velocity variables and photometric variables
(ellipsoidal variations) we conclude that the binary fraction is close to two-thirds
in agreement with \citet{maxted01}.

About 17 per cent of our sample shows strong double lined composite spectra 
and about 33 per cent have noticeable 
Ca\,{\sc ii} H\&K ($\lambda3933$ and $\lambda3968$ \AA), 
Mg\,{\sc i} ($\lambda5183$ \AA) or
Na\,{\sc i} D ($\lambda5890$ and $\lambda5896$ \AA) absorption lines. 
Fe\,{\sc i} and CH molecular lines blend at
$\sim$$\lambda4300$ \AA\ and notable in F and G type stars. 
This strong feature is often referred in the literature as the $G$ band.
We found $G$-band absorption in composite spectra.
Although, these lines
might have an interstellar origin, their simultaneous presence and a 
flat flux distribution
can indicate a composite spectrum.

The high incidence rate of binaries and their
importance in subdwarf evolution in addition to the fact that subdwarfs
in composite spectra binaries cannot be modelled as single stars,
inspired us to include binary decomposition in {\small XTGRID}. 
In double-lined binaries both components can be examined
simultaneously. In particular, the F and G companions are relatively easy
to identify because they have characteristic spectral features different from
subdwarfs and a comparable optical brightness.
Late G and K companions present a bigger challenge because they
have significantly lower contribution and weaker lines.

A library of low- and medium-resolution observed
spectra was collected for spectral decomposition 
from the {\small MILES} (\citealt{cenarro07}, \citealt{sanchez06})
and {\small HILIB} \citep{pickles98} libraries with 
resolutions $\Delta\lambda = 2.3$ \AA\ and R = 500, 
respectively. For visual inspections the spectral library of \citet{silva92}
was also used. 
Altogether, 946 spectra between $\lambda3525$ and $\lambda7500$ \AA\ at 2.3
\AA\ resolution were included in our decomposition library from the {\small MILES} data base.
This library provides a wide range of spectral types at sufficiently high
resolution and good fluxing. 
{\small HILIB} templates were used to calculate synthetic colour indices in the optical
and infrared.

We searched for the best-fitting template spectrum by interpolating in temperature, 
surface gravity, 
metallicity (as [Fe/H]) and flux ratio 
$({\rm F}_{\lambda,comp.}/({\rm F}_{\lambda,comp.}+{\rm
F}_{\lambda,sd}))$ 
of the secondary component along with
{\small TLUSTY} model parameters for the primary. 
The decomposition was performed at 
every third iteration and during the three consecutive iterations when fitting
convergence was examined. 
The observed data were fitted with the linear combination of the
subdwarf model and a secondary template. 
At the beginning, a 50 per cent flux contribution was
assumed for F and early G type companions and the interpolation step size was 0.1
times the flux difference between spectra. 
At every decomposition iteration the interpolation
step size was halved and after the fifth iteration it was set to 0.001. 
To avoid trapping, all template spectra were compared to the observed data at
every iteration, but interpolation was performed only between the best-fitting
template and its neighbours.

The decomposition of late G and K companions required a
more elaborated method. 
Their most significant effect on the flux distribution
is a slight flattening of the flux in the red part of the spectrum and 
relatively strong Na D lines.
In this case, the initial flux ratio of the secondary was set 
to 15 per cent at the
long wavelength end of the spectrum.
Fitting started with a pre-defined section size to
consider only spectral lines in the beginning, and the continuum was 
progressively considered by gradually increasing the section size up to three 
times the starting value.
If the secondary contribution decreased below 10 per cent, the spectrum was sampled
with the smallest possible sections, which was set to three times the size of
a resolution element. 
This strategy is necessary only for late-type companions
and does not change the results for F and early G
stars, which can be modelled based on spectral lines.  
The fitting windows were reset to their input configuration in the last step
and kept fixed during error
calculations, because changing the size of fitting sections can introduce
systematic shifts.
In the case of late companions a high SNR ($\ >100$), medium to high resolution data, and good fluxing are critical.  

\begin{figure*}
\begin{center}
 \includegraphics[width=0.7\textwidth,clip=,angle=-90]{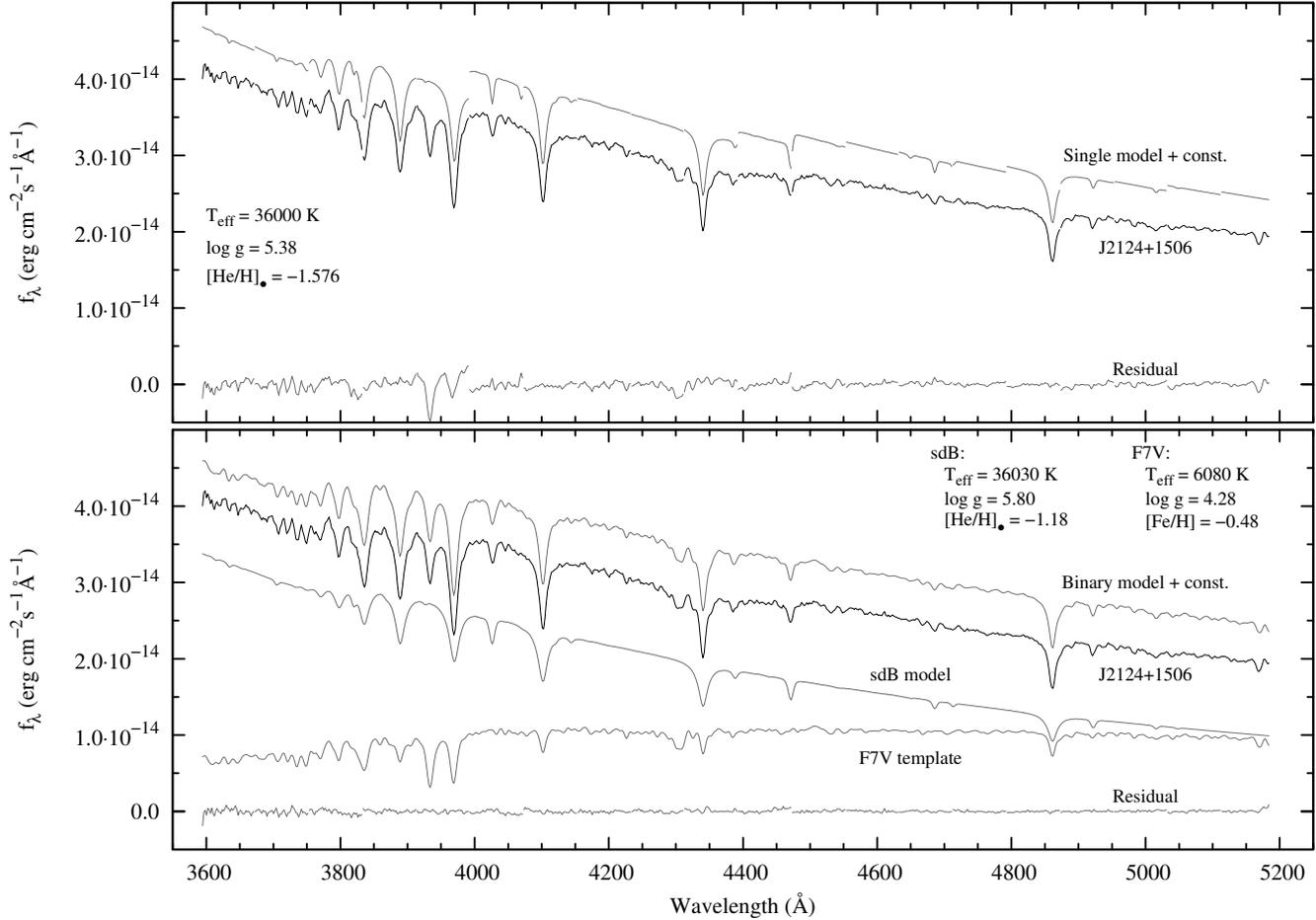}
\caption{\small 
 Binary decomposition of J2124+1506. The single model (top panel) can
 not reproduce the composite spectrum, 
 the G-band ($\sim$$4300$ \AA), Ca\,{\sc ii} H\&K (3933 and 3968 \AA) and 
 Mg\,{\sc i} (5183 \AA) lines are
 clearly in the residual. The jagged model also indicates that the
 theoretical continuum is
 inconsistent with the observation. When the spectrum is modelled with an
 additional cool MS component (bottom), 
 all major spectral features are reproduced well. The
 subdwarf temperature in this case was adjusted mildly, while there was
 significant change in the surface
 gravity and He abundance. A CNO abundance could not be measured in the spectrum.
\label{Fig:2124}}
\end{center}
\end{figure*}

We consider that our semi-empirical approach in
characterizing the cool star is not just faster, but 
less ambiguous than working with synthetic spectra for both components. 
To check if our interpolation returns real stellar spectra, 
the final templates were always
compared to those calculated with the same parameters by 
{\small MILES} 
Webtools\footnote{\url{http://www.iac.es/proyecto/miles/pages/webtools.php}}
\citep{vazdekis03}.
Companion spectral types were determined using the {\small MILES}
classifications.
We show an example for our fit and spectral decomposition of 
the sdB-F7V binary J2124+1506 in Figure \ref{Fig:2124}.

To resolve double-lined close binaries of similar luminosities, individual 
radial velocities would be helpful. However, such a study would require
high-resolution
spectra and templates or calculated spectra for both components. 
A productive
method to resolve these binaries would be to obtain both ultraviolet and optical
spectra.

\subsection{Spectral fitting}

We used the same initial model with: $T_{\rm eff}=$ 40\,000
K, 
$\log g=$ 5.6 cm s$^{-2}$, $[{\rm He/H}]_\bullet= -1$ and 
$[{\rm (C,N,O)/H}]_\bullet= -2$
to fit our observations. Maximum relative
changes per iterations were limited to $\pm$5 per cent for $T_{\rm eff}$, $\pm$2
per cent
for $\log g$ and $+$100 per cent, $-$50 per cent for all abundances in order to maintain a stable
{\small TLUSTY} convergence. 
We fitted only fluxed spectra with $80$ \AA\ segments (each having six to 16
resolution elements depending on the resolution) using the entire spectral range.
In order to decrease model atmosphere calculation time we used 30 depth
points. 
Our tests showed this simplification affects model convergence
before it would significantly change the emergent flux at low resolution. 
Model atmospheres
were calculated in non-LTE radiative equilibrium without convection. 
Detailed
profiles of H and He lines were included, but rotational broadening was
not. 
A high rotational velocity is not typical for subdwarf stars.
Fitting of $\sim$200 observed spectra took $\sim$800 h with six processors (15.2
GHz total) and needed the calculation of about 25\,000 model atmospheres. 

\noindent All fits are available in our online 
catalogue\footnote{\url{http://pleione.asu.cas.cz/~nemeth/work/galex/}}.

\subsubsection{Known issues}

The piecewise normalization procedure implemented in {\small XTGRID} can
reduce the effects of continuum inconsistencies between the theoretical and
observed spectra as well as the effect of interstellar reddening. 
However,
the method creates discontinuities in the fit residuals of stars that suffer
from large reddening. 
Such fractures in the residuals may divert the
fitting procedure and introduce systematic shifts in the derived parameters. 
Therefore, we performed extensive tests to discover if a wavelength dependent
extinction correction would change atmospheric parameters significantly.
We included the extinction function from 
\citet{cardelli89} in {\small XTGRID}, and implemented an iterative reddening
determination in the program. 
After recalculating some heavily reddened spectra, we did not find
systematic differences. 
Our quoted error bars are larger than the effects of our normalisation on
the parameters.
We estimated the
interstellar extinction from the final fits and listed the associated 
$E$($B-V$) colour excesses in Table \ref{Tab:1} and \ref{Tab:6}. 
These values are consistent with the dust
maps of \citet{schlegel98}
that provide an upper limit to the extinction coefficient toward each target.

Our binary decomposition 
method tends to attribute interstellar absorption lines along with other spectral lines
to the template and fits the primary in this environment. 
This can cause
degeneracy for binaries with similar components, like in J0716+2319. 
A single star fit provided a subluminous
star with $T_{\rm eff}=12940\pm300$ K and $\log g=5.34^{+0.01}_{-1.43}$ and 
slightly subsolar abundances.
The apparent Ca and Mg lines, the H line profiles, the very asymmetric error in 
gravity and the infrared excess suggested the
presence of a companion. 
Binary decomposition provided a much better fit with a B-A1V pair
with $T_{\rm eff}=11140\pm180$ K, $\log g=4.39^{+0.06}_{-0.22}$ primary
and $T_{\rm eff}=9310$ K, $\log g=3.67$ secondary.

Bright companions, like A stars, outshine sdB stars in the optical. 
Atmospheric parameters of 
relatively cooler subdwarfs in such binaries can be measured only with large
uncertainty. 
In J2349+4119 the $T_{\rm eff}=7940$ K,
$\log g=3.44$, A4V companion contributes 87 per cent of light in the B photometric band.
The determination of $T_{\rm eff}=11730$ K and $\log g=5.8$ for the primary is less
certain. 
Radial velocity measurements would help finding the true nature of this
binary. 

In unresolved composite spectra {\small XTGRID} can attribute subdwarf metal
lines to the lines of the companion, leading to incorrect CNO abundances. 
While C and/or N is relatively abundant compared to O and show lines in the
spectrum, this affects the O abundance determination in the first place. 
Companion $G$-band absorption overlaps with the strongest O\,{\sc ii} lines
($\lambda4273-\lambda4322$ \AA) of cool subdwarfs in composite spectra binaries. 
These lines are important in the O abundance determination 
and their attribution to secondary features 
can, in some case, lead to an underestimated O abundance of subdwarfs. 
Contrarily, {\small XTGRID} might
overestimate the O abundance of sdO stars in unresolved composite spectra
binaries.
Our current O abundances are only preliminary determinations, a complete study
requires higher-dispersion spectra.

\begin{figure*}
 \includegraphics[width=0.6\linewidth,clip=,angle=-90]{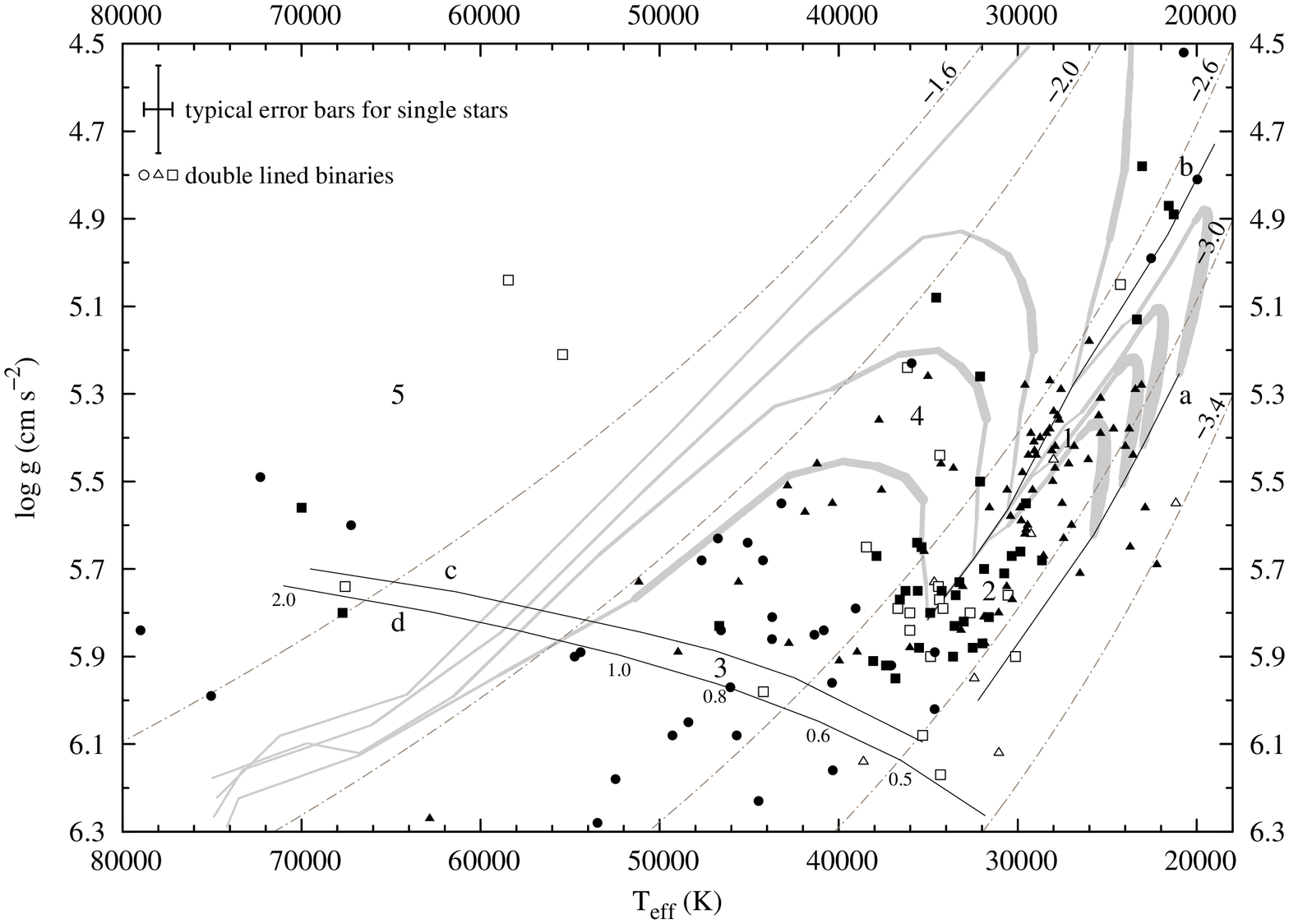}
 \caption{{\small
$T_{\rm eff}$$-$$\log g$ diagram.
 The ZAEHB and TAEHB \citep{dorman93} are
 marked with "a" and "b",
 respectively. The HeMS is taken from \citet{paczynski71a} and
 \citet{divine65}
 and are marked with labels "c" and "d", respectively. Stellar masses are
 marked along the HeMS in M$_{\odot}$.
 The grey
 lines are theoretical evolutionary tracks from \citet{dorman93}
 for stellar masses from top to bottom: 0.480, 0.475, 0.473 and 0.471
 M$_{\odot}$. Line
 widths are proportional to evolutionary time-scales. 
 Typical error bars are
 shown in the upper left corner. He-rich $([{\rm He/H}]_\bullet>-1)$ stars are
 indicated with filled circles, He-poor $([{\rm He/H}]_\bullet<-2.2)$ stars
 with filled triangles and
 stars with $-2.2\le[{\rm He/H}]_\bullet\le-1$ are with filled squares. 
 Subdwarfs in
 resolved composite spectra binaries are shown with open symbols following
 the same classification. 
 Iso-luminosity curves for 
 ${\rm L}/{\rm L}_{\rm edd}=-1.6, -2.0, -2.6, -3.0 {\rm\ and} -3.4$ 
 are shown with dash-dotted lines. These are characteristic values in the
 luminosity distribution.}
\label{Fig:hrd}}
\end{figure*}

The puzzle of the N\,{\sc iii}/C\,{\sc iv} $\lambda$4640--$\lambda$4665 \AA\ blend still
challenges the present spectral modelling. 
We observed the blend in about half of the hot He-sdO stars over
a wide range of temperature and our fits do not reproduce it satisfactorily.
We revised our line list and updated the atomic data for the C\,{\sc iv}
$\lambda4658.3$ \AA\ multiplet with lines between the
1s$^2$5f(2F$^\circ$)--1s$^2$6g(2G), 
1s$^2$5g(2G)--1s$^2$6h(2H$^\circ$) and
1s$^2$5g(2G)--1s$^2$6f(2F$^\circ$) levels 
from the T\"{u}bingen NLTE Model-Atmosphere 
Package (TMAP; \citealt{werner99}) available at the
German Astrophysical Virtual Observatory
page\footnote{\url{http://astro.uni-tuebingen.de/~TMAD/}}. We noted a line
strength inconsistency of the C\,{\sc iii} $\lambda4070$ \AA\ multiplet with the
C\,{\sc iv}
$\lambda4658.2$ \AA\ multiplet as well as with the 
C\,{\sc iv} $\lambda5801.3$ and
$\lambda5812$ lines. 
Our models show too strong C\,{\sc iii} and too weak C\,{\sc iv} lines
compared to the observations, indicating an
underestimated temperature or an overestimated gravity. 
The effects of a change in gravity are low on metal lines, hence only a change
in temperature could provide better fits.  
However, the ionisation balance shows
that the C\,{\sc iv} ion is dominant over C\,{\sc iii} above $\sim$42\,000 K, 
hence an increase in the temperature would decrease the 
C\,{\sc iii} line strength, but would not
increase the C\,{\sc iv} opacity. 
Such changes in temperature or gravity would be inconsistent
with the H and He line strengths as well. 
The high C and N abundance of these stars
implies a high probability of heavier elements, in particular the iron-peak
elements, which can be accounted for the missing opacities. 
\citet{haas96},
\citet{werner98} and recently \citet{otoole06a} found high abundances of heavy elements in
hot subdwarf stars. 
A review of the atomic data input is necessary and deserves a separate study.

\section{Properties of hot subdwarfs}\label{Sec:properties}

Our sample is large enough and appropriate 
for revisiting the distribution of stars in the $T_{\rm
eff}-\log g$ and $T_{\rm eff}-{\rm He}$ planes as well as to look
for possible correlations in the He and CNO abundances. 
Such diagrams are very important tools in tracking subdwarf evolution and pulsational
modes,
because with the help of a
model atmosphere program, stellar temperature, gravity and element abundances
can be derived without any assumptions on distances or stellar luminosities. 
A comparison of the distribution of stars in these diagrams 
is also useful to test our fitting method. 

Our spectral classification follows
the simple scheme of Paper I: Subdwarfs with $T_{\rm eff}<40\,000$ K 
showing dominant H lines and weak He\,{\sc i} are classified as
sdBs; with $T_{\rm eff}>40\,000$ K and weak lines of He\,{\sc ii} are sdOs. 
Stars with dominant He\,{\sc i} lines are He-sdBs, while those with dominant
He\,{\sc ii} lines
are classified as He-sdOs. 
Stars with $T_{\rm eff}<20\,000$ K and $\log g<5$ are listed
as BHB stars and with $\log g<4.5$ are classified as MS B stars. 
Using this categorization, out of the 180 stars we found 166 subdwarfs, 
124 of which are classified
as sdB and 42 as sdO stars. The sdB/sdO number ratio of $\sim$3 is
the same figure as previously determined \citep{heber09}. 
We found five He-sdB stars
(or $\sim$4 per cent) among
the 124 sdB and 26 He-sdO (or $\sim$62 per cent) out of the 42 sdO stars. 
In the 29 resolved spectroscopic binaries only four (namely: J0710+2333,
J1602+0725, J2020+0704 and J2038-2657) have sdO primaries and two binaries
(J0047+0337 and J2331+2815) are in the sdB/sdO transition region. 
No He-sdO subdwarfs were found in composite spectra
binaries. 
One system (J0716+2319) is a MS binary and one (J2349+4119) is a possible 
low mass pre-WD--MS binary. 
The remaining 21 composites are classified as sdB--MS binaries.

\subsection{The $T_{\rm eff}-\log g$ plane}
\label{Sec:tefflogg}

The properties of our hot subdwarfs in the
temperature-gravity plane are shown in Figure \ref{Fig:hrd}.
The distribution of sdB stars follow the theoretical EHB and He-sdO stars
are scattered near the He main-sequence (HeMS),
with some stars along post-EHB evolutionary tracks. The post-EHB tracks
of \citet{dorman93} for subdwarf masses from top to bottom: 0.480, 0.475, 
0.473 and 0.471 ${\rm
M}_{\odot}$ are shown in grey. The zero-age EHB (ZAEHB) and terminal-age EHB
(TAEHB) are marked with "a" 
and "b", respectively. 
The HeMS is taken from \citet{paczynski71a} and \citet{divine65}
and are labelled with "c" and "d", respectively. 
Stars with 
$[{\rm He/H}]_\bullet>-1$ are marked with filled circles; 
most of these spread along the HeMS. 
Stars with $[{\rm He/H}]_\bullet<-2.2$ are indicated with filled triangles 
and gather at lower temperatures and gravities than stars with 
$2.2\le[{\rm He/H}]_\bullet\le-1$ that are shown with filled
rectangles and crowd around $T_{\rm eff}\approx35\,000$ K and $\log
g=5.8$. 
Resolved composite spectra binaries are marked with open symbols
using the same scheme. Dash-dotted lines show iso-luminosity fractions, where
the luminosity is expressed as the fraction of the Eddington-luminosity
calculated from the effective temperature and surface gravity:
\begin{equation}
\log \frac{{\rm L}}{{\rm L}_{\rm Edd}} = 4\times\log T_{\rm eff}-\log
g -
15.118.
\end{equation}
This distance independent quantity statistically separate stars. 
Assuming a constant stellar mass for sdB
stars, these curves are also proportional to their absolute luminosities. 

Figure \ref{Fig:hrd} reveals five distinct regions in the 
$T_{\rm eff}-\log
g$ diagram. 
The cooler, He-poor sdB stars (no. 1) are found around 28\,000 K and $\log g=5.45$. 
These are potential
slow pulsators ($P\approx40-170$ min). 
The hotter sdB stars (no. 2) near 33\,500 K and $\log g=5.8$ are on 
average 10 times
more He abundant and possible rapid pulsators ($P\approx 2-6$ min). 
The hot and He-rich sdO stars (no. 3) are found mostly between 40\,000 and 
55\,000 K near $\log g=5.9\pm0.4$. 
Although numerous He-sdO stars are above the theoretical HeMS, as
theoretical models predict, 
an asymmetric scatter toward high gravities is apparent in our data and 
noted in the literature \citep{ostensen09}.
He-weak sdO stars (no. 4) are located above both the
EHB and HeMS.
Finally, the least defined population: possible progenitors of 
low-mass WDs, pre-WDs, post-Asymptotic Giant Branch (post-AGB) stars and
central stars of planetary nebulae (CSPN), that might cross this 
region evolving towards their cooling tracks (no. 5) are also
distinguished. 
The lack of stars along sections of fast evolution, in particular between
$-2.0<{\rm L}/{{\rm L}_{\rm Edd}}<-1.6$ 
is also notable in Figure \ref{Fig:hrd}.
Furthermore, it is immediately seen in the
figure that He abundance, surface temperature and gravity are 
correlated.

While sdB stars are assumed to evolve more or less along the indicated
evolutional tracks, He-sdO stars in lack of a H envelope 
are expected to leave the HeMS towards higher
temperatures and gravities after He exhaustion.
However, this fast evolution cannot explain the observed clustering of 
high gravity He-sdO stars.
\citet{hu08} 
found that stars with
M$_{\rm ZAMS}>2$ M$_\odot$ 
can ignite
core He-burning in non-degenerate conditions and
after the EHB such stars can develop a He-burning shell and a different
structure than canonical subdwarfs. 
An observational property of these stars would be their clustering below the
theoretical HeMS.
From an observational point of view, incomplete line broadening parameters of He lines can also be 
responsible for overestimating gravity in such stars. 
More observations are needed to better characterize these high gravity He-sdO
stars. 

A known shift in sdB temperatures by about +2000 K compared to the
blue edge of the theoretical EHB band 
\citep{charpinet07} is also noticeable in Figure \ref{Fig:hrd}.
Similar shifts have been observed in previous analyses with various
modelling and fitting techniques in volume limited samples. 
Hence, this discrepancy deserves attention from the theoretical side. 
\citet{jeffery06} 
showed that including more iron-peak elements in the
models would shift the theoretical EHB towards the observations. 

\begin{figure}
\centering
 \includegraphics[width=\linewidth,clip=,angle=0]{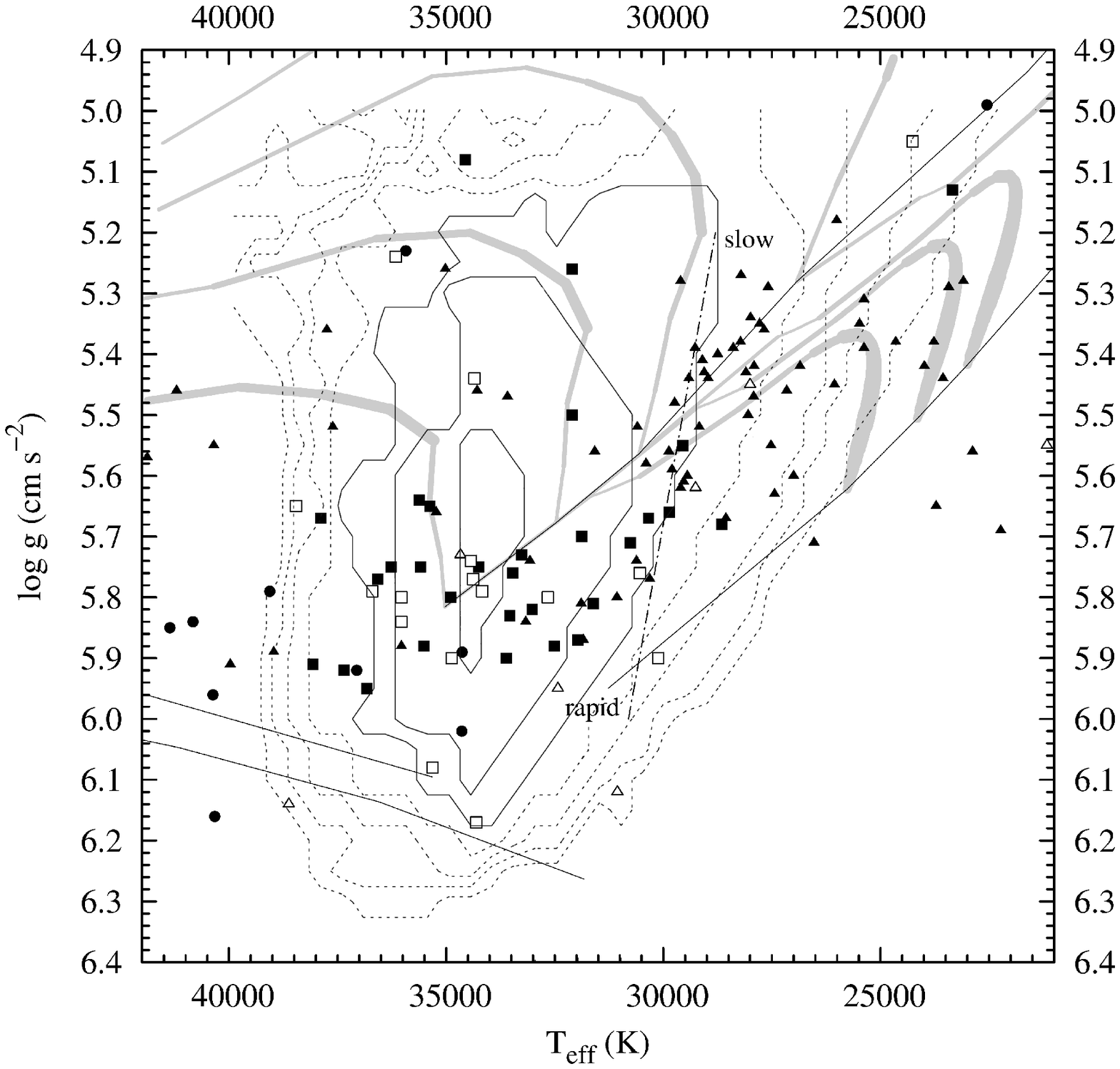}
 \caption{{\small
 Subdwarfs in the theoretical instability region. Contours were 
 rendered from \citet{charpinet11} and are overplotted on our data.
 Rapid ({\it p}-mode) pulsators are concentrated
 within the subdwarf instability region \citep{charpinet01},
 indicated by the contours, driving is most
 efficient in the three highest contours (solid lines). The dash-dotted line
 separates slow and rapid pulsators. This boundary was found observationally 
 based on an independent dataset \citep{charpinet10}. }
\label{Fig:instabil}}
\end{figure}

The contours in Figure \ref{Fig:instabil} mark 
the theoretical pulsational instability region by \citet{charpinet01} 
and are overplotted on our sample. 
The pulsations are driven by the iron-peak
opacity bump. 
\citet{charpinet09} and \citet{hu11} calculated
models with equilibrium Fe abundance between radiative
levitation and gravitational settling.
The contours in Figure \ref{Fig:instabil} were reproduced from \citet{charpinet11} and 
show the number of
pulsating $\ell$ = 0, {\it p}-modes. 
The outermost contour 
corresponds to a single mode and the
highest in the centre to seven. 
Driving is most efficient in the three innermost contours.
Possible rapid 
({\it p}-mode) pulsators
are located around 33\,500 K and $\log g=5.7$ while presumed slow ({\it
g}-mode) pulsators are around 28\,000 K and $\log g=5.4$. 
The boundary that separates slow and rapid pulsators is marked with the
dash-dotted
line and was found independently \citep{charpinet10}. 
We refer to these
groups as possible pulsators, although variable and non-variable stars
co-exist in their observed parameter space. 
\citet{ostensen10} 
found a $\sim$10 per cent incidence rate for rapid pulsators in
their survey for pulsating hot subdwarfs with the Nordic Optical
Telescope. 
Slow pulsators show a larger incidence rate, reaching $\sim$75 per cent according
to \cite{green03}, 
which was recently confirmed by the Kepler sdB sample
\citep{ostensen11}.
In between these two
groups are the hybrid (or mixed-mode) pulsators that show both slow and rapid modes.   
Based on the distribution of stars in Figure \ref{Fig:instabil} many of our
subdwarfs are
possible pulsating candidates and a deep asteroseismic follow-up would be
reasonable. 
Based on the locations of known pulsators we sorted out and flagged
candidates for pulsation studies in Table \ref{Tab:1}.  
We would like to note that the {\sl GALEX} sample presented here is 
a consistent set, the relative position of the stars to each
other are free of major systematic shifts. 

\begin{figure*}
\begin{center}
 \includegraphics[width=0.45\linewidth,clip=,angle=-90]{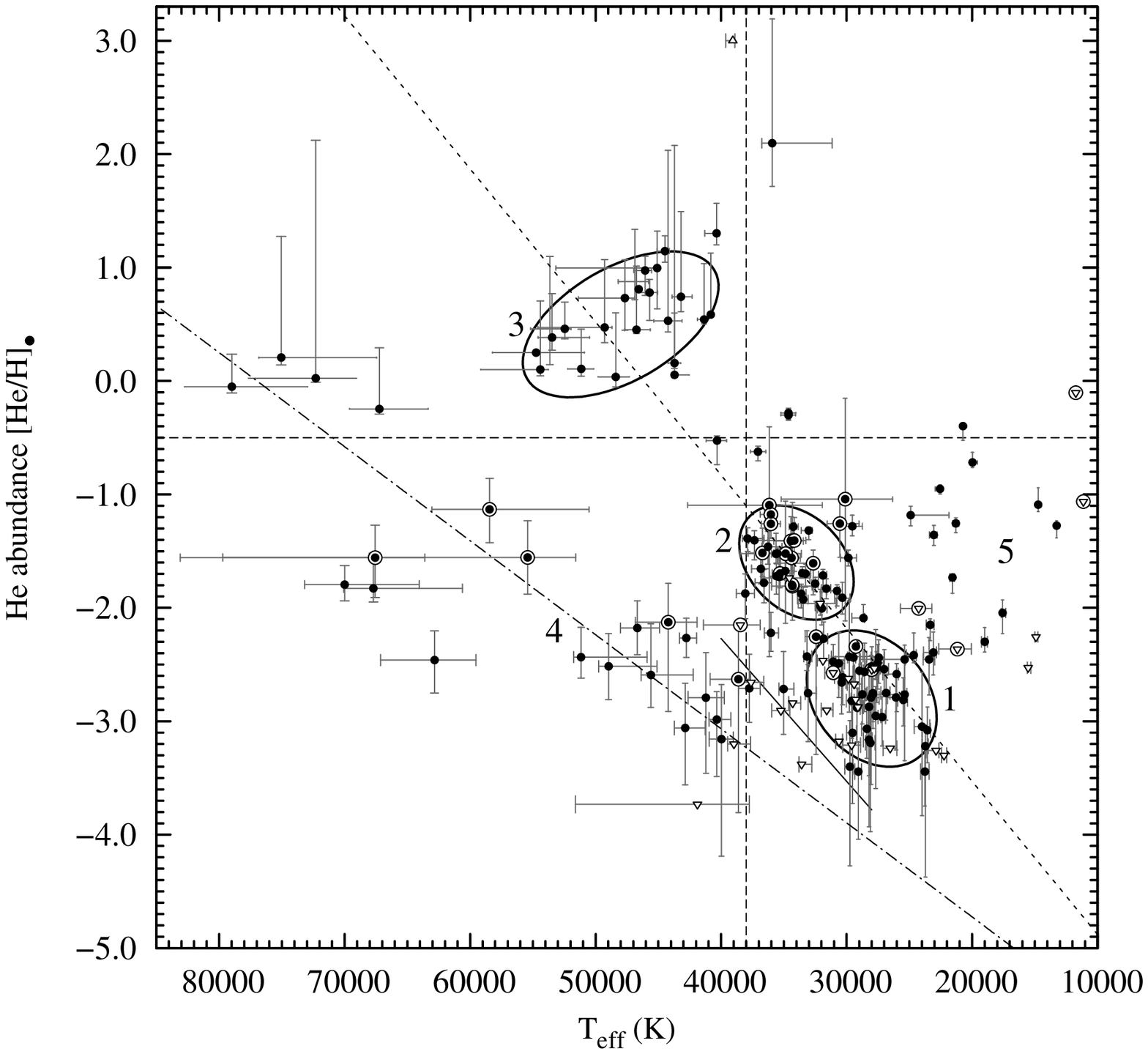}
 \includegraphics[width=0.45\linewidth,clip=,angle=-90]{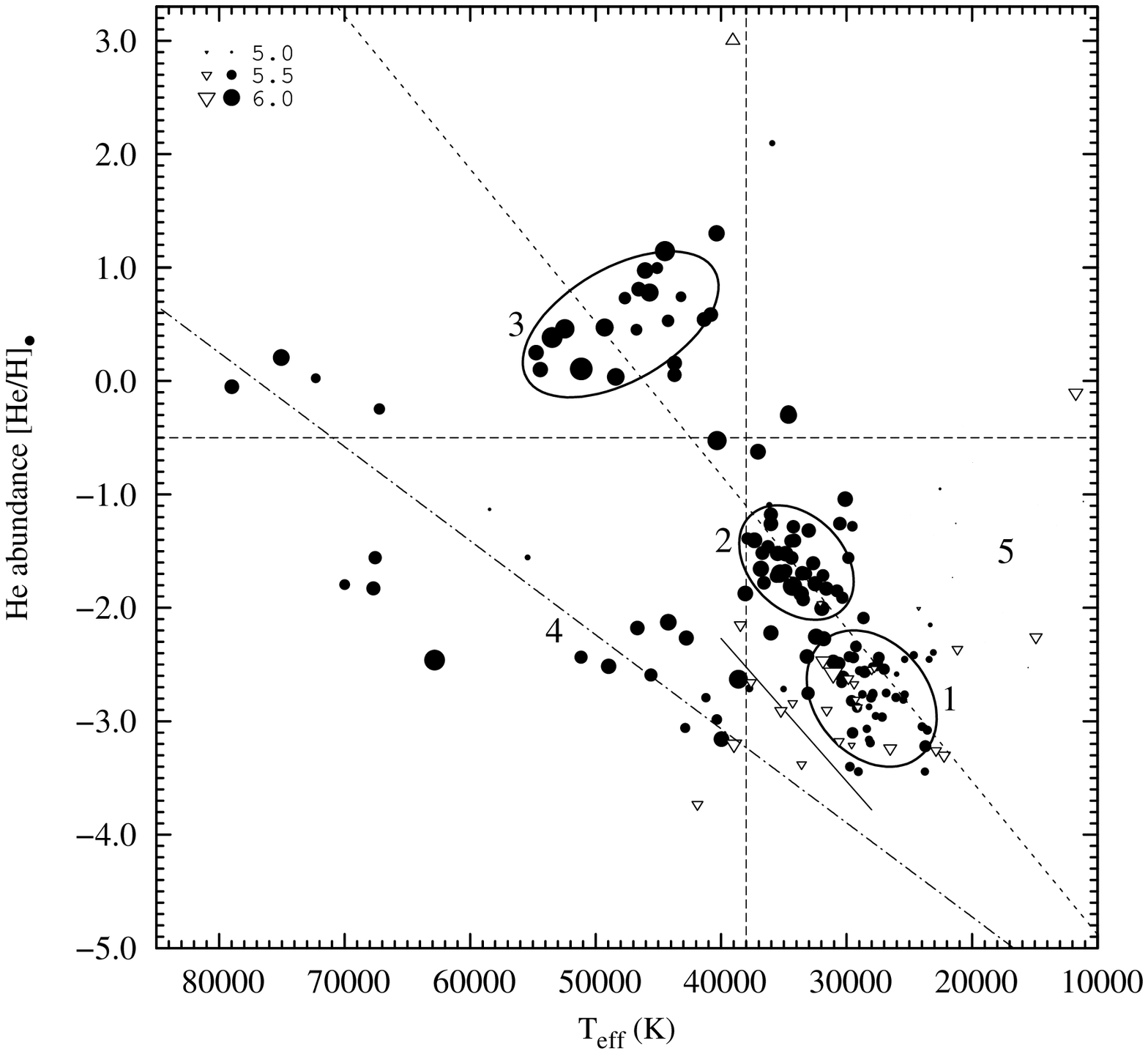}
 \caption{{\small
{\it Left}: He abundance versus effective temperature. Two trends of increasing
 abundance with effective temperature can be seen. Also remarkable are the
 clumping of possible slow (\#1, near $28\,000$ K and $[{\rm He/H}]_\bullet=-1.5$) and rapid
 sdB pulsators (\#2,
 near $33\,500$ K and $[{\rm He/H}]_\bullet=-2.7$) and the hot He-rich sdOs (\#3). The sequence of
 He-weak
 subdwarfs (\#4 and dash-dotted line) follow a similar correlation with effective temperature
 as more He rich stars (short dashed). The short full line is the best fit from
 \citet{edelmann03} for the He-weak sequence. 
 Suspected progenitors of low-mass WDs and BHB stars
 are also possible in this region of the HRD (\#5). 
The pure He atmosphere 
 of J0851-1712 is an extreme case, we determined only a lower limit of its He
 abundance. Upper limits
 are indicated with open down triangles and binary stars are encircled. {\it Right:} 
 Same data without error bars and point sizes are proportional to surface
 gravities (point size $=\log g-4.8$) as shown in the upper left corner.
 Indicated regions are intended to guide the eye and are not
 derived from the data. Interesting to note the distribution of composite
 spectra binaries as well as surface gravity correlations.}
\label{Fig:het}}
\end{center} 
\end{figure*}

\subsection{The $T_{\rm eff}-{\rm He}$ diagram}

\begin{figure*}
\centering
 \includegraphics[width=0.6\linewidth,clip=,angle=-90]{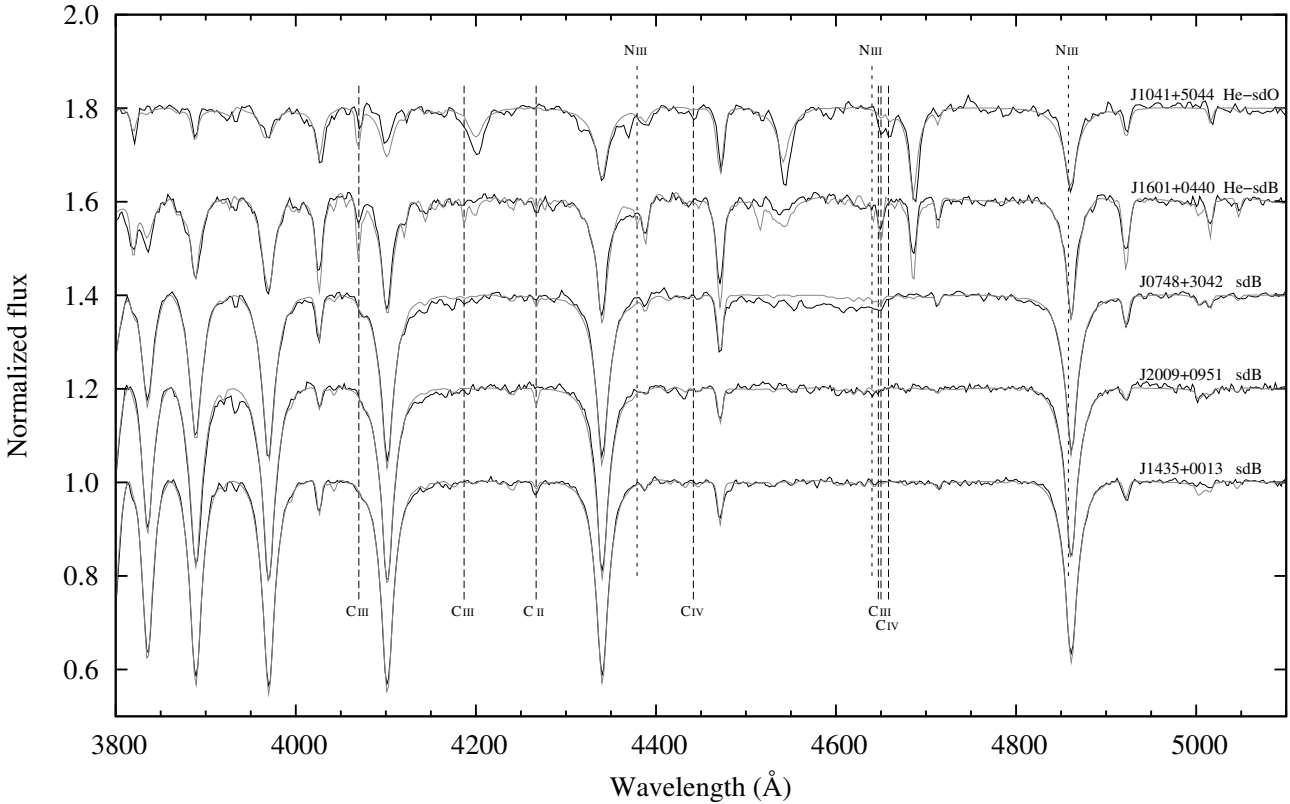}
 \caption{Sample fits (grey lines) of five subdwarf spectra (black lines)
 with increasing effective temperatures from bottom to top. The strongest C
 and N lines are marked. \label{Fig:spectra}}
\end{figure*}

The He abundance in hot subdwarf stars is another fundamental parameter 
and must be investigated along with surface temperature and gravity. 
The panels of Figure \ref{Fig:het} show the distribution of stars in the 
$T_{\rm eff}-{\rm He}$
diagram and their correlations with surface gravities.
The five groups introduced in Section \ref{Sec:tefflogg} are
separated in the $T_{\rm eff}-{\rm He}$ plane as well. 
The He-poor (no. 1) and He-rich (no. 2) sdB
stars, the sequences of He-sdO (no. 3) and He-weak sdO (no. 4)
stars are marked in Figure \ref{Fig:het} as well as BHB stars and possible WD
progenitors (no. 5). 
The two $T_{\rm eff}-{\rm He}$ correlation sequences first described by
\citet{edelmann03} are also remarkable.
In general, the He abundance increases with temperature.
Our He detection limit is at
$[{\rm He/H}]_\bullet\approx-3$ and does not allow to 
trace the two helium sequences
deeper, although some data suggest a
continuation of the He-weak sequence. 
Similarly large and homogeneously modelled samples like the Hamburg Quasar Survey
(HS, \citealt{edelmann03}), Supernova Ia Progenitor surveY (SPY,
\citealt{lisker05}, \citealt{stroer05}) 
and Sloan
Digital Sky Survey (SDSS, \citealt{hirsch08}) confirm this observation
\citep{otoole08}. In
Figure \ref{Fig:het} we plot the best fit line (short dashed) for sdB stars 
from the HS sample
\citep{edelmann03} for the He-rich sequence:
\begin{equation}
[\rm {He/H}]_\bullet=
-3.53+1.35\left(\frac{T_{\rm eff}}{10^4 {\rm K}}-2.00\right).
\end{equation}
This independent trend fits our sdB data as well, in particular for He-rich
sdBs and passes through the region of He-sdO stars. We also plot their
regression (full line) for the He-weak sequence. However, we found a
different trend (dash-dotted line) for these stars, possibly because the
temperature range of the He-weak sequence was under-represented in the HS
survey ($T_{\rm eff}<40\,000$ K). Separating He-weak stars at 
\begin{equation}
[{\rm He/H}]_\bullet<1.08\times10^{-4} T_{\rm eff}-6.64,
\end{equation}
and fitting them in a similar manner we got:
\begin{equation}
[\rm {He/H}]_\bullet= 
-4.26+0.69\left(\frac{T_{\rm eff}}{10^4 {\rm K}}-2.00\right).
\end{equation}
Although, it can be seen from these fits that the distribution of subdwarf 
stars is
much more complex and cannot be satisfactorily described with simple linear
trends, hence we omit their significance test.
Multiple breaks can be found in the He abundance distribution. There is
a clear cut in temperature (vertical dashed line) in Figure \ref{Fig:het}, 
separating sdB and sdO stars 
somewhere between 36\,000 and 40\,000 K depending on
the He abundance. To the right of this line are the two, relatively well 
defined
groups of sdB stars, while to the left are the hot, either He-weak or 
He-rich sdO stars.
A He abundance split is also remarkable at the line 
$[{\rm He/H}]_\bullet=-0.5\pm0.5$
(horizontal dashed line), 
in particular over 40\,000 K, where no stars can be found. 
Such a gap can be seen in the SPY data as well.
Observationally this corresponds to a region where stars are expected to either never enter or pass 
through quickly during their evolution.
The major formation theories in Section 
\ref{Sec:sdBd} also predict a lack of stars in this region.
Similarly to the $T_{\rm eff}-{\log g}$ distribution, possible rapid and slow
sdB variable stars are separated according to their He abundance. 
Rapid pulsator candidates with $[{\rm He/H}]_\bullet\approx-1.5$ 
have, on average, an order of magnitude higher He abundance
then their slow pulsating counterparts with 
$[{\rm He/H}]_\bullet\approx-2.7$. 
A region of photospheric mixing and He enrichment near
$T_{\rm eff}\approx38\,000$ K at $\log g\approx6$ was 
predicted by \citet{groth85} and recently
reviewed by \citet{bertolami08}. 
Our data show some abundance extremes in this
region with $[\rm {He/H}]_\bullet>1$ for He-sdO stars, culminating at 
the location of 
J0851-1712 in which we could determine only a lower limit of the He
abundance.
Toward higher temperatures from the sdB/sdO
border the 
He abundance in He-sdO stars shows a decrease with temperature.
The He abundance in these stars
shows a correlation with mass
along the HeMS.

Below the sdB/sdO border,
a few He-sdB stars in our data show a He overabundance, but in general we
can conclude that the
He abundance is proportional to the effective temperature in sdB stars, 
and no stars can be found with
similarly high He abundances like in He-sdO stars. 
In the proximity of the
sdB/sdO transition both He-rich and He-poor sdB and sdO stars can be found. 
The case of He-weak sdO stars is simpler.
They show a correlation with effective temperature similar to sdB stars 
and are connected to sdB stars at the sdB/sdO border. 
Their binary frequency and surface gravity diversity suggests that the 
He-weak sequence is a mixture of multiple populations. We discuss this point 
further in Section \ref{Sec:sdBd}.

While parameter errors of temperature and gravity show gradual increase to 
higher temperatures,
the error budget for the He abundance seems to be three-fold: 
sdB and BHB stars have well defined
atmospheric parameters with symmetric error bars. He-weak sdO stars
have significantly larger errors and show asymmetries in temperature. 
He-sdO stars show a rather symmetric temperature, but 
very asymmetric He abundance errors; our method tends to
stop at the lower limit. 
Probably because at such high temperatures and He abundances the
spectrum is He dominated and hydrogen lines are difficult to trace with
low-resolution spectroscopy. 
In such conditions He becomes the reference element and the
$\chi^2$ is less sensitive for the change of the shallow He lines.
Beyond the low resolution of our spectra, the incomplete chemical composition
of our models can also be blamed for the
larger parameter errors of He-sdO stars.

By comparing the panels in Figure \ref{Fig:het} further differences
between the two He sequences become apparent: there is a strong correlation between surface
temperature, gravity and He abundance of sdB and He-sdO stars, but this cannot be
said about He-weak sdO stars. 
Among these stars a much weaker correlation and scattered high
gravity stars can be found. 
Also notable are the low
temperature sdB and BHB stars that have a wide distribution in He abundance
and low surface gravities. 
Some of our subdwarfs show a low He
abundance at a relatively high gravity, these are possible progenitors of
low-mass WDs.

\subsection{CNO trends}

\begin{figure}
\centering
 \includegraphics[width=0.7\linewidth,clip=,angle=-90]{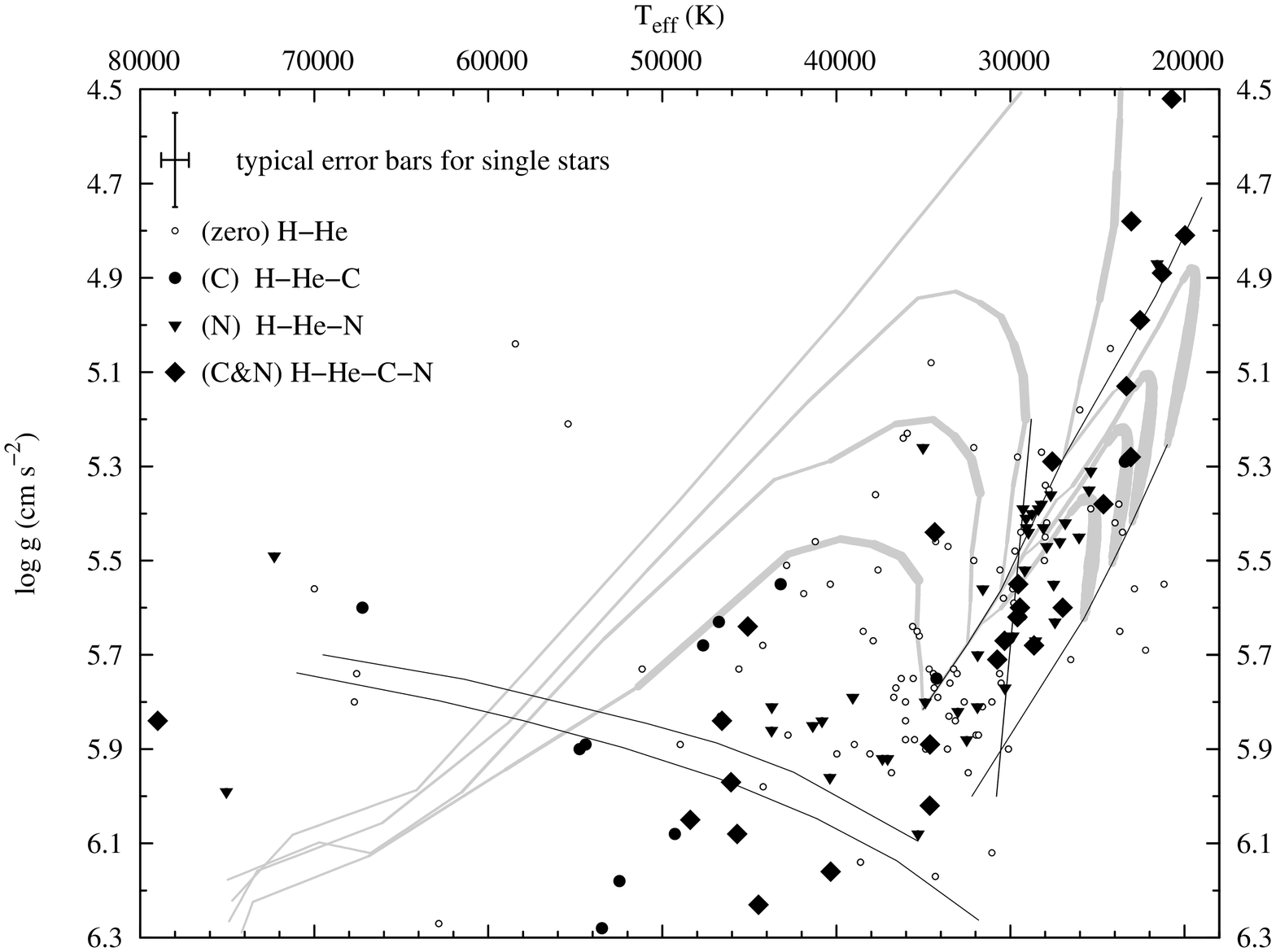}
 \includegraphics[width=0.7\linewidth,clip=,angle=-90]{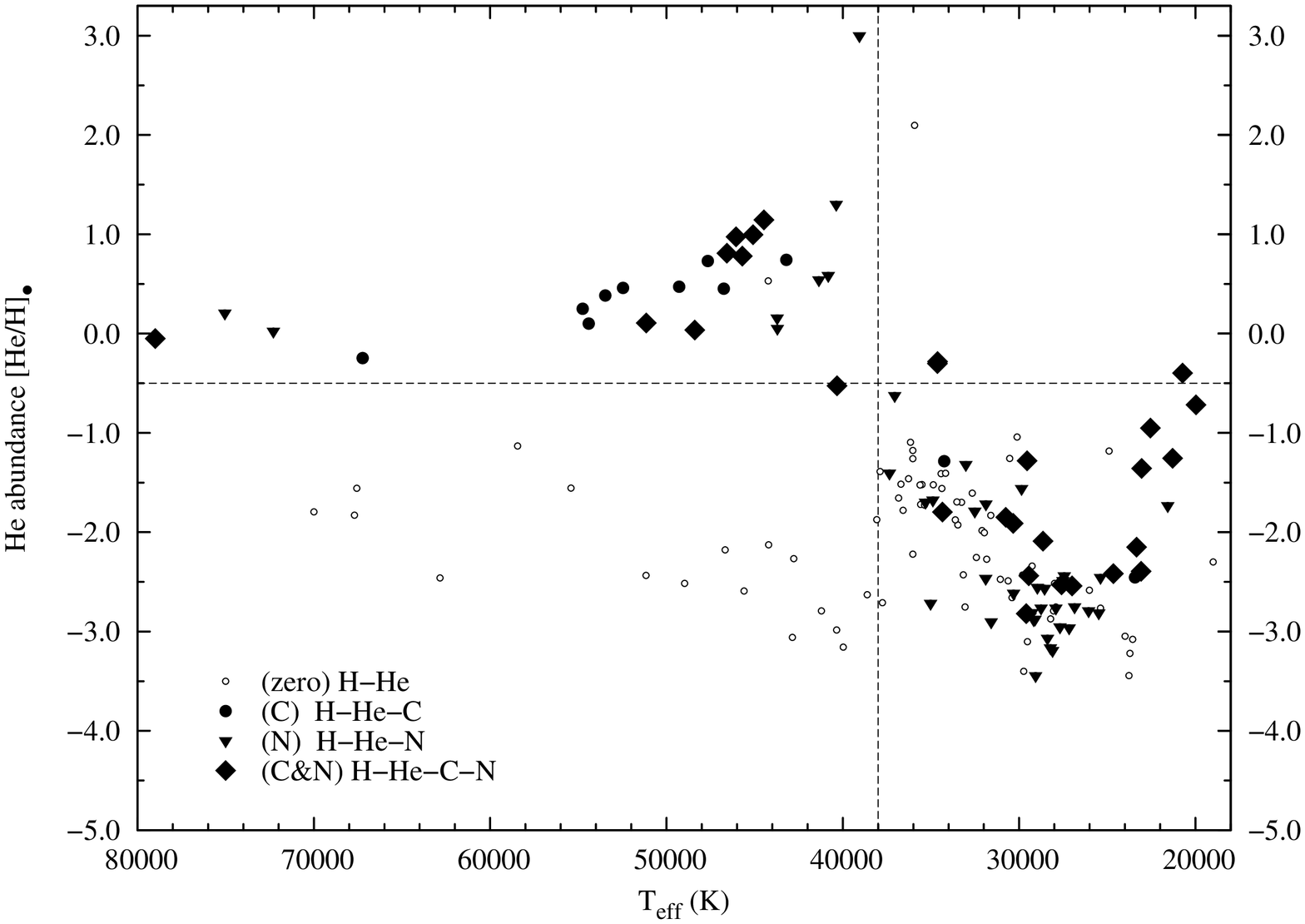}
 \caption{ Metal abundances in the $T_{\rm eff}-\log g$ and 
  $T_{\rm eff}-{\rm He}$ diagrams. \label{Fig:metal}}
\end{figure}

Metal abundance patterns are difficult to obtain from optical
spectra, because the vast majority of CNO lines are in the ultraviolet and
CNO lines in the optical are relatively insignificant. 
In spite of the
weakness of these lines, our analysis provided C and N abundances 
for a number of sdB and sdO stars.
Our metal abundance determination is
based on the global effects of a specific chemical composition on the entire spectrum
as well as on individual lines. 
We adopted this method 
mainly because in low-dispersion optical spectroscopy weak metal lines are heavily
blended and hard to identify, but, thanks to the flux blocking of UV lines
the overall shape of the energy distribution is also affected by the
chemical composition.
Sample fits for various subdwarf types are shown in Figure
\ref{Fig:spectra}. 
Following the C\&N 
analysis of sdO stars by \citet{stroeer07}, we summarise our
results for the {\sl GALEX} sample in Figure \ref{Fig:metal}. 
In the $T_{\rm eff}-\log g$ diagram the coolest EHB and BHB stars
($T_{\rm eff}<24\,000$ K) show
both C and N abundances. 
Cooler sdB stars between 24\,000 K and 30\,000 K show a
higher N abundance, but no C in general.
In a narrow range around 30\,000 K both C and N
were found in some stars. 
The majority of sdB stars over 30\,000 K show much
less diversity, in most of these stars only He was found similarly to post-EHB
stars.
He-sdO stars show a diversity of surface abundances, but a pattern can be
outlined. N-class stars are concentrated below $T_{\rm eff}<45\,000$ K. 
He-sdO stars with higher temperature and surface gravity
belong to the C-class, while a group of C\&N-class objects can be found 
in a narrow strip at $\sim$45\,000 K.
The C-class seems to belong exclusively to hotter He-sdO stars. 
Finally, the hottest sdO stars ($T_{\rm eff}>60\,000$ K) show
a N overabundance in their spectra.
The evolutionary status of these stars is an interesting question. 
They might belong to evolved He-sdO stars moving away from the HeMS as well as to He-weak sdO stars
moving towards the HeMS in the temperature-gravity diagram \citep{husfeld89}.

Based on the $T_{\rm eff}-{\rm He}$ graph, we can confirm the 
conclusions of \citet{stroeer07} that He-weak sdO stars do not show metal 
traces.
We found only a few stars around $T_{\rm eff}\approx40\,000$ K near
$[{\rm He/H}]_\bullet\approx-0.5$ where a small clumping was observed in the SPY
sample. 
However, in our data, the trend of decreasing He abundance with increasing
effective temperature
can be traced down to $[{\rm He/H}]_\bullet>0$ in contrast with
$[{\rm He/H}]_\bullet>1$ found by \citet{stroeer07}.

Figure \ref{Fig:metal} gives only a qualitative picture of the C and N distribution,
because these graphs do not reflect element abundances.
To further investigate the C and N abundance distributions and 
possible C--He, N--He and C--N abundance correlations we refer to 
the panels of Figure \ref{Fig:cnoabn_nl}, where 
only positive determinations are shown. In Appendix A we
repeat these figures supplemented with error bars and upper limits.

\begin{figure*}
\centering
\subfigure
  {\includegraphics[width=0.40\linewidth,clip=,bb=67 154 551 700,angle=-90]{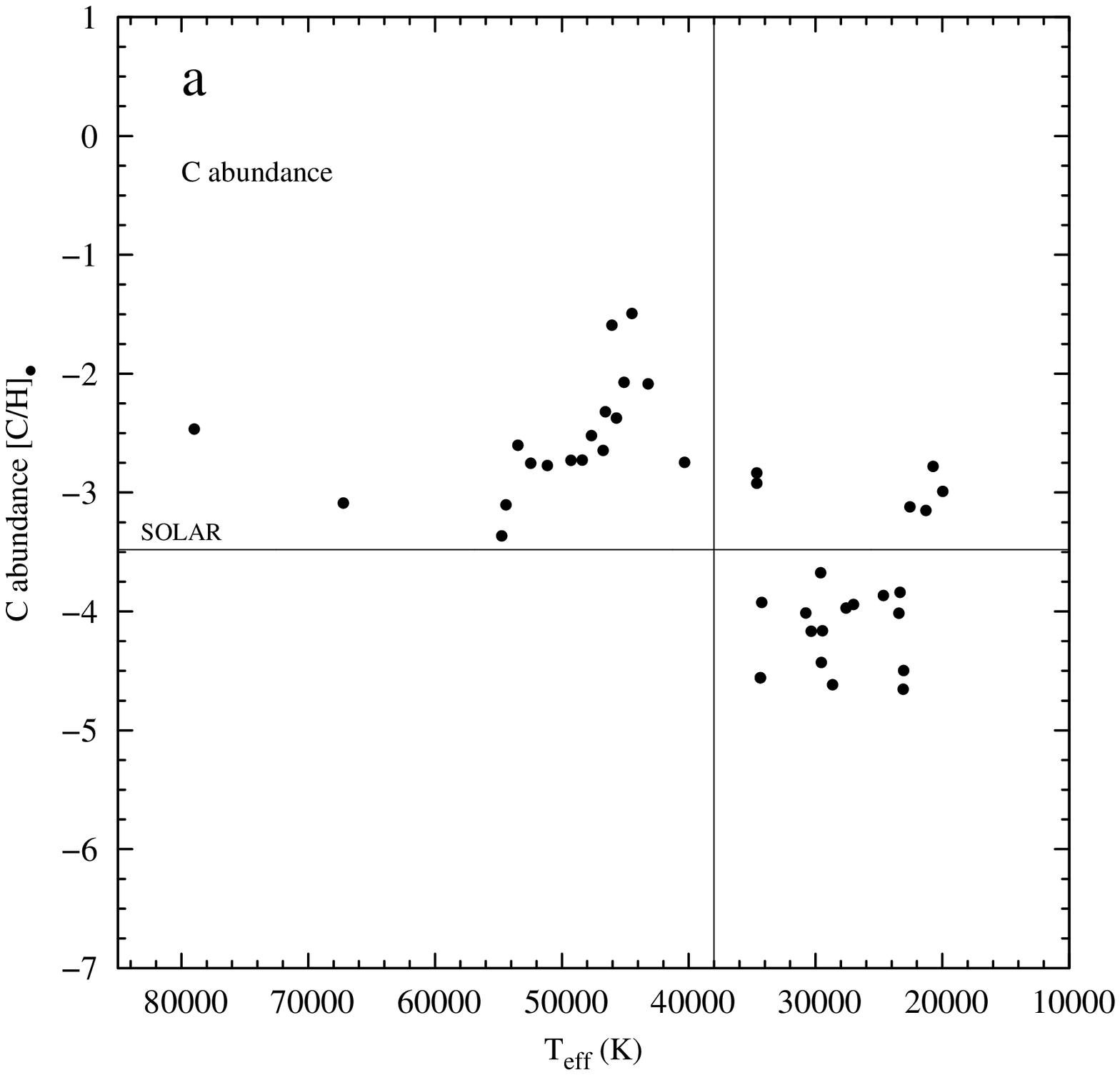}
     \label{fig:first_sub}    }
\subfigure 
{\includegraphics[width=0.40\linewidth,clip=,bb=67 154 551 700,angle=-90]{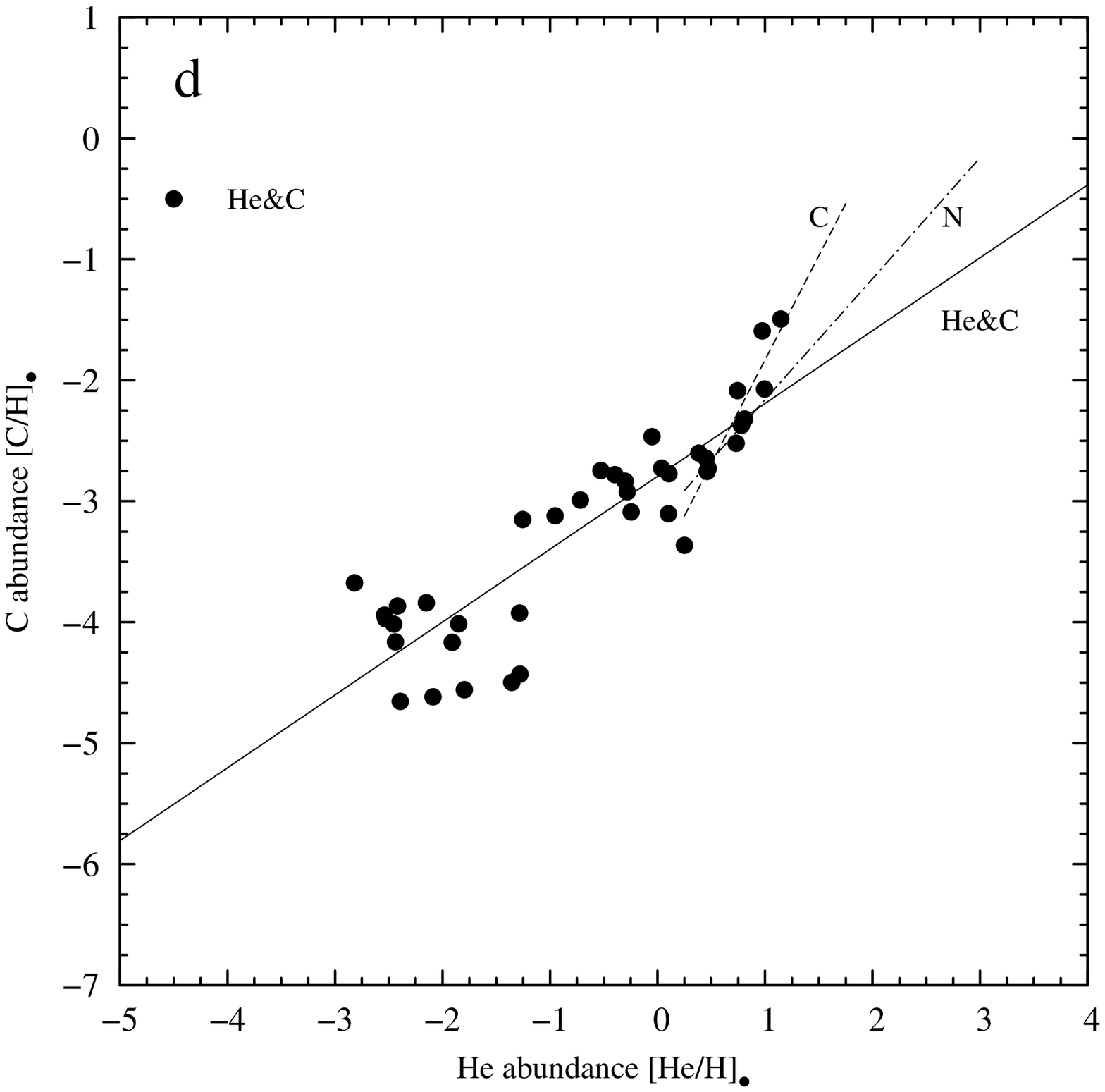}
      \label{fig:second_sub}  }
\\
\subfigure
{\includegraphics[width=0.40\linewidth,clip=,bb=67 154 551 700,angle=-90]{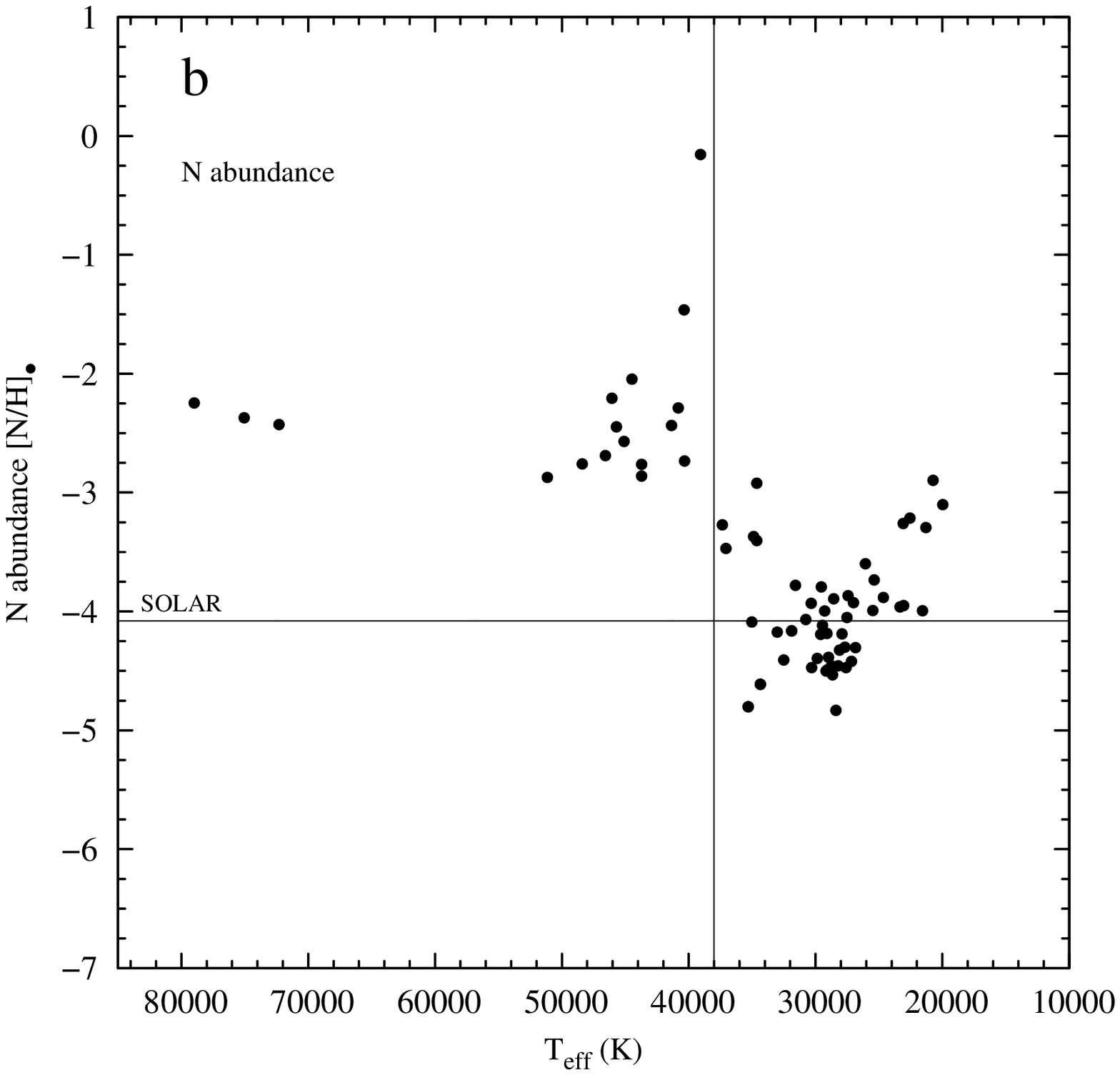}
      \label{fig:third_sub}    }
\subfigure
{\includegraphics[width=0.40\linewidth,clip=,bb=67 154 551 700,angle=-90]{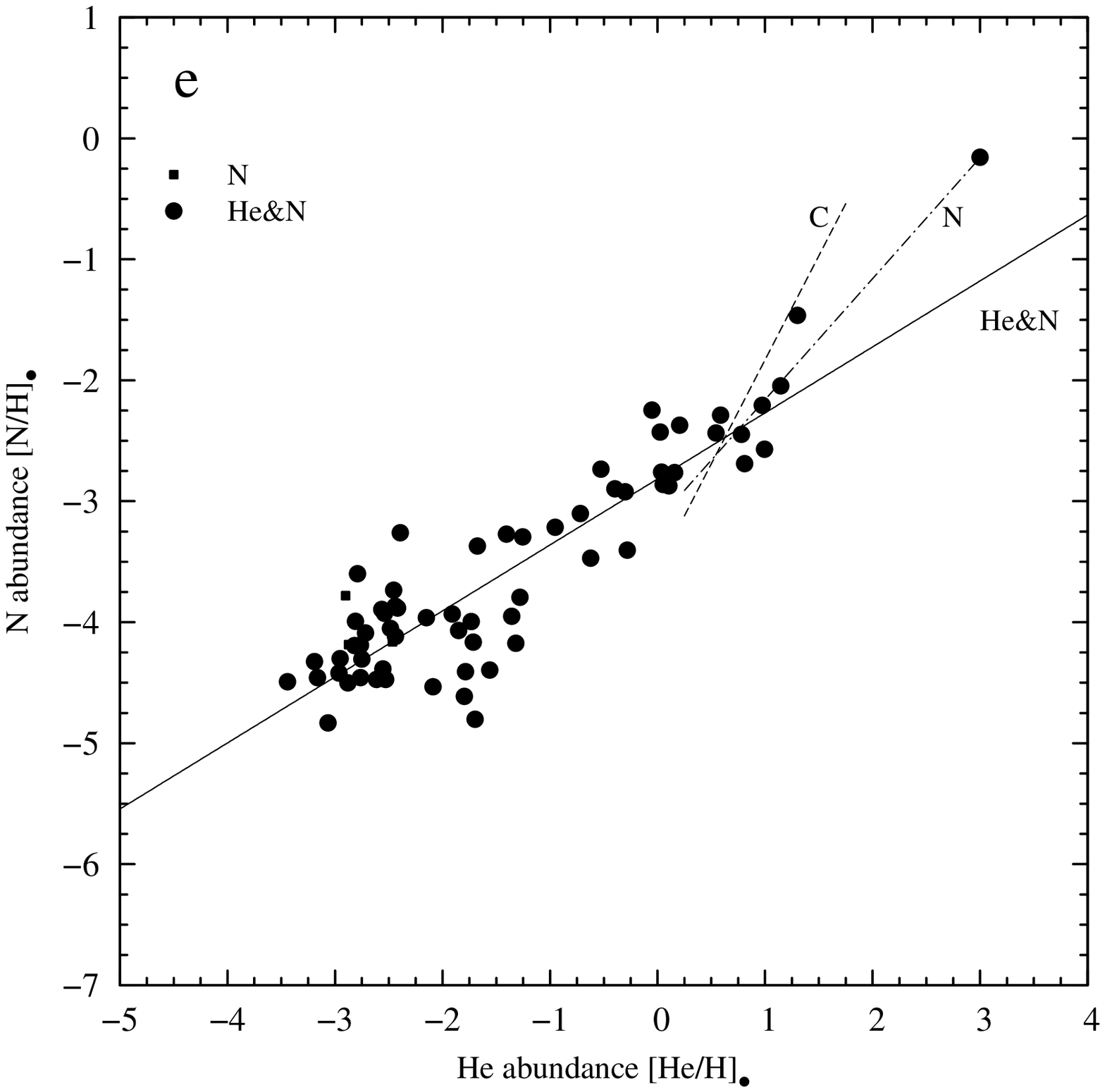}
      \label{fig:third_sub}    }
\\
\subfigure
{\includegraphics[width=0.40\linewidth,clip=,bb=67 154 551 700,angle=-90]{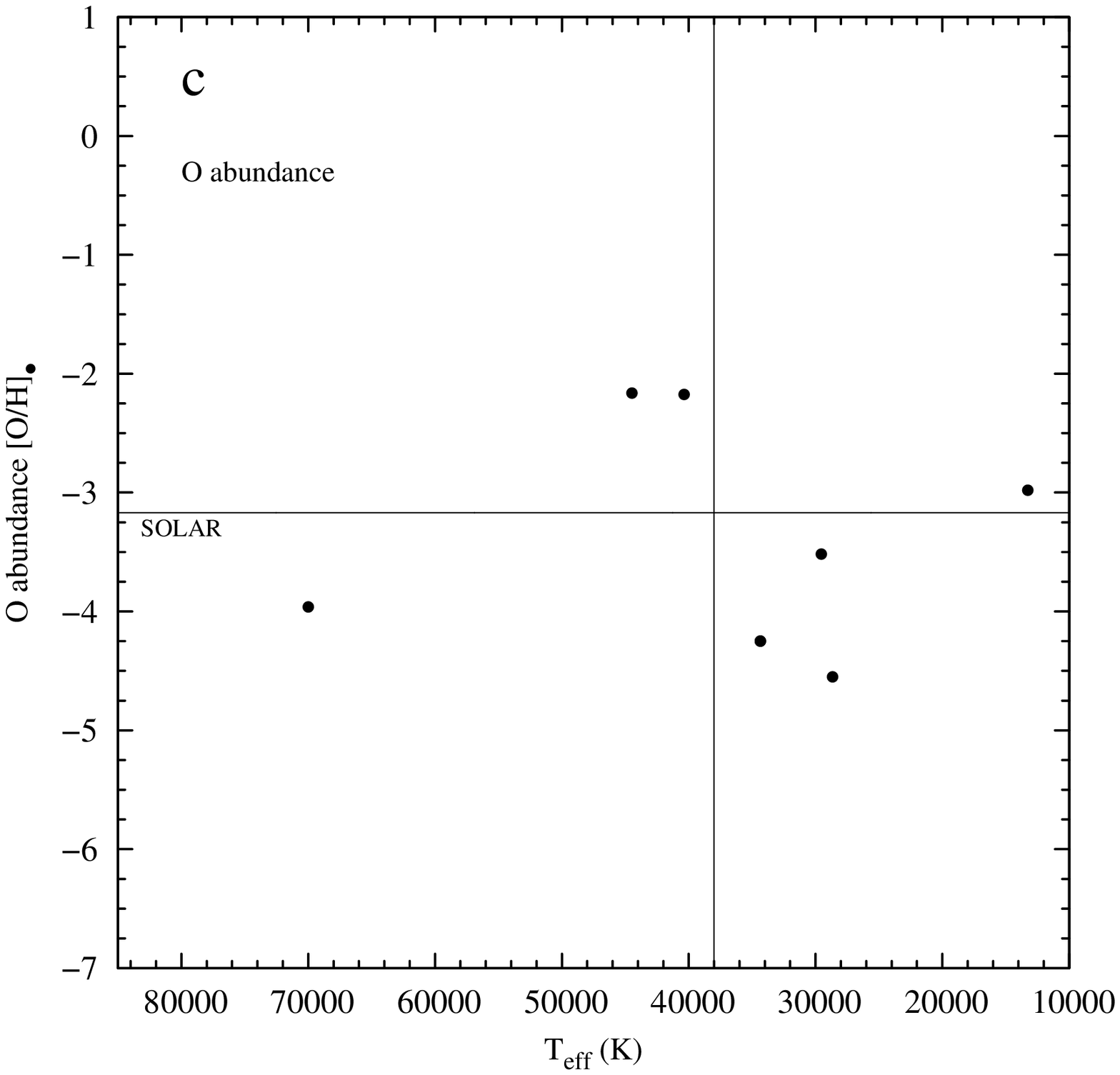}
      \label{fig:third_sub}    }
\subfigure
{\includegraphics[width=0.40\linewidth,clip=,bb=67 154 551 700,angle=-90]{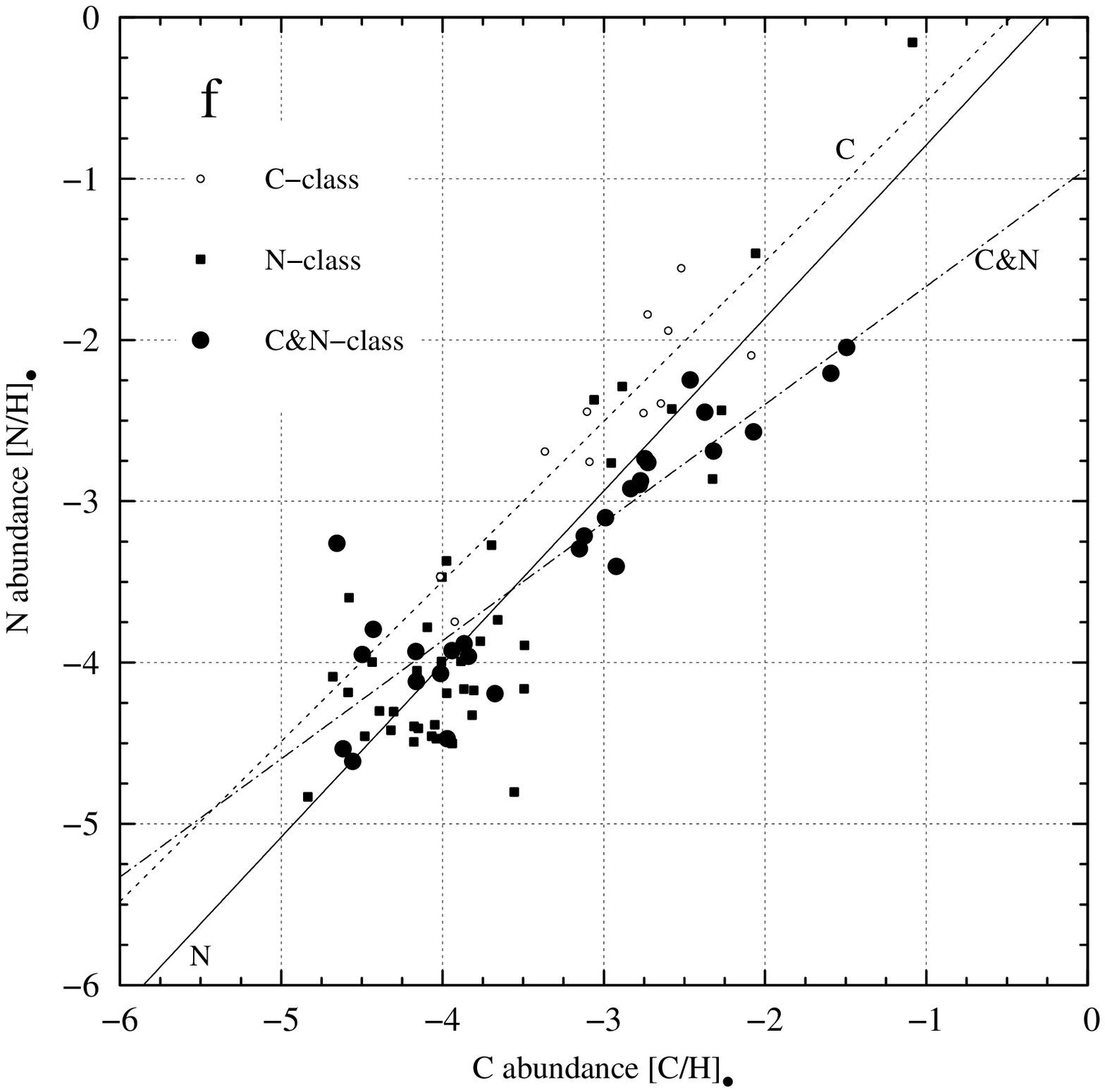}
      \label{fig:third_sub}    } 
\caption{Panels {\it a}, {\it b} and {\it c} show the measured C, N and O
 abundances with respect to effective
 temperature. 
 A vertical line at 38\,000 K separating sdB and sdO stars is overplotted as
 well as the solar abundance for each element with a horizontal line. 
 Panels
 {\it d} and {\it e} show correlations between the C, N and He abundances.
 In panel {\it f}, the C--N abundance correlations are shown, trends are
 fitted for N-rich stars (full line), C-rich stars (dashed line) and both C-
 and N-rich stars (dash-dotted line).
 }
\label{Fig:cnoabn_nl}
\end{figure*}

Panels {\it a}, {\it b} and {\it c} of Figure \ref{Fig:cnoabn_nl} 
show CNO trends with respect to
temperature. The approximate border between sdB and sdO stars at
$T_{\rm eff}\approx38\,000$ is shown by a vertical line and solar
abundances from \citet{grevesse98} are indicated with horizontal dashed lines. 
Out of the 180 stars, we measured a He    
abundance in 149; 68 showed N, 39 showed C, but only 7 showed O in their spectra. 
A quick look reveals that the bulk of sdB stars have subsolar C and O abundances and
roughly solar N abundances. 
He-sdO stars show higher abundances, strongly correlated
with effective temperature. 
Interpreting the abundance distribution in the direction of increasing temperature: 
both C and N show a similar abundance in cooler sdB or BHB stars in the range 
$20\,000 {\rm\ K}<T_{\rm eff}<23\,000 {\rm\ K}$ of about $[{\rm
C/H}]_\bullet, [{\rm
N/H}]_\bullet\approx-3$.
In sdB stars between
$23\,000 {\rm\ K}<T_{\rm eff}<34\,000 {\rm\ K}$ a lower abundance 
of about $[{\rm C/H}]_\bullet, [{\rm N/H}]_\bullet\approx-4$ was measured. 
Above $T_{\rm eff}\approx35\,000$ K across the sdB/sdO border a significant
abundance increase of C and N is observed reaching a peak near 
$T_{\rm eff}\approx43\,000$ K at $[{\rm C/H}]_\bullet, [{\rm N/H}]_\bullet\approx-1$,
over this temperature the abundance of both
elements decrease steadily with temperature.
The hottest stars indicate a gradual abundance increase over
$T_{\rm eff}\approx60\,000$ K. 
Only a few stars without He in their spectra show C or N.
These trends are very similar to
those found in the He abundance distribution and similar observations were 
made in the SPY data (\citealt{lisker05}, \citealt{stroeer07}). 

In panels {\it d} and {\it e} of Figure \ref{Fig:cnoabn_nl} the C
and N abundances are plotted with respect to the He abundance. The 39 stars that
have C determination have He as well. Out of the 68 stars showing N, 64 show
a He abundance. 
The stars in
which both C and N were found simultaneously are fitted with a full line. 
A similarly complex C and N abundance pattern can be outlined from this data as in the case of
He. 
In sdB stars both C and N show large variations, although a higher N
abundance in cooler sdB stars is apparent.
Towards higher He abundances, starting at 
$[{\rm He/H}]_\bullet\approx-1$, He-sdB stars nicely connect sdB and He-sdO stars
\citep{ahmad03} and
show a higher C and N abundance than sdB stars. 
Unlike in sdB stars, the He abundance in He-sdO stars decreases with
increasing effective temperature. 
Therefore, in the direction of increasing He abundances, we observe the hottest He-sdO stars
first, and finally, we see the most He abundant and cooler He-sdO stars.
We observed a similar trend for the C and N abundances as well.
Due to this different population, an abundance drop is observed at 
$[{\rm He/H}]_\bullet\approx0$,
which is followed by a steady abundance increase with increasing He
abundance.
These trends reach a peak at 
$[{\rm C/H}]_\bullet, [{\rm N/H}]_\bullet\approx-1$
and were fitted with a dashed line in panel 
{\it d} for the C abundance: 
\begin{equation}
[\rm {C/He}]_\bullet=1.72(\pm0.23)\ [{\rm He/H}]_\bullet-3.55(\pm0.17)
\end{equation}
and with a dash-dotted line in panel 
{\it e} for the N abundance: 
\begin{equation}
[\rm {N/He}]_\bullet=1.00(\pm0.13)\ [{\rm He/H}]_\bullet-3.16(\pm0.18),
\end{equation}
respectively.
These fits suggest that He-sdO stars with a low He abundance are either
C-rich or N-rich. 
Over $[{\rm He/H}]_\bullet\approx0.55$ 
both C and N is present in the spectra and C is more
abundant. 
Finally, the most He abundant He-sdO stars show only N in their spectra.
The hottest sdO stars complicate the trend,
but it can be seen that towards higher He abundances (lower temperature) the N 
abundance progressively remains
below the C abundance. 
The ratio of these two elements has important
connections with formation theories. 
In the case of the extremely He-rich
star J0851-1712 we could not determine an upper limit to the He abundance 
to support
this trend. For such He dominated stars and He-sdO stars in general, 
it might worth
considering changing the reference element to He because abundances relative
to H may show extreme values merely due to the vanishing H in the atmosphere.

In panel {\it f} the N abundance is plotted with respect to the C abundance. 
The
three lines show the linear fits to stars that show a C (dashed line), 
N (full line) and both C and N abundances
(dash-dotted line). 
While data for C or N-rich stars
independently show about the same abundances in He-sdO stars, 
stars near the C--N peak that show both C and N simultaneously suggest
a C to N ratio of about $\sim$3. 
Considering the asymmetric errors of our abundance determination (see
Appendix A) this number is a lower limit. 
The same trend predicts a N overabundance in the coolest sdB stars.
We found equilibrium abundance at 
$[{\rm C/H}]_\bullet, [{\rm N/H}]_\bullet\approx-3.5$
that
corresponds to $[{\rm He/H}]_\bullet\approx-1.25$; just above the group of 
possible short period sdB pulsators in the temperature--He-abundance diagram. 
C\&N-class objects are C-rich above this He abundance and N-rich below. 
Based on the number of detections the C and N abundance peaks of He-sdO
stars are also recognisable in panel {\it f} near
$[{\rm C/H}]_\bullet\approx-2.5$ and 
$[{\rm N/H}]_\bullet\approx-2.3$, respectively.
We note that our C abundance determination in some cases
has asymmetric errors and the N abundance has large errors
making the derivation of fine details difficult. 
However, our observations confirm the C\&N classification scheme of He-sdO stars.
Interestingly, the most He abundant star (J0851-1712) 
showed a N abundance, but we could not
determine a C or O abundance. It might belong to the N-class. 
\citet{zhang12} showed that WD mergers can produce such abundance patterns. 
Their slow mergers are predicted to be N-rich and
populate the low-mass end of the HeMS around $T_{\rm eff}\approx41\,000$ K.
In turn, fast mergers are expected to be C-rich with higher masses around
$T_{\rm eff}\approx45\,000$ K. Their composite model retains both
elements with N overabundance in low-mass and C overabundance
in high-mass He-sdO stars. 
Such C enrichment can be explained by the standard hot-flasher scenario
as well, for He-sdO stars it predicts a C/N ratio in the range of $\sim$0.85 
to $\sim$9.5
\citep{bertolami08} in agreement with our observations. 

In Appendix A we show upper limit determinations and error bars. 
These graphs show even more similarities with the He abundance
distribution, but a deeper look reveals some differences as well.
The appearance of the C-weak and N-weak
sequences is remarkable. 
The C--N peak shows a structure: C shows a peak
at $T_{\rm eff}\approx46\,000$ K and overabundant
compared to N up to $T_{\rm eff}\approx55\,000$ K, while N shows a peak
at $T_{\rm eff}\approx42\,000$ K and decreases with increasing
temperature faster than C. 
This structure has already
been observed in He-sdO stars \citep{hirsch09} and was investigated in the case
of double He-WD mergers \citep{zhang12}. 
From the model analysis of
\citet{hirsch09} we found a mean temperature of $\sim$44\,600 K
for C-rich and $\sim$40600 K for N-rich He-sdO stars, in agreement with our
results.
In the hottest He-sdO stars ($T_{\rm eff}>60\,000$ K) we were able to determine N
and found a slight anti-correlation with the He abundance unlike at lower
temperatures. 
Our O determination is not yet reliable to derive such trends, but
a correlation with the He abundance is possible.  

The boundary 
that separates rapid and slow sdB pulsators 
(dash-dotted line in the top panel of Figure \ref{Fig:metal}) 
correlates with the observed abundance patterns. 
Suspected rapid pulsators near 33\,500 K are He-rich and do not show metal traces, 
while slow pulsators around 28\,000 K are He-poor and many show a N abundance.
We found 27 stars in the temperature-gravity region of known rapid pulsators. 
Out of the 27 stars, we found a N abundance in six ($\sim$22 per cent). 
Similarly, we found 23 stars at the location of known slow pulsators and 
16 ($\sim$70 per cent) of these show a N abundance. 
In between these
groups, where hybrid pulsators are expected, 
we found both C and N in most of the stars. 
Interestingly, flashless evolutional tracks by 
\citet{hu08}
connect the group of C\&N class sdB stars with C\&N class high gravity He-sdO stars. 
Recently,
\citet{geier12}
found a He isotope anomaly ($^3$He is strongly enriched in the atmosphere) 
in eight stars in a sample of 46 sdB stars. 
These stars are situated in a narrow strip in the temperature-gravity plane
between 27\,000 and 31\,000 K, where we see C\&N class sdB
stars in our survey.
No connection has been found in earlier studies between surface abundance
patterns and pulsational instability 
\citep{otoole06b}.  
However, we observed different He, C, and N contents 
along the EHB that might play a role and need further 
investigations.

\subsection{Binaries}

\setcounter{table}{3}
\begin{table*}
\caption[3]{Parameter shifts during decomposition. He abundance
determination from upper limit to positive identification (normal font),
upper limit before--after (italic font) and positive identification before--after
(bold font) are distinguished. Companion absolute V magnitudes are derived from
spectral types and from binary decomposition for subdwarfs. Absolute
magnitudes of binary components allowed to estimate distances.
}
\begin{tabular}{crrrr@{$-$}lccr}
\hline
GALEXJ                 &\multicolumn{1}{c}{$\Delta{T}$}                  & 
                        \multicolumn{1}{c}{$\Delta{\log g}$}             & 
                        \multicolumn{1}{c}{$\Delta[{\rm He/H}]_\bullet$} &
                        \multicolumn{2}{c}{\hspace{-0.1cm}Type}                         &
                        M$_{\rm V,sd}$                                          & M$_{\rm V,comp}$ & d (pc)\\
\hline
004759.6$+$033742B & 2420        & 0.90  & -0.347      & sdB&F6V   & 3.95 & 3.6 & 750 \\
011525.9$+$192249B & 2860        & 0.96  & {\it 0.265} & sdB&F2V   & 3.65 & 2.9 & 1340 \\
011627.2$+$060314B & 1610        & 1.06  & {\bf 0.327} & sdB&F6V   & 4.36 & 3.6 & 1040 \\
020447.1$+$272903B & 1580        & 0.30  & {\bf 0.090} & sdB&G0IV  & -    & 2.7 & - \\
021021.8$+$083058B & 3730        & 0.77  & 0.406       & sdB&F2IV  & -    & 1.6 & - \\
022454.8$+$010938B &    0        & 0.46  & {\it 0.309} & sdB&F4V   & 4.02 & 3.3 & 780\\ 
071029.4$+$233322B & 8010        & -0.12 & {\bf 0.424} & sdO&F6V   & 2.64 & 3.6 & 2240 \\
071646.9$+$231930B & -1800       & -0.95 & {\it 0.286} & B&A1V?    & 1.57 & 1.0 & 1690 \\
101756.8$+$551632B & 2480        & 0.41  & {\it 0.668} & sdB&F5V   & 4.06 & 3.5 & 730 \\
110541.4$-$140423B & -           & -     & -           & sdB&F6V   & 5.02 & 3.6 & 440 \\
141133.4$+$703736B & 1800        & 0.75  & {\it 0.669} & sdB&F0.5V & 3.17 & 2.6 & 1640 \\
151325.7$+$645407B & 610         & 0.93  & {\bf 0.417} & sdB&G0V   & 5.17 & 4.4 & 990 \\
152513.0$+$605321B & 150         & 0.59  & {\bf 0.211} & sdB&G0V   & 4.33 & 4.4 & 710 \\
160209.1$+$072509B & 70          & 0.20  & {\bf 0.464} & sdO&G0V   & 3.83 & 4.4 & 1820 \\
161902.7$+$483144B & 5730        & 0.87  & -           & sdB&F0V   & 3.46 & 2.6 & 1710 \\
173651.2$+$280635B & -           & -     & -           & sdB&F7V   & 4.71 & 3.9 & 390 \\
175340.5$-$500741B & -470        & 0.75  & {\bf 0.703} & sdB&F7V   & 4.52 & 3.9 & 780 \\
202027.2$+$070414B & -590        & 0.93  & {\bf 0.690} & sdO&F3V   & 3.33 & 3.1 & 2060 \\
202059.8$-$225001B & -950        & 0.15  & {\bf 0.121} & sdB&G0V   & 4.38 & 4.4 & 430 \\
202216.8$+$015225B & 2300        & 0.77  & {\bf 0.441} & sdB&F6V   & 4.34 & 3.6 & 1020 \\
203850.2$-$265747B & -1550       & 0.14  & 0.606       & sdO&G3.5III  & 1.95 & 0.95 & 1830 \\
210031.7$+$145213B & 2370        & 0.76  & {\bf 0.498} & sdB&F5.5IV-V & 4.54 & 3.5 & 970 \\
212424.3$+$150619B & -270        & 0.42  & {\bf 0.396} & sdB&F7V   & 4.00 & 3.9 & 940 \\
213730.9$+$221908B & -1770       & 0.48  & {\bf -0.009}& sdB&G1V   & 4.03 & 4.5 & 1010 \\
214022.8$-$371414B & -840        & 0.77  & {\bf 0.239} & sdB&F6V   & 4.15 & 3.6 & 870 \\
222758.5$+$200623B & 2540        & 0.85  & {\bf 0.930} & sdB&F5V   & 4.98 & 3.5 & 300 \\
232917.9$+$325348B & 3850        & 1.27  & -0.067      & sdB&F4V   & 4.37 & 3.3 & 930 \\
233158.9$+$281522B & -290        & 0.38  & {\it 0.994} & sdB&F8V   & 3.33 & 4.1 & 1750 \\
234903.2$+$411925B & -1150       & 1.37  & -           & B?&A4V    & 3.60 & 1.7 & 480 \\
\hline
Average          & 1306        & 0.67            & {\bf 0.418} \\
Median           & 610         & 0.76            & {\bf 0.420} \\
\hline
\end{tabular}
\label{Tab:shifts}
\end{table*}

\begin{figure}
\centering
\includegraphics[width=0.65\linewidth,clip=,angle=-90]{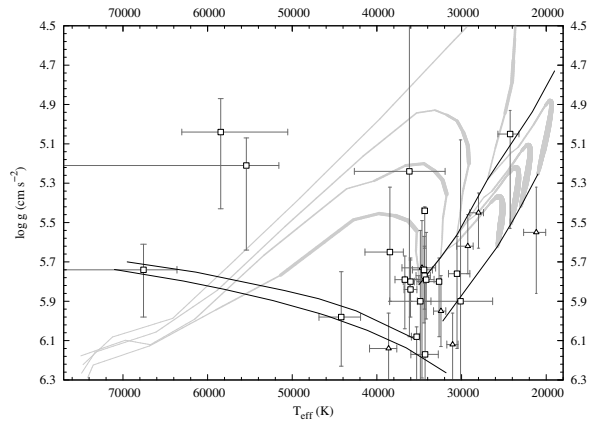}
 \caption{{\small
 Same $T_{\rm eff}$$-$$\log g$ diagram as Figure \ref{Fig:hrd}, here showing
 only composite spectra binaries. A crowding of composites can be observed
 near the location of possible rapid pulsators at $T_{\rm eff} =33\,500$ K, $\log
 g=5.8$. }
\label{Fig:hrdbin}}
\end{figure}

Figure \ref{Fig:hrdbin} shows the distribution of subdwarfs in composite spectra
binaries in the temperature-gravity plane.  
We did not find He-sdO stars in such binaries. 
Based on the Ca infrared triplet \citet{cuadrado01} and \citet{cuadrado02}
found that companions from composite spectra are in the range of 
0.8-1.2 ${\rm M}_{\odot}$ F--G type MS or subgiant stars. 
We confirm this observation and list the results of our binary decomposition
in Table \ref{Tab:shifts}.
Monochromatic flux ratios are listed in Table 
\ref{Tab:bvr}
for the
companion stars at the effective
wavelengths of the broadband Bessell filters 
\citep{bessell05}. 
In parentheses we also give predicted flux ratios that were derived 
from the synthetic subdwarf spectrum and companion template outside our observed spectral range.

\begin{figure}
\centering
\includegraphics[width=0.7\linewidth,clip=,angle=-90]{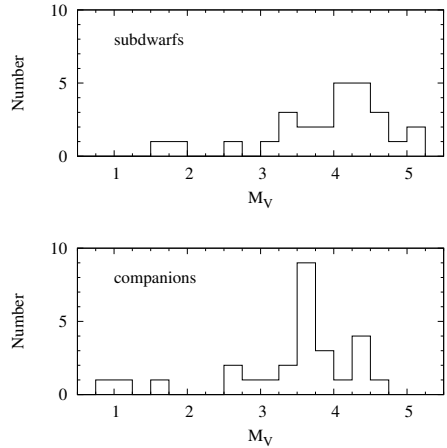}
 \caption{{\small
 Absolute V magnitude distributions of hot subdwarfs and their companions.}
\label{Fig:magdist}}
\end{figure}

Before starting spectral decomposition we always performed a single star fit 
and used the results in the input of the binary model.
This allowed us to estimate the shifts in subdwarf atmospheric
parameters during decomposition with respect to a single star fit.
We emphasise that the relative shifts in Table \ref{Tab:shifts} are specific to our modelling and
fitting method that does not make assumptions for the companion
types. 
Temperature shifts show
large variations with an average of $\sim$$1300$ K increase. 
Surface gravity
shows a stronger correlation, with an average 0.7 dex increase. 
Similar strong correlation can be observed in the He abundance shifts.
On average a 0.42 dex increase with a small scatter 
is indicated by the spectral decomposition of
the 14 stars in which He was measurable in both the single star and binary fits. 
Spectral decomposition helped to achieve a He abundance measurement in four
cases while in six cases it confirmed an upper limit measurement.
We adopted absolute V magnitudes based on the spectral type and luminosity
class of the companions from \cite{gray92}. 
Using the flux ratios from Table \ref{Tab:bvr} we
calculated subdwarf absolute V magnitudes and approximate distances.
The absolute magnitude distributions of subdwarfs and their companions are
shown in Figure \ref{Fig:magdist} and confirm a consistent spectral
decomposition. 
F type MS stars are the easiest companions to find, hence these are the most
frequent in our sample.

\setcounter{table}{4}
\begin{table}
\begin{center}
\caption[2]{
Monochromatic flux ratios 
$({\rm F}_{\lambda,comp.}/({\rm F}_{\lambda,comp.}+{\rm F}_{\lambda,sd}))$ 
of binary companions at the effective
wavelengths of the broadband Bessell filters. 
Flux ratios outside our observed spectral ranges were calculated from
synthetic and template spectra, these predicted flux ratios are listed 
in parentheses.}
\begin{tabular}{ccccc}
\hline
GALEXJ                 &  $U$   &  $B$   &  $V$   &  $R$   \\
                       &3663 \AA&4361 \AA&5448 \AA&6407 \AA\\
\hline
004759.6$+$033742B & 0.227 & 0.434 &(0.580)&(0.672)\\
011525.9$+$192249B & 0.331 & 0.551 &(0.667)&(0.737)\\
011627.2$+$060314B & 0.286 & 0.522 &(0.669)&(0.749)\\
020447.1$+$272903B & 0.058 & 0.153 & 0.271 & 0.371 \\
021021.8$+$083058B & 0.385 & 0.616 &(0.735)&(0.799)\\
022454.8$+$010938B & 0.333 & 0.537 &(0.660)&(0.735)\\
071029.4$+$233322B & 0.073 & 0.180 & 0.292 & 0.382 \\
071646.9$+$231930B & 0.564 & 0.637 & 0.629 & 0.633 \\
101756.8$+$551632B & 0.328 & 0.510 & 0.626 & 0.699 \\
110541.4$-$140423B &(0.454)& 0.685 & 0.787 & 0.838 \\
141133.4$+$703736B & 0.345 & 0.547 & 0.628 & 0.683 \\
151325.7$+$645407B & 0.334 & 0.529 & 0.671 & 0.757 \\
152513.0$+$605321B & 0.170 & 0.328 & 0.484 & 0.591 \\
160209.1$+$072509B & 0.094 & 0.231 &(0.373)&(0.478)\\
161902.7$+$483144B & 0.341 & 0.575 & 0.689 & 0.755 \\
173651.2$+$280635B &(0.285)& 0.523 & 0.678 & 0.753 \\
175340.5$-$500741B & 0.277 & 0.496 &(0.640)&(0.724)\\
202027.2$+$070414B & 0.182 & 0.404 &(0.552)&(0.652)\\
202059.8$-$225001B & 0.165 & 0.338 &(0.496)&(0.594)\\
202216.8$+$015225B & 0.295 & 0.522 &(0.664)&(0.743)\\
203850.2$-$265747B & 0.244 & 0.496 &(0.715)&(0.806)\\
210031.7$+$145213B & 0.347 & 0.594 &(0.723)&(0.792)\\
212424.3$+$150619B & 0.189 & 0.370 &(0.522)&(0.622)\\
213730.9$+$221908B & 0.112 & 0.232 &(0.393)&(0.508)\\
214022.8$-$371414B & 0.279 & 0.484 &(0.625)&(0.712)\\
222758.5$+$200623B & 0.481 & 0.700 &(0.797)&(0.846)\\
232917.9$+$325348B & 0.365 & 0.605 & 0.729 & 0.797 \\
233158.9$+$281522B & 0.090 & 0.199 & 0.330 & 0.430 \\
234903.2$+$411925B & 0.768 & 0.866 & 0.853 & 0.869 \\
\hline
\end{tabular}
\label{Tab:bvr}
\end{center}
\end{table}

We did not find line asymmetries indicating a measurable radial velocity
difference between the components in any of our stars. 
Radial velocity did not show variations in the much larger, low-resolution 
sample of similar composite spectra binaries
 of \cite{green08}, which suggests
that orbital periods must range from at least a few months to few years. 
Recently, \citet{ostensen12b}
measured radial velocity variations and
detected orbits of eight sdB--F/G binaries. 
They found orbital periods in the range of 500-1200 days.

The effects of spectral decomposition of subdwarfs with F and early G
type companions
were found to be a positive shift in the subdwarf temperature, gravity and He abundance. 
In the
case of late G and early K-type companions, due to the lower flux contribution, 
we expect lower shifts.  
A shift to higher gravities, hence to lower
luminosities of some of these sdB stars is possible. 
A flux flattening and
strong Na D lines can also indicate binarity. 
The high SNR allowed us to detect signs
of such companions in the far red part of the spectra. 
However,
such faint stars do not contribute enough flux next to a 40-300 times more
luminous subdwarf to carry out a proper decomposition. 
Due to the lack of strong H lines in their spectra and their low
contribution, these stars are not expected to significantly affect the derived 
subdwarf atmospheric parameters.
The binary population with late MS (K and M dwarf) companions is high according to
radial velocity surveys and probably similarly high with WDs. 
Such companions are too faint to show resolvable spectral signatures. 

\begin{figure}
\centering
\includegraphics[width=0.9\linewidth,clip=,angle=-90]{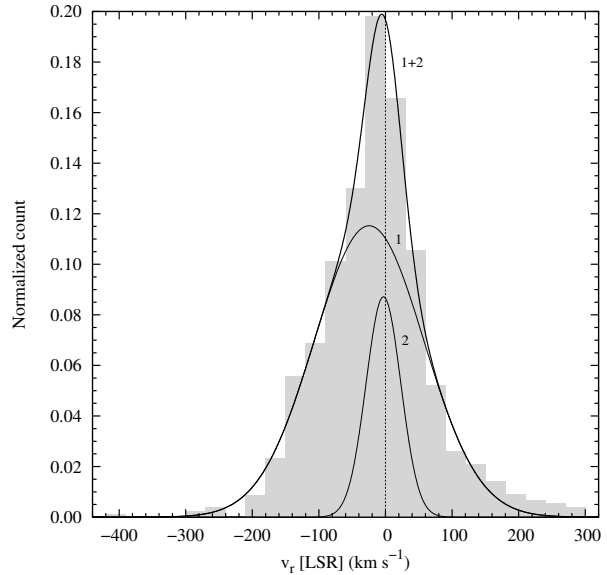}
 \caption{{\small
 Velocity distribution relative to the kinematic LSR.}
\label{Fig:radvel}}
\end{figure}

Because close binaries are
usually found by radial velocity surveys we checked the absolute shifts of our
spectra relative to synthetic spectra.
Figure \ref{Fig:radvel} shows the radial velocity distribution in the {\sl GALEX}
sample with respect to the kinematic local standard of rest (LSR) frame. 
We found a two component Gaussian velocity distribution with 
velocity dispersions 
$\sigma_1=82$ and $\sigma_2=26$ km s$^{-1}$,
and mean velocities
$v_1=-25$ and $v_2=-3$ km s$^{-1}$, respectively.
Half of the stars in the catalogue show a radial velocity $v_r({\rm LSR})>47$ 
km s$^{-1}$ and 42 show $v_r({\rm LSR})>100$ km s$^{-1}$ which, 
in some case, may be evidence of binarity. 
These stars are distinguished with the "rv" flag in Table \ref{Tab:1}.
Two subdwarfs 
(J0321+4727 and J2349+3844, \citealt{kawka10}) were
found in close binaries and J1717+6757 is a
double-degenerate binary \citep{vennes11b} out of the 42 kinematically
peculiar stars. 
J0716+2319 shows binarity and
will be discussed in a forthcoming paper.
Interestingly, four out of the 29
composites show a notable radial velocity.
In fact, some stars can be hierarchical triples \citep{maxted01}, these
require further observations.

\begin{figure*}
\includegraphics[width=8.5cm,clip=,angle=-90]{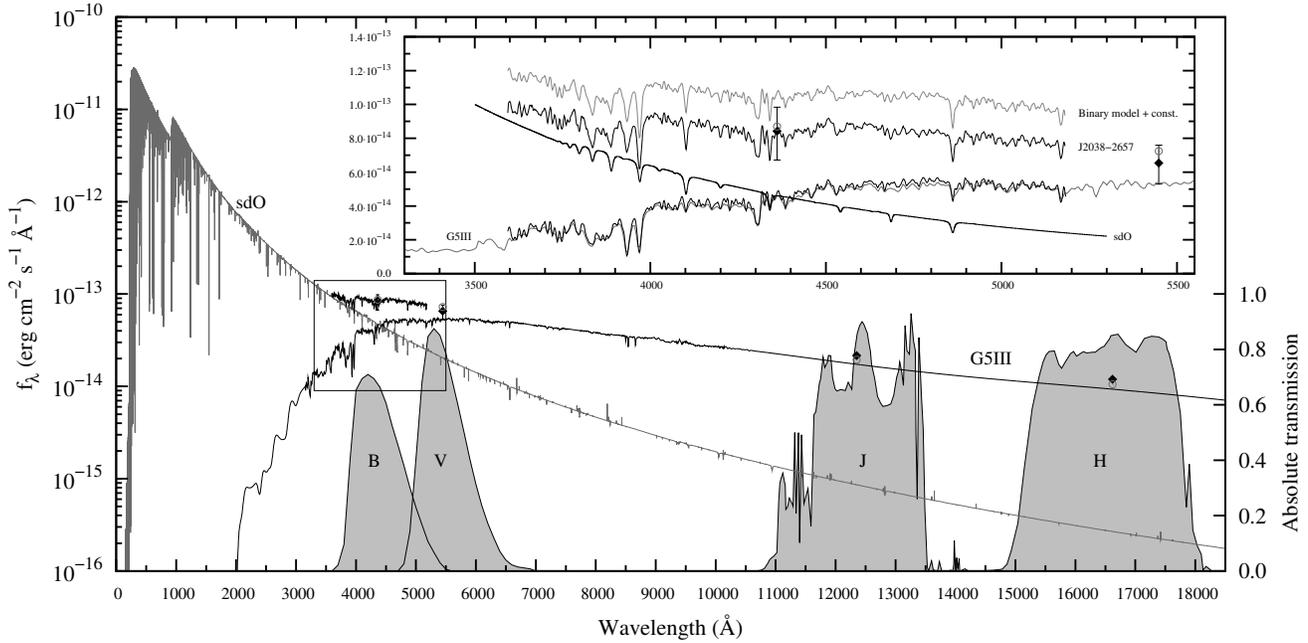}
 \caption{\small
Spectral energy distributions of the components in J2038-2657. The {\small TLUSTY} SED for the sdO star
 (grey) and the G5III template from the {\small HILIB} library are marked. Bessell
 $B$, $V$ and {\small 2MASS}
 $J$ and $H$ filter transmission
curves cut out the contribution of each star to the synthetic photometry.
{\it Inset:} Magnified part of the spectrum where decomposition was performed. The
{\small HILIB} G5III template is in grey, overplotted with our interpolated {\small MILES}
template in black, their match is excellent. The G5III and the sdO spectra
adds up nicely and this binary model fits the observation. The binary model is
shifted up for clarity. At the effective wavelengths of Bessell $B$, $V$ and
{\small 2MASS} $J$ and $H$ filters we show {\small GSC2.3.2} (quick-$V$) and {\small 2MASS} $J$ and $H$
photometry with black diamonds. Data for the Bessell $B$ filter is the
associated optical source magnitude form {\small 2MASS}. Grey circles show our
synthetic photometry.
\label{Fig:2038}}
\end{figure*}

We found eight composite spectra binaries among the 27 stars in the empirical region of rapid pulsators, 
while only one binary out of the 23 stars in the region of slow pulsators.
The higher frequency of composite spectra binaries among possible rapid pulsators
is obvious and supports the importance of binary evolution in subdwarf formation.
A possible correlation between binarity and rapid pulsations requires further investigations.

Figure \ref{Fig:2038} shows the
SEDs of the components in the sdO--G3.5III binary J2038-2657. 
The sdO dominates the flux in the UV, 
the subgiant emerges in the optical and dominates the infrared. The inset
shows our observation, the components were smoothed to the correct spectral resolution.
Our recent, high-resolution time-resolved spectroscopic follow-up 
revealed variable H$_\alpha$ emission in this binary.

\section{Luminosity and mass distribution}\label{Sec:luminosity}

\begin{figure}
\begin{center}
 \includegraphics[width=0.8\linewidth,angle=-90]{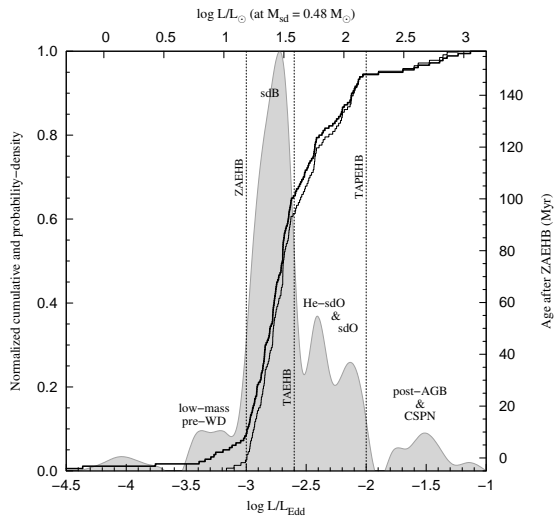}
 \caption{Luminosity distribution of 180, mostly hot subdwarf stars from the
 {\sl GALEX} survey. The locations of the ZAEHB, TAEHB and terminal age post-EHB
 (TAPEHB) are marked. Stellar luminosities are expressed in solar
 luminosity on the top axis assuming a subdwarf (sdB, sdO) mass of 0.48
 M$_\odot$. He-sdO stars show a wide mass distribution.
\label{Fig:lumfun1}}
\end{center}
\end{figure}

\begin{figure}
\begin{center}
\includegraphics[width=1.35\linewidth,clip=,angle=-90]{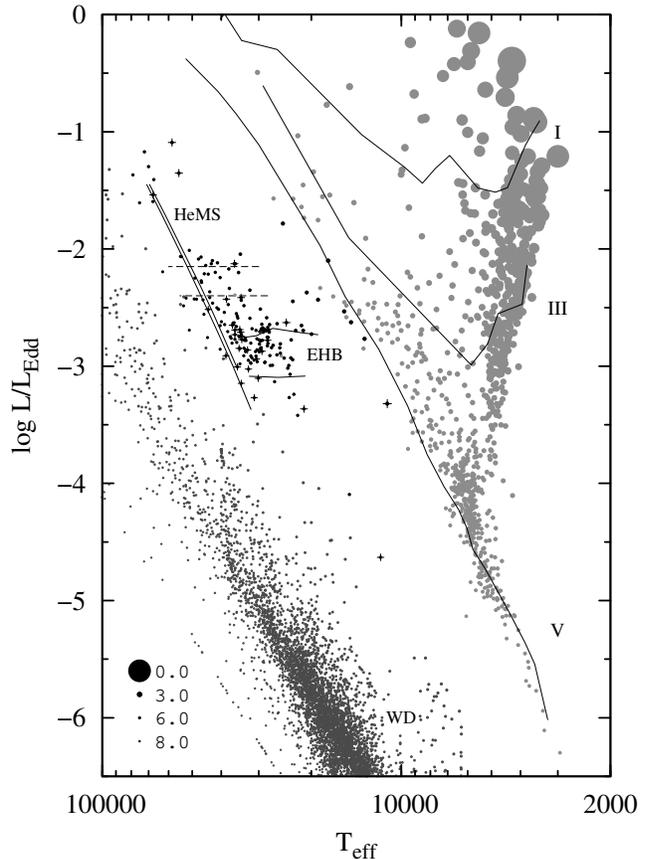}
 \caption{ $T_{\rm eff}-\log ({\rm L}/{\rm L}_{\rm Edd})$ distribution of
 stars in the {\small MILES} library (light grey), WD sample ($\log g>6$) from SDSS DR4
 (\citealt{eisenstein06}, dark grey) and the {\sl GALEX} subdwarf sample (black).
 The EHB, HeMS and WD sequences and three luminosity classes from \citet{kurucz93} are marked
 in the figure. Point sizes are proportional to the inverse of surface
 gravities (point size $= 2/(\log g+1)$) for the sake of clarity.  
 Composites from the {\sl GALEX} sample
 are marked with crosses. \label{Fig:lumt}}
\end{center}
\end{figure}

The {\sl GALEX} 
sample presented here is one of the largest homogeneously analysed subdwarf sample
to date, therefore
we describe the cumulative luminosity distribution function (CLDF) 
of the 180 stars shown by the
thick histogram in Figure \ref{Fig:lumfun1}. 
The bottom axis is the Eddington luminosity-fraction
calculated from the derived effective temperatures and surface gravities.
The shaded area shows the probability density 
function calculated from the smoothed CLDF.
The luminosity is also indicated in solar
units on the top axis, assuming an average stellar mass of 0.48 ${\rm M}_{\odot}$. 
This scale is valid only for sdB stars and their descendant sdO stars, and
represents a lower limit for
He-sdO stars that have progressively
higher masses toward higher temperatures along the HeMS
(\citealt{divine65}, \citealt{paczynski71a}).
Again, we found five well separated sections in the luminosity distribution. 
We identify the lowest luminosity stars
below $\log({\rm L}/{\rm L}_{\rm Edd})<-3.5$ with possible
progenitors of extremely low mass WDs such as J0805-1058. 
The group between
$-3.5<\log({\rm L}/{\rm L}_{\rm Edd})<-3$ consists of 
possible BHB stars, like J0746+0610, J1738+2634 and J2153-7004. 
About 55 per cent of our sample have luminosities between $-3<\log({\rm L}/{\rm L}_{\rm Edd})=-2.6$, 
we identify these with most of the sdB and some of the low-mass He-sdO stars. 
At higher luminosities the slope of the CLDF is defined by sdO and He-sdO stars 
up to $\log({\rm L}/{\rm L}_{\rm Edd})=-2$. 
Finally, the most luminous sdO, post-AGB, or CSPN stars are found at
$\log({\rm L}/{\rm L}_{\rm Edd})>-1.6$.
After passing their maximum light, these stars
become WDs and cross the subdwarf region on their way to the lowest luminosities. 
Two peaks at higher luminosities are noticeable.
We identify the peak at $\log({\rm L}/{\rm L}_{\rm Edd})\approx-2.5$ with
He-weak post-EHB sdO stars and at 
$\log({\rm L}/{\rm L}_{\rm Edd})\approx-2.15$ with He-sdO stars.
These groups can be seen to separate in Figure \ref{Fig:lumt}, 
where analogously to the HRD, the distribution of
subdwarfs, WDs \citep{eisenstein06} and stars from the {\small MILES} library 
are shown in the
Eddington-luminosity fraction versus temperature plane. 
The He content and binary
frequency on the He-weak sdO sequence seems to be 
different for the two sdB groups and suggests that
He-rich sdB stars reach higher temperatures in their post-EHB evolution than
He-poor sdB stars.

To estimate the mass-distribution 
in our sample we compared our observed CLDF to a synthetic CLDF calculated
from the 
subdwarf evolutionary tracks of \citet{dorman93}.
This method looks for the most probable mass-distribution to
reproduce the CLDF similarly to the method of \citet{zhang10}. 
The Dorman evolutionary tracks were calculated for eight
core-masses at eight different surface He abundances with 
various envelope thicknesses, giving 137 tracks altogether.
However, they do not support the entire observed 
temperature and gravity range.
The tracks were calculated for the HB with the EHB
at its hot extreme that, along with the HeMS, is only partially represented. 
For this reason, we fitted only stars with 
$-3.2<\log({\rm L}/{\rm L}_{\rm Edd})<-1.7$ and excluded low 
temperature and high gravity He-sdO
stars ($T_{\rm eff}<45000$ K and $\log g>5.9$).
This way we also excluded possible low-mass pre-WDs in the sample 
that are not supported by the tracks.
The CLDF of the remaining stars is shown by the thin 
histogram in Figure \ref{Fig:lumfun1}.
Next, we trimmed the evolutional tracks to the observed 
temperature and gravity range. 
This step was necessary, because due to the approximately constant
luminosity along the HB and EHB, a wide range of temperatures 
and gravities can contribute to a given point in the CLDF, 
causing a degeneracy in our solutions. 
Then, from the effective temperature, surface gravity 
and the age from the zero-age horizontal branch (ZAHB) of the Dorman-models we calculated 
evolutionary tracks in the Eddington-luminosity fraction
versus age plane. 
Next, we interpolated these tracks to a new and common age axis. 
Finally, we applied a $\chi^2$ minimisation to determine the 
mass-distribution in the sample 
using the normalised weighted average of these tracks 
(synthetic CLDF) and fitted the observed CLDF. 
Such an analysis also allows for estimating the evolutionary times on the
EHB and post-EHB phases.

We assumed that our sample is complete in the observed luminosity range and well-distributed,
representing the true frequencies of all subdwarf types. 
Our main goals were to describe the slopes and the break points of
the CLDF in general. 
The models reproduce both sdB and sdO stars reasonably well, although they
predict a slower evolution and higher luminosities for post-EHB stars. 
From the theoretical models we found an average EHB lifetime of $\sim$100 Myr, 
followed by $\sim$50 Myr post-EHB evolution, in contrast with \citet{zhang10} 
who found 160 and 20 Myr, respectively.
The observed CLDF also suggests that the post-EHB evolution is about 50 per cent of
the EHB lifetime.
However, large discrepancies are expected if the observed sample is not
complete or not fully described by these models. 
More recently, \citet{ostensen12a} found $\sim$80 Myr EHB lifetime followed
by $\sim$62 Myr post-EHB evolution in the case of the He-rich sdB pulsator KIC 1718290. 

We found that our subdwarfs have a well defined mass distribution around
${\rm M}=0.52\pm0.02$ M$_\odot$. 
A small contribution of stars with 
${\rm M}=0.48\pm0.01$ M$_\odot$ and a shallow distribution 
up to $0.6$ M$_\odot$ was also found. 
We would like to note that the $\sim$2000 K systematic shift found for sdB
stars comparing to these theoretical models would shift the CLDF for sdB stars 
by about $\sim$0.1 unit to lower luminosities.
This adjustment would shift the mass distribution of
sdB stars to considerably lower masses by about 0.1 M$_{\odot}$. 

By applying theoretical models to the SPY sdB and sdO data \citet{zhang10}
derived the number of sdO stars evolved form sdB stars to be
observationally 12 per cent. Using our spectral classification criteria and 
the class separation in the temperature He abundance diagram (Figure
\ref{Fig:het}) we found
120 sdB and 15 sdO stars, or about 12.5 per cent that are possible post-EHB stars.

\section{Modelling homogeneity}\label{Sec:homogeneity}

\begin{figure}
\begin{center}  
\includegraphics[width=0.8\linewidth,clip=,angle=-90]{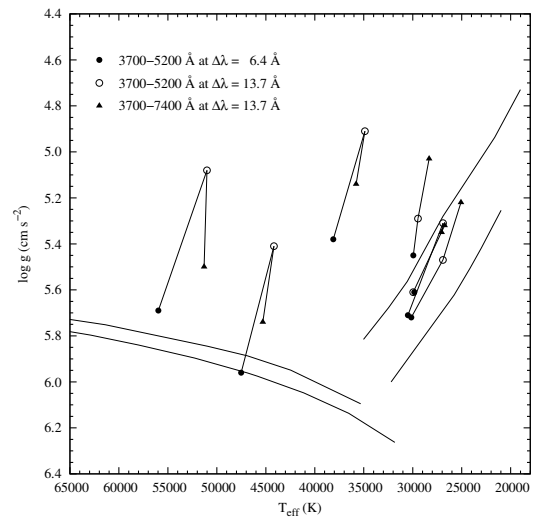}
 \caption{\small
 Systematic shifts in the temperature-gravity diagram 
 between NTT EFOSC2 observations at different resolutions and/or spectral
 coverage. 
 Data points of the same stars are connected.
 We measured $\Delta\lambda=6.4$ \AA\, for grism \#7
 ($3700-5200$ \AA) and
 $\Delta\lambda=13.7$ \AA\, for \#11
 ($3700-7400$ \AA). 
 The different spectral window and low-order variations 
 in the resolution can result, at worst, in $\Delta{T_{\rm eff}}\approx5000$ K 
 and ${\Delta{\log g}}\approx0.6$ systematic shifts.
\label{Fig:systematics}}
\end{center}
\end{figure}

Systematic shifts inevitably arise when data from different instruments are 
modelled with various model atmosphere codes. 
To decrease such shifts in the atmospheric parameters of subdwarfs some
aspects need consideration, such as: various modelling assumptions and
implementations, different atomic data, various fitting
methods and error estimations can all introduce systematic differences.
Therefore, independent model atmosphere analyses on large and homogeneous datasets,
like the HS, SPY and SDSS data would be desirable with various modelling
and fitting codes to find systematic offsets and their reasons.

Beyond homogeneous modelling and fitting techniques, consistent
atmospheric parameters would require homogeneous data as well (similar SNR,
resolution, spectral range). 
This requirement is difficult to comply due to observing time restrictions. 
For this reason a careful reduction is important, in particular, to 
apply a correct dispersion solution because 
even small departures from linearity or an incorrect radial
velocity correction can change the parameters considerably. 
We found the effects of instrumental resolution and its variation along the
dispersion axis to be very important. 
Therefore, we measured
the resolution of our spectra in the middle of the available spectral range using arcs. 
In spite of a careful reduction and analysis, Figure
\ref{Fig:systematics} shows systematic shifts between different resolution
data. 
We conclude that it is
important in the future 
to obtain higher resolution data and to measure and model 
the variations of spectral resolution for
each spectrum in the entire spectral range. 
A combination of the $\chi^2$ and
residual minimisation (or equivalent width fitting) will be implemented in
{\small XTGRID} to overcome the small scale variations caused by spectral
convolutions. 
We found that 
high series members of H lines (from H$_{\delta}$ to the Balmer-jump) are 
important in temperature and
gravity determination and data quality is essential in this spectral range. 
Figure \ref{Fig:systematics} shows that the observed spectral range 
can also be a source of systematic deviations, especially the
inclusion of the ${\rm H}_{\alpha}$ and ${\rm H}_{\beta}$ lines, 
which form higher in the atmosphere and may show non-LTE effects.

\begin{figure}
\begin{center}  
\includegraphics[width=0.68\linewidth,clip=,angle=-90]{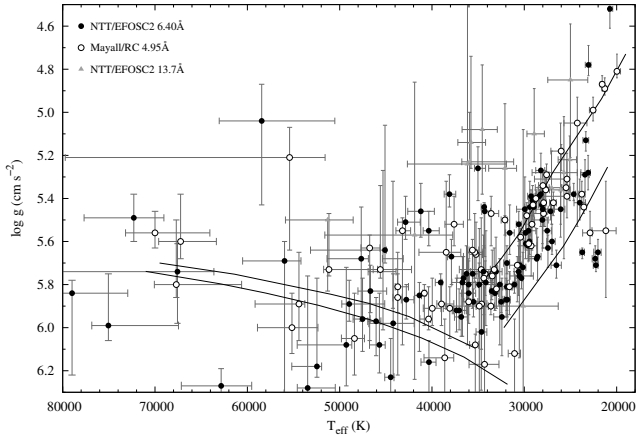}
 \caption{\small
$T_{\rm eff}-\log g$ diagram indicating different instruments and resolutions
 for all of the 191 observations.
 There is no correlation between instrumental setups and derived atmospheric
 parameters for the higher resolution data, but only less accurate parameters
 could be derived from the low-resolution observations.
\label{Fig:insyst}}
\end{center}
\end{figure}

Systematic shifts that arise from different resolutions or different spectral coverage 
are represented in Figure 
\ref{Fig:systematics} 
where we compare the parameters derived from
spectra taken with grism \#7 and \#11 with the EFOSC2 spectrograph on the NTT.
Shifts from the inclusion of the 
H$_{\alpha}$ 
line or due to the strong blending
of spectral lines in low-resolution spectroscopy can be seen. 
This problem affects 16 observations of 14 stars in our
sample, seven of these stars, shown in Figure \ref{Fig:systematics}, have higher
resolution observations as well, hence we regard the effects on our final results negligible. 
The higher resolution NTT and Mayall spectra provide consistent atmospheric
parameters as shown in the temperature--gravity plane in Figure
\ref{Fig:insyst}.
We conclude that spectral resolution affects the derived atmospheric
parameters and low resolution data provides systematically lower temperatures
and gravities for sdO stars. 
These trends are not so remarkable in sdB stars, but considerable shifts can
be found.  
For detailed subdwarf
modelling in the optical, not just a high signal-to-noise, but medium to high resolution
data is necessary ($\Delta\lambda<3$ \AA, or ${\rm R}>3000$).

\section{Subdwarf populations}\label{Sec:sdBd}

The observed distribution of subdwarf stars can be explained in the framework of both 
the canonical formation theory (\citealt{han03a}, \citealt{han03b},
\citealt{mengel76}) 
and the hot-flasher scenario (\citealt{bertolami08}, \citealt{lanz04},
\citealt{dcruz96}), with an
interplay between atmospheric diffusion and stellar winds \citep{unglaub08}. 
Both theories have three formation channels that we describe in the context
of our observations. 

\begin{figure*}
\centering
\includegraphics[width=8cm,clip=,angle=0]{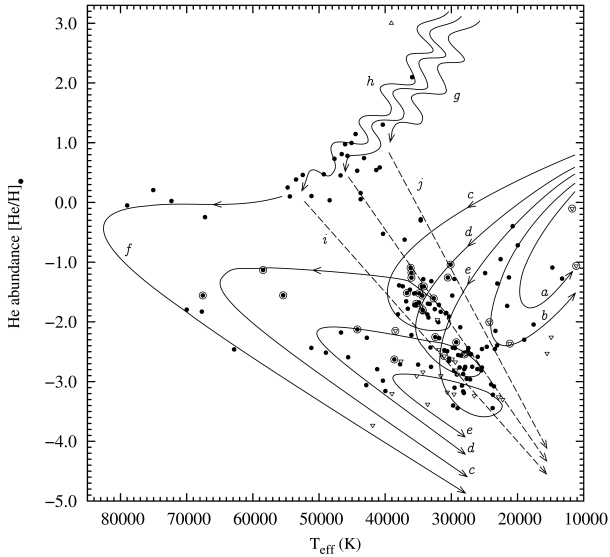}
\hspace{1cm}
\includegraphics[width=8cm,clip=,angle=0]{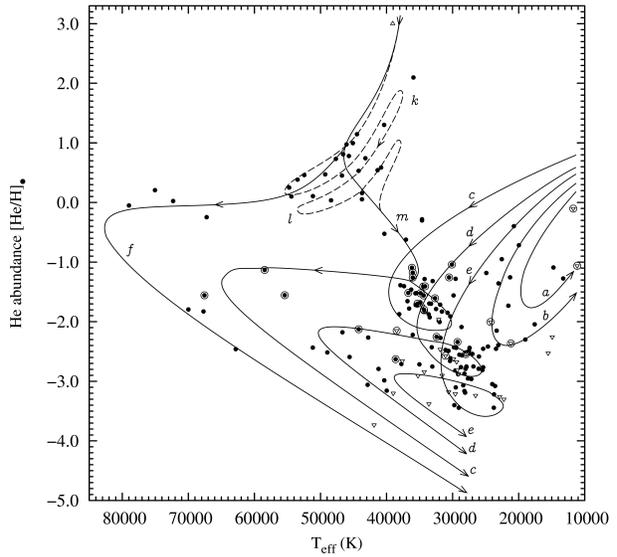}
 \caption{{\small
 Possible evolutionary sketches for sdB, sdO and He-sdO stars based on our
 observations and the predictions of the canonical ({\it left}) and hot-flasher
 ({\it right}) scenarios.  }
\label{Fig:evol1}}
\end{figure*}

\begin{figure}    
\centering
\includegraphics[width=8cm,clip=,angle=0]{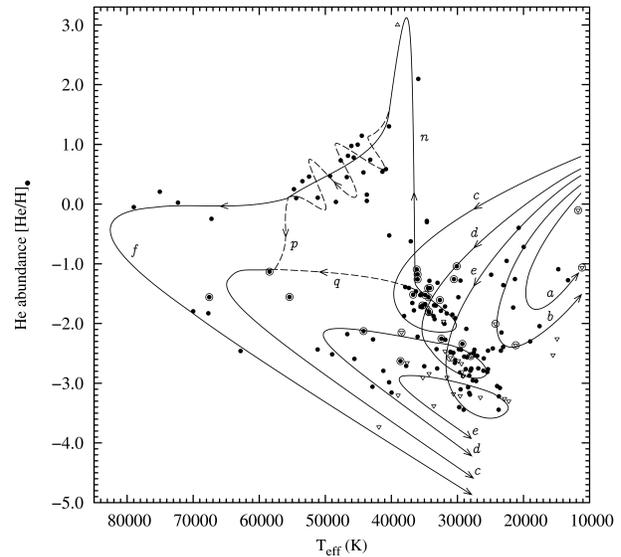}
 \caption{{\small
 Evolutionary sketches for sdB, sdO and He-sdO stars based on our
 observations and theoretical predictions with atmospheric mixing and stellar
 winds. 
 }
\label{Fig:evol2}}
\end{figure}

In Figure \ref{Fig:evol1} we present evolutionary sketches in the 
$T_{\rm eff}-{\rm He}$ diagram of the canonical and hot-flasher formation
models, 
and in Figure \ref{Fig:evol2} we
show predictions if diffusion and stellar wind may change the observed He
abundance. 

First, we describe the branches of the canonical formation.  
The {\it common-envelope channels} \citep{han03a} predict subdwarfs with low-mass MS (CE-1
channel) or WD (CE-2 channel) companions in close binaries with $0.1-10$ days
orbital periods. 
Such binaries are usually discovered in radial velocity surveys, because 
in these systems the low-mass companion is hardly visible next to a bright subdwarf. 
These subdwarfs are
predicted to have a very thin H-rich envelope and a sharp mass distribution
peaked around $0.46$ M$_{\odot}$. 
Our observations confirm a population near 28\,000 K and $\log g=5.45$. 

Hotter and more
helium abundant sdB stars near 33500 K and $\log g=5.8$ are predicted by the
{\it Roche-lobe overflow channel} (ROFL-1, \citealt{han03a}). 
These stars are expected in wide binaries with
$1-2\ {\rm M}_{\odot}$\ companions, $400-1500$ days orbital periods and 
with thin H envelopes. 
The location of possible rapid pulsators in our survey coincides with this
population and shows a higher composite frequency. 
Interestingly, the rapid pulsator class-prototype, EC 14026-2647 itself, and
the three rapid pulsators discovered later are all in very similar composite
spectra binaries \citep{odonoghue97}. 

He-sdO stars may form by {\it WD binary mergers} \citep{zhang12}. 
This channel would intrinsically explain 
single subdwarf stars that are expected to show He dominated atmospheres
enriched in CNO-processed nuclei. 
Following the He core-flash these stars evolve through
small loops (few 1000 K) in the temperature-gravity diagram of
\citet{zhang12} 
and reach their locations at various temperatures
depending on their C and N abundance. 
Binary merger channels have been proposed with low-mass stars as well as
with substellar
companions \citep{soker98}. 
The merger theory assumes a large
rotational velocity of the progeny that is not generally observed in single subdwarf stars.
\citet{geier09} found $v_{rot}\sin{i}<10$ km s$^{-1}$ for $\sim$100 single
sdB stars. 
However, \citet{lanz04} found large rotational velocities for two C-rich sdB stars 
(PG 1544+488 and JL 87) and \citet{hirsch08} found two out seven sdO stars
having a projected rotational velocity in excess of 20 km s$^{-1}$. 
Recently, \citet{geier11b} found EC 22081-1916 a single sdB star that is a fast rotator with
$v_{rot}\sin{i}\approx150$ km s$^{-1}$ and most probably a merger product of
a WD and a substellar companion. 
In close binaries tidal spin-up can cause such high rotational velocities. 
[Note added in proof: After this paper has been accepted, \cite{vennes12}
reported
that one of our targets (J1411-3053) is a fast rotating sdB star
($v_{\rm rot}{\sin}i=164\pm5$ km s$^{-1}$) in a short period sdB--WD binary.]
A targeted radial and rotational velocity study would be helpful to constrain
the properties of our subdwarfs and investigate the merger channel.

The hot-flasher scenario \citep{bertolami08} also predicts three major groups of subdwarfs. 
In the {\it early hot-flasher} channel 
the core-flash occurs after the star leaves the RGB, during
the constant luminosity phase and hot subdwarfs are predicted with standard
H/He envelopes. 
Possible rapid pulsators in our catalogue show such atmospheres; they are He
enriched and have notably less CNO in their atmospheres. 
The {\it late hot-flasher (shallow mixing)} 
theory assumes a core-flash early after
the star reached the WD cooling track. 
This concept predicts a He and N
enriched H atmosphere, similar to possible slow pulsators 
or cooler He-sdO stars in our sample. 
The {\it late hot-flasher (deep mixing)} predicts strong He, 
C and N enrichment, exactly as we found in the case of hotter He-sdO stars. 
All these features can be observed in
Figure \ref{Fig:metal} and \ref{Fig:cnoabn_nl} as well. 
In the hottest sdO stars we
found signs of increasing abundances of C and N with temperature.

Since the evolution from the primary He core-flash to the EHB or HeMS is about 40 times  
shorter than the evolution on the EHB, stars are expected to group around the
EHB or the HeMS. 
Such an accumulation is not observed for He-sdO stars.
This may suggest that these subdwarfs evolve through the HeMS in large numbers 
without slowing down, which is physically impossible. 
Hence we conclude that
they show more diverse atmospheric parameters before
settling on the EHB as sdB stars or starting their evolution towards the WD
cooling sequence.

The left panel of Figure \ref{Fig:evol1} shows evolutionary sketches in the
$T_{\rm eff}-{\rm He}$ diagram based on our
observations and the predictions of the aforementioned canonical studies. 
The distribution of sdB and He-weak sdO stars can be explained 
according to the canonical theory. 
Convection and rotation on the HB are
significant below 11\,500 K. 
BHB stars show a decreasing He abundance with surface temperature until core He
burning is on.
After core exhaustion, these stars evolve towards the AGB and lower surface
temperatures. 
Along this evolution the He abundance is expected to increase as shown by
lines {\it a} and {\it b} in Figure \ref{Fig:evol1}.
In sdB stars He is sinking to about 24\,000 K \citep{otoole08}.
Over 24\,000 K the increasing UV flux starts to bring He to the surface and
steadily increases the He abundance until about 36\,000--38\,000 K where core He
burning stops. 
In stars with a lower He abundance (thicker H envelopes and
larger total mass) this happens at lower temperatures. 
Canonical models then predict a fast evolution to higher temperatures. 
After passing a maximum temperature, stars cool and He sinks again, these
post-EHB stars rapidly evolve to WDs.
We sketched and labelled this evolution with lines {\it c}, {\it d} and {\it e}. 

The canonical formation of He-sdO stars \citep{zhang12} predicts different channels
depending on the merger type. 
Assuming He-WD binaries 
it predicts C and N enriched He-sdO stars in agreement
with our observations. 
In Figure \ref{Fig:metal} He-sdO stars
separate in temperature according to their C and N abundance. 
If these stars belong to the HeMS, this separation suggests a different
average stellar mass as well. 
Following the He core-flash the slow merger model includes accretion from a debris
disc and predicts N-rich low-mass He-sdO stars (line {\it g}). 
The fast merger model involves a period of coronal accretion and predicts higher mass
C-rich subdwarfs (line {\it h}). 
The composite model predicts a similar abundance pattern, but a broader mass
distribution where low-mass He-sdO stars are enriched in N and high mass stars
are enriched in C, hence this evolution proceeds between the slow and fast merger
tracks.
After core He exhaustion they may evolve
directly to lower temperatures and abundances as indicated by the dashed
lines {\it i} and {\it j}, or reach a maximum temperature
before joining the WD cooling tracks (line {\it f}).

The right panel of Figure \ref{Fig:evol1} shows evolutionary sketches based on 
the hot-flasher scenario of \cite{bertolami08}. 
This predicts an evolutionary link between He-sdO, the hotter sdB and He-weak sdO stars
(line {\it m}).
The different binary frequency of 
He-sdO and sdB stars implies a mixed sdB population, 
therefore we repeat the sketches for the canonical evolution 
(lines {\it c}, {\it d} and {\it e}) that may contribute to sdB stars.
The early hot-flasher and late hot-flasher (shallow mixing) channels may
also contribute to the mixed sdB population.
The hot-flasher formation of He-sdO stars predicts larger loops (dashed
lines) from He shell-flashes 
than the canonical theory, extending to 10\,000 K. 
These loops might cover the entire temperature range of He-sdO stars in our
sample. 
Along these loops stars may cross the wind limit and show
changing enhancements of C and N. 
In light of the hot-flasher scenario, our observations imply that 
stars after a He shell-flash may show 
C enriched atmospheres ({\it l}) that may transform to N-rich CNO processed
atmospheres in their cooler phase ({\it k}). 
After these transients stars either move
towards higher temperatures (line {\it f}) or settle on
the EHB among suspected rapid pulsators and join the canonical evolution of
the hotter sdB stars (line {\it c}).

Between 36\,000 and 40\,000 K a photospheric convective
mixing is predicted over $\log{g}\approx5.9$ increasing the He abundance until no H is
measurable \citep{groth85} as shown by line {\it n} in Figure \ref{Fig:evol2}. 
This mixing makes the tracking of formation theories difficult and might be responsible 
for the atmospheric properties of He-sdO stars. 
These stars may cross the wind limit periodically and transient He enhancements 
(dash-dotted line) may occur while stars drift to higher temperatures. 
With this evolution to higher temperatures the C/N ratio is steadily increasing
according to our observations. 
Convection ceases around 45\,000--52\,000 K and He sinks again \citep{otoole08}.  
As a result, He-sdO stars may return to the canonical evolution indicated by line 
{\it c}
through dashed line {\it p}, or reach a maximum temperature and follow line
{\it f}. 
The gap between 40\,000 and 54\,000 K at 
$[{\rm He/H}]_{\bullet}\approx-1.5$ coincides with the He-sdO region in our
sample suggesting that
most of the He-rich sdB stars may either pass through very fast (dashed line
{\it q}), or suffer a 
huge metal abundance increase upon entering this region. 
A larger sample could help to statistically confirm the existence of 
this empty region. 
A period of deep atmospheric mixing can also explain the high $^3$He concentrations
observed at some subdwarf stars \citep{heber09}. 
Due to a high temperature gradient in
subdwarf atmospheres the {\it pp}-chain can produce this isotope relatively
close to the photosphere. 
Because the equilibrium abundance of $^3$He is much higher
at low temperatures, $^3$He produced in deeper layers may accumulate
and replace $^4$He in the atmosphere if diffusion brings it to the surface. 

\begin{figure}
\centering
\includegraphics[width=0.7\linewidth,clip=,angle=-90]{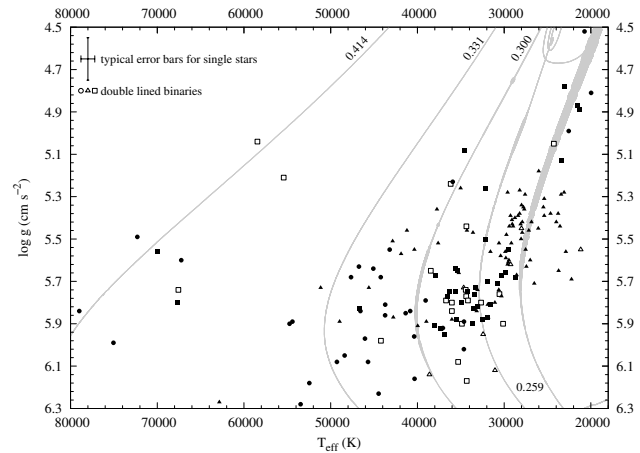}
 \caption{{\small Low-mass He WD evolutional models from
 \citet{driebe98}. Tracks for 0.259, 0.300, 0.331 and 0.414 M$_\odot$ are
 overlaid
 and labelled with the stellar mass. The 0.259 M$_\odot$ track crosses this
 region twice following a He shell-flash.}
\label{Fig:driebe}}
\end{figure}

Progenitors of low-mass WDs evolve through the parameter region occupied by subdwarfs. 
Figure \ref{Fig:driebe} shows evolutionary tracks for different stellar 
masses taken from \citet{driebe98}. 
Evolutional time-scales are inversely proportional to line widths in order to 
emphasise regions where an accumulation of stars is expected. 
Such low-mass pre-WDs are conceivable in our
sample, in particular among He-weak sdO stars, but, in general, 
the distribution of our subdwarfs does not confirm a significant fraction of
such low-mass, non-core-He burning stars.

These interpretations can qualitatively relate our observations to
the various subdwarf evolutional theories, but deeper investigations are necessary on a 
larger sample to estimate the
contribution of the channels of different subdwarf formation theories as well
as the role of diffusion and mass-loss.

\section{Catalogue of subluminous stars}\label{Sec:catalogue}

The catalogue in Table \ref{Tab:1} and \ref{Tab:6} contains 191 observations of 180
subluminous field stars sorted according to increasing right ascension. 
In resolved
binaries the primary and the companion are labelled with $A$ and $B$ suffixes, respectively. 
Our observations are typesetted with bold and {\small MILES} templates
with normal font.  
We list {\sl GALEX} $N_{\rm UV}$ and {\small GSC2.3.2} $V$ magnitudes taken from the 
{\small GSC} quick-$V$
photometry.
Atmospheric parameters like effective temperature, surface gravity, 
He and CNO abundances are listed for subdwarfs and 
only effective temperature, surface gravity and metallicity for companions.  
The interstellar reddening was estimated towards each target and we list
the values of the derived E($B-V$) colour excess.
The $V$--$J$ and $J$--$H$ colour
indices were calculated from {\small 2MASS} photometry. 
In parenthesis, we give synthetic colour indices calculated from
{\small TLUSTY} SEDs and {\small HILIB} templates in a similar fashion like in Figure \ref{Fig:2038}. 
Because of the higher uncertainties in the $V$ magnitudes the $J$--$H$ colour
indices are more reliable. 
Our comments are 
abbreviated with flags and these are explained in Table \ref{Tab:flags}. 
We performed a cross-correlation of our subdwarf list with 
SIMBAD\footnote{\url{http://simbad.u-strasbg.fr/simbad/}}
and 
The Subdwarf Database\footnote{\url{http://catserver.ing.iac.es/sddb/}} 
(\citealt{ostensen04}, \citealt{ostensen06}).
In SIMBAD a search radius of 1 arcminute was used. 
Resolved stars are marked with
one of their common identifiers in the last column. 
Ambiguous identifications
are appended with a question mark. 
Out of the 180 stars we found 32 in the PG, 16 in the First Byurakan Survey
(FBS, \citealt{mickaelian08}), 6 in the HS, 3 in the Edinburgh-Cape (EC,
\citealt{kilkenny91}) and 2 in the
Hamburg-ESO Survey (HE, \citealt{wisotzki91}) catalogues.  
We cross-correlated our data with
the list of subdwarf pulsators in \citet{ostensen10} as well, but no match was
found. 
A cross-correlation with \citet{geier11a} in search for
close-binaries revealed only the two known short-period binaries: J0321+4727
and J2349+3844, already discussed in \citet{kawka10}. 

\setcounter{table}{5}
\begin{table}
\begin{center}
\caption[3]{Flags that are used in the comments of Tables \ref{Tab:1} and \ref{Tab:6}.}
\begin{tabular}{ccp{60mm}}
\hline
Flag      & & Meaning \\
\hline 
c         &-& Contamination, scattered light from a nearby star. \\
t         &-& See text; further details can be found in Section \ref{Sec:idividuals}. \\
\#7, \#11 &-& Spectral resolution; \\
          & & \#7: ESO/NTT/EFOSC2, grism \#7, $\Delta\lambda=6.4$ \AA. \\
          & & \#11: ESO/NTT/EFOSC2, grism \#11, $\Delta\lambda=13.7$ \AA. \\
          & & If resolution is not given: NOAO/Mayall/RC, grating KPC-10A, $\Delta\lambda=4.95$ \AA. \\
d         &-& Binary decomposition degeneracy: 
              components are either similar type, or a high metallicity single line 
              binary. \\
b         &-& Spectral signatures of a cool companion. \\
rv        &-& Radial velocity with respect to the kinematic 
              LSR is over 100 km s$^{-1}$.  \\
RV        &-& Radial velocity curve proves a close binary,
              references are given in footnote.\\
f         &-& Poor fluxing below 4500 \AA.  \\
ca, na    &-& Ca\,{\sc ii} H\&K or Na\,{\sc i} D lines; indicating either a cool companion or
              interstellar absorption. \\
he        &-& He\,{\sc i}/He\,{\sc ii} line inconsistency.\\
M         &-& Data is taken from the {\small MILES} library.\\
C\&N      &-& C\&N $\lambda4640-\lambda4665$ \AA\ blend.\\
s         &-& Possible slow pulsator\\
r         &-& Possible rapid pulsator\\
\hline
\end{tabular}
\label{Tab:flags}
\end{center}
\end{table}

\subsection{Comments on individual stars and overlaps with other catalogues}
\label{Sec:idividuals}

The new analysis confirms our previous results on 
{\it PG 0314+146} (He-sdO), {\it PG 0838+133} (He-sdO), {\it PG 1432+004}
(sdB) and {\it PG 1629+081} (sdB). The new fitting method provides
consistent results with Paper I.

{\it J0047+0958} (HD 4539) was observed with NTT and is one of the two 
subdwarfs in the {\small MILES} library observed with the 2.5m Isaac Newton Telescope at 
2.3 \AA\ resolution. It is the brightest star in our sample with $V=10.24$ magnitude. 
\citet{cenarro07} determined $T_{\rm eff}=25\,200$ K and $\log g=5.40$ 
in good agreement with our $T_{\rm eff}=24\,650^{+590}_{-200}$ K and $\log g=5.38^{+0.03}_{-0.05}$. 
Then, we refitted the {\small MILES} spectrum with our method and found
$T_{\rm eff}=24\,000^{+180}_{-380}$ K and $\log g=5.21^{+0.03}_{-0.04}$. From
the higher resolution data we derived slightly lower temperature and gravity
for this star. 

{\it J0059+1544} (PHL 932) was suspected to be the central star of a
planetary nebula. Therefore, 
it has been targeted by numerous studies to
reveal its connection. \citet{frew10} derived $T_{\rm eff}=33\,490\pm73$
K, $\log g=5.81\pm0.02$ and
$[{\rm He/H}]_\bullet=-1.58\pm0.03$ and concluded
it is an sdB star exciting a H\,{\sc ii} region in the dense ISM. 
\citet{napiwotzki99} derived $T_{\rm eff}=35\,000\pm900$ K, $\log g=5.93\pm0.12$
and $[{\rm He/H}]_\bullet=-1.53\pm0.05$ and a mass of $0.28\pm0.01 {\rm M}_{\odot}$.
Our analysis confirms a slightly lower He abundance, 
we measured $T_{\rm eff}=33\,530^{+190}_{-310}$ K, $\log
g=5.83^{+0.04}_{-0.05}$ and $[{\rm He/H}]_\bullet=-1.69^{+0.06}_{-0.04}$.

{\it J0321+4727} and {\it J2349+3844} have been analysed before
by \citet{kawka10}.
For J0321+4727 they obtained 
$T_{\rm eff}=29\,200\pm300$ K, 
$\log g=5.5\pm0.1$ and 
$[{\rm He/H}]_\bullet=-2.6\pm0.1$ 
in agreement with our measurements:
$T_{\rm eff}=27\,990^{+460}_{-400}$ K, 
$\log g=5.34\pm0.07$ and 
$[{\rm He/H}]_\bullet=-2.52^{+0.17}_{-0.22}$, although our new temperature
is slightly lower. 
For J2349+3844, we derived a lower temperature. Fitting only H$_\alpha$,
\citet{kawka10} measured  
$T_{\rm eff}=28\,400\pm400$ K,
$\log g=5.4\pm0.3$ and
$[{\rm He/H}]_\bullet=-3.2\pm0.1$, while we measured 
$T_{\rm eff}=23\,770^{+330}_{-350}$ K,
$\log g=5.38^{+0.05}_{-0.06}$ and
$[{\rm He/H}]_\bullet=-3.44^{+0.25}_{-0.30}$. 
Both stars show radial
velocity variations and are confirmed binaries. 
Phase variation is
a possible source of discrepancies in measured atmospheric parameters.

{\it J0507-2802} (HE 0505-2806) and {\it J0657-7324} (CPD-73 420). 
For both
stars we have multiple observations showing variable composite spectra and
a significant IR excess. 
Changing composite features are also visible in
their spectra.
An inspection of their fields revealed crowding; both stars are visual
binaries. 
The composite spectra is most probably the result of 
contamination from the nearby, but possibly independent star. 
Atmospheric turbulence can scatter light in the slit from these stars depending on
observing conditions. 
In such cases the decomposition is important to remove
the contamination and improve the subdwarf parameter determination, although, 
it is difficult due to the variability and chromatic aberration
of the scattered light. 
Reliable parameters for the nearby star cannot be
derived. 

{\it J0639+5156} was observed by \citet{vuckovic12} in a photometric
campaign to look for bright pulsating subdwarfs. 
This star is the brightest rapid pulsator to date with a main pulsating
period of 260 s and amplitude of $\sim$13.5 mmag in the R filter. 
A slow pulsation mode with $\sim$2 mmag at 1765 s period suggests the star is
a hybrid pulsator. 

{\it J0806+1527} was observed by \citet{baran11} and found to be a hybrid
pulsator with two rapid and two slow modes. 

{\it J0851-1712} (TYC 6017-419-1). 
We derived $T_{\rm eff}=39\,060$ K and $\log g=5.79$ for the
star, placing it at the end of the EHB on the HeMS. 
We did not detect H in
its atmosphere while the C/N ratio remained below 0.12 and the O/N ratio
below 0.54 by number
with increasing He abundance. We classify this star as a N-class object.
The C/N ratio of 0.05 and O/N ratio of 0.2 by number is 
typical for CNO processed material. Peculiar abundances can be the result of 
mixing, diffusion and mass-loss processes \citep{unglaub01}. 
The atmospheric parameters of the star 
correspond to the region of photospheric convective mixing in the 
$T_{\rm eff}-\log{g}$ diagram. 
High resolution and high signal-to-noise observations could help improving 
the CNO abundance determination of this exotic star.

{\it J0934-2512.} Our previous analysis in Paper I found peculiar
He\,{\sc i} 
$\lambda$4471/He\,{\sc ii} $\lambda4686$ line ratios inconsistent with other Balmer
line spectra with similar He abundances. 
He\,{\sc ii} lines are relatively
strong for the temperature of the star. 
We verified that the relevant atomic data (energy levels, oscillator
strengths, line widths) in our model atoms and line lists are up-to-date
for these He lines. Therefore, the cause for the observed line strength
inconsistencies remains to be determined. 
Our new analysis showed that the star is in the post-EHB evolutional phase in the
$T_{\rm eff}-\log g$ diagram, but belongs to the possible short period sdB pulsators in
the $T_{\rm eff}-{\rm He}$ diagram. 
The spectrum indicates
a radial velocity that does not exclude binarity. 
To test for a subdwarf companion we repeated our fit starting with a single star model
and included a companion from the spectral library compiled from 
all other subdwarfs modelled in this work. 
We found the spectrum might be dominated by an sdB star and
the He\,{\sc ii} line can be the signature of a hot, but faint companion. An
underestimated temperature, gravity and He abundance could have similar
effects, but would affect the entire spectrum.
The true nature of the He line and surface gravity
inconsistency would require high resolution spectra and radial velocity
measurements. 
To a lesser extent such inconsistencies can be observed in six other sdB stars
with various atmospheric parameters. 
Because of a possible degeneracy in the spectral
decomposition of such binaries they are listed as single stars with flag "he"
in Table \ref{Tab:1} referring to He line inconsistency.

\input{table_incl.tex}

{\it J1831+0851} is possibly an imminent WD progenitor. 
It lies well below the HeMS at a
gravity higher than typical for He-sdO stars. Also, its
He abundance is not comparable to its spectral class. 

{\it J1845-4138}. We found
$T_{\rm eff}=35\,930^{+\ 840}_{-4770}$ K, $\log g=5.23^{+0.27}_{-0.23}$
and $[{\rm He/H}]_\bullet=2.1^{+1.1}_{-0.4}$ in good agreement with our previous
analysis. This star shows an extremely high He abundance compared to its
location in the $T_{\rm eff}-{\log g}$ diagram. It is in the region of
post-EHB evolution, where He-weak sdB stars are, but shows a He abundance
typical for He-sdO stars. Its surface gravity is also inconsistent with its
location. A high-dispersion spectrum is needed for further analysis. No
infrared excess was found.

{\it J1902-5130} (CD-51 11879, LSE 263) was analysed by \citet{husfeld89}.
They found 
$T_{\rm eff}=70\,000\pm2500$ K,
$\log g=4.9\pm0.25$, 
$[{\rm H/He}]_\bullet<-1.0$, 
$[{\rm C/He}]_\bullet=-4.0\pm0.5$ and
$[{\rm N/He}]_\bullet=-2.4\pm0.3$.
Our analysis confirms the temperature and the He dominated, N-rich
atmosphere, but provided a 0.5 dex higher surface gravity. 
Missing or incomplete atomic data and broadening parameters in the earlier
analysis can be accounted for such discrepancy.
Due to upper limit determinations and asymmetric errors, a direct comparison
with our results is not possible. 

{\it J2043+1034} (LSIV+10 9) was analysed by \citet{dreizler93} who found
$T_{\rm eff}=44\,500\pm1000$ K,
$\log g=5.55\pm0.15$
and mass fractions ($\beta_{\rm x}=m_{\rm x}/m_{\rm total}$):
$\log\beta_{\rm He}=-0.001$, 
$\log\beta_{\rm C}=-1.85\pm0.1$,
$\log\beta_{\rm N}=-2.45\pm0.15$ and
$\log\beta_{\rm O}=-2.40\pm0.15$.
Our analysis resulted in 
$T_{\rm eff}=45\,700^{+1000}_{\ -630}$ K,                   
$\log g=6.08^{+0.13}_{-0.24}$,
and mass fractions:
$\log\beta_{\rm He}=-0.023$,    
$\log\beta_{\rm C}=-2.69$,
$\log\beta_{\rm N}=-2.71$ and
$\log\beta_{\rm O}<-2.12$.
We derived a notably higher gravity and lower abundances, in particular for
He and C. 
This can be due to the incomplete composition of our models and that we
fitted
the entire spectrum not just selected lines. 
The relatively lower quality of our fits for He-sdO stars also suggests the
importance of heavier metals in their modelling.   

{\it J2245+3221} (HS 2242+3206) was fitted by \citet{edelmann03}. 
Their
analysis of H$_{\beta}-$H$_{10}$, He\,{\sc ii} 4686, He\,{\sc i} 4471 and
He\,{\sc i} 4026 lines
provided $T_{\rm eff}=29\,300\pm1000$ K, $\log g=5.65\pm0.15$ and
$[{\rm He/H}]_\bullet=-2.8\pm0.2$. Our
data covered the optical range from 3700 to 7150 \AA\ and we derived
$T_{\rm eff}=29\,530^{+470}_{-320}$ K, $\log g=5.61^{+0.08}_{-0.09}$ and
$[{\rm He/H}]_\bullet=-3.1^{+0.29}_{-0.62}$ in good agreement.

\section{Conclusions}\label{Sec:conclusions}

We presented a homogeneously modelled sample of 124 sdB and 42 sdO stars and
determined non-LTE atmospheric parameters 
by considering H, He and CNO opacities. 
With binary decomposition
we were able to derive accurate atmospheric parameters for the members in
$\sim$30 composite spectra binaries. 
Our study shows that sdB stars concentrate in two groups in the
$T_{\rm eff}-{\log g}$ and $T_{\rm eff}-{\rm He}$ diagrams, 
suggesting two typical H envelopes with different masses and compositions.
He-sdO and He-weak sdO stars also clearly separate in the $T_{\rm eff}-{\rm
He}$ diagram and He-weak sdO stars 
can be related to sdB stars based on their He abundance
and binary frequency. 

The developed method for spectral decomposition can reliably determine atmospheric
parameters in composite spectra binaries. 
The method can also be used to remove the spectral contamination caused
by merging stars from subdwarf spectra. 
Such cleaning process and homogeneous modelling 
on large samples are necessary to provide accurate atmospheric parameters for 
evolutional, pulsational and diffusion theories. 
We investigated the impact of binarity on subdwarf atmospheric parameters 
and showed that unresolved composite spectra binaries can 
cause significant shifts. 
Such systematic shifts can easily overwhelm non-LTE or abundance
effects
and we consider binary decomposition necessary for precision spectroscopy
of hot subdwarfs with a notable IR excess. 

We did not find a significant sdB--A binary frequency with our current
selection criteria. 
However, a joint analysis of ultraviolet and optical measurements would
be necessary to spectroscopically distinguish the components in such
binaries. 
Recently, \citet{girven12} investigated the companions of hot subdwarfs 
and found that F, G and K type MS companions outnumber A and M-type stars.

Our abundance analysis of He-sdO stars confirms the C\&N classification scheme of
\citet{stroeer07}: cooler stars show N enriched atmospheres, hotter
He-sdO stars are C-rich while we detected both C and N mostly in between these two
groups.
The C\&N-class objects suggest a C/N ratio of $\sim$3 in He-sdO stars.  
The spectra of cooler sdB stars show a N enrichment while hotter sdB stars
do not show notable CNO abundances in general.

The observed luminosity distribution suggests a shorter EHB ($\sim$100 Myr)
and longer post-EHB ($\sim50$ Myr) evolution than previous studies found. 
The luminosity distribution of sdB stars is sharply peaked at
$\log({\rm L}/{\rm L}_\odot)\approx1.5$ assuming an average sdB mass of
$0.48$ M$_\odot$. 
We found two peaks in the luminosity distribution of sdO (post-EHB) and He-sdO stars.

We reviewed our results in terms of the current evolutional theories.
Our study does not give satisfying answers for the relative 
contributions of the
various channels, but shows evidences that
He-sdO and He-rich sdB stars are related. 
The He-weak sdO sequence
and the group of He-rich sdB stars are mixtures of different populations. 

A binarity analysis of the hottest He-sdO stars
($T_{\rm eff}>60\,000$ K) could provide interesting results about the relation of
these stars to cooler He-sdO or He-weak sdO stars. 
In this region the sequences of evolved He-sdO stars and He-weak sdO
stars overlap. 
A much larger sample could help clarifying the surface gravity 
diversity observed at these stars, which currently suggests a connection to
the He-weak sequence.

Further follow-up is necessary at higher resolution and high signal-to-noise
to improve CNO determination in the optical and better understand subdwarf 
stars, and the contributions of various
formation and evolution channels.
With higher quality data it may become possible to derive reliable O
abundances that could offer further clues in understanding these stars. 

In addition to subdwarfs, 
our survey recovered six WDs that will be discussed in a forthcoming paper.

\section{The sd1000 Collaboration}\label{Sec:future}

\begin{figure}
\begin{center}  
\includegraphics[width=1\linewidth,clip=,angle=0]{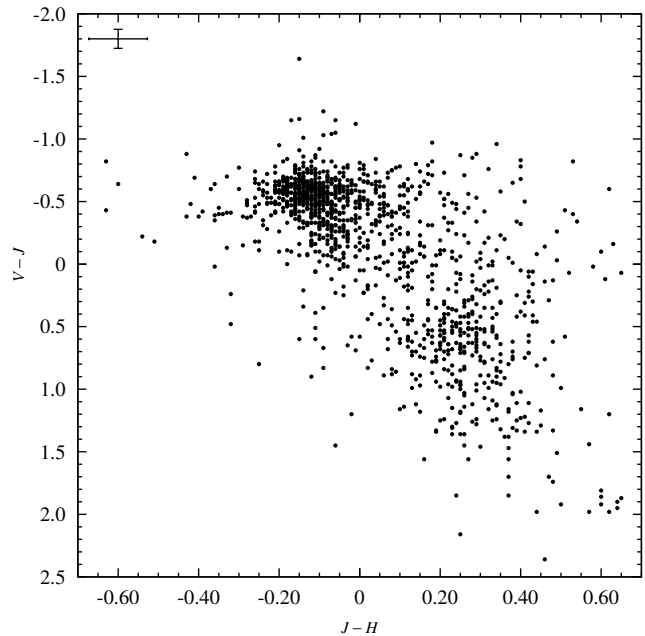}
 \caption{\small
$V$$-$$J$ - $J$$-$$H$ diagram for $\sim$1900 hot subdwarf stars in The 
Subdwarf Database \citep{ostensen06}, data were collected from
various sources. Typical error bars for colour indices 
were calculated from average photometric errors.
\label{Fig:sddb}}   
\end{center}
\end{figure}

A similar study on a larger sample could provide a
more complete picture of hot subdwarfs. 
For this reason we 
would like to extend this pilot study and initiate a cooperative work under the name 
"SD1000 Collaboration", dedicated to process $\sim$1000 hot subdwarf spectra in a
homogeneous way. 
Such an analysis can provide important observational constraints for
theoretical studies and help the advance of subdwarf research. 
The Subdwarf Database currently holds 2399 stars and
about 1900 of these have $V$, $J$ and $H$ photometric measurements. 
Figure \ref{Fig:sddb} shows the colour indices of these stars. 
About $\sim$800 stars in the data base are brighter than
$V=15$ and $\sim$1150 are brighter than $V=16$ providing a rich source of bright
targets for the project. 

However, recording new, high-quality optical spectra (R $>3000$ or
$\delta\lambda<3$ \AA, SNR $>100$) for $\sim$1000 stars would require about
60--80 nights on 4-m class telescopes what is not practically feasible.
Acquiring data from public archives could be a way, although, data reduction of various
observing runs can introduce a significant overhead and not all data are
available in such archives. 
Moreover, it is important to maximise the
number of spectra to number of instruments ratio in the sample to
decrease instrumental biases. 
The most efficient strategy would use 
contributed data for the bulk and obtain new observations for a small
subset to fill the sample.
Therefore, the success of this project depends strongly on contributed data
and we look for collaborators. 
We estimate that enough observations ($\sim$2000 stars) have been taken over the last 30 years
to compile the target sample.

Model atmosphere
analysis would require $\sim$2000 hours with single star models and 8 processors.
Using binary decomposition and estimating about 20 per cent composite spectra in the sample
would extend the analysis to $\sim$2400 hours, which is still manageable.

\section*{Acknowledgments}
We acknowledge support from the Grant Agency of the Czech Republic
(GA \v{C}R P209/10/0967) and from the Grant Agency of the Academy of
Sciences of the Czech Republic (IAA 300030908, IAA 301630901). 
We would like to thank our referee Roy {\O}stensen for his valuable 
comments and suggestions.
This research has made use of the SIMBAD database 
and the VizieR catalogue access tool,
operated at CDS, Strasbourg, France.

\label{lastpage}

\bsp

\newpage

\begin{appendix}
\section{CNO abundance correlations}

\begin{figure*}
\includegraphics[width=0.42\linewidth,bb=67 154 551 700,clip=,angle=-90]{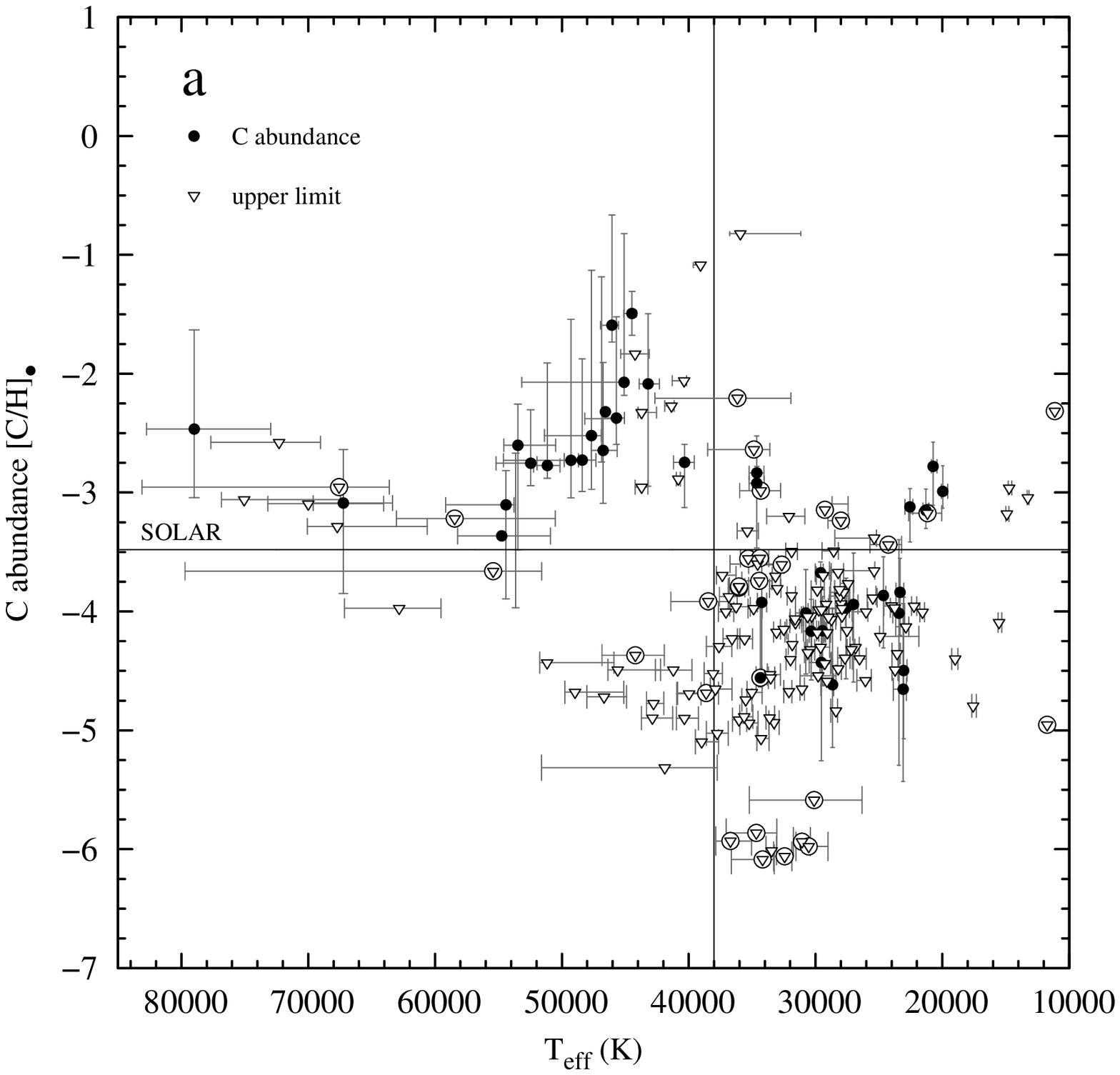}
\includegraphics[width=0.42\linewidth,bb=67 154 551 700,clip=,angle=-90]{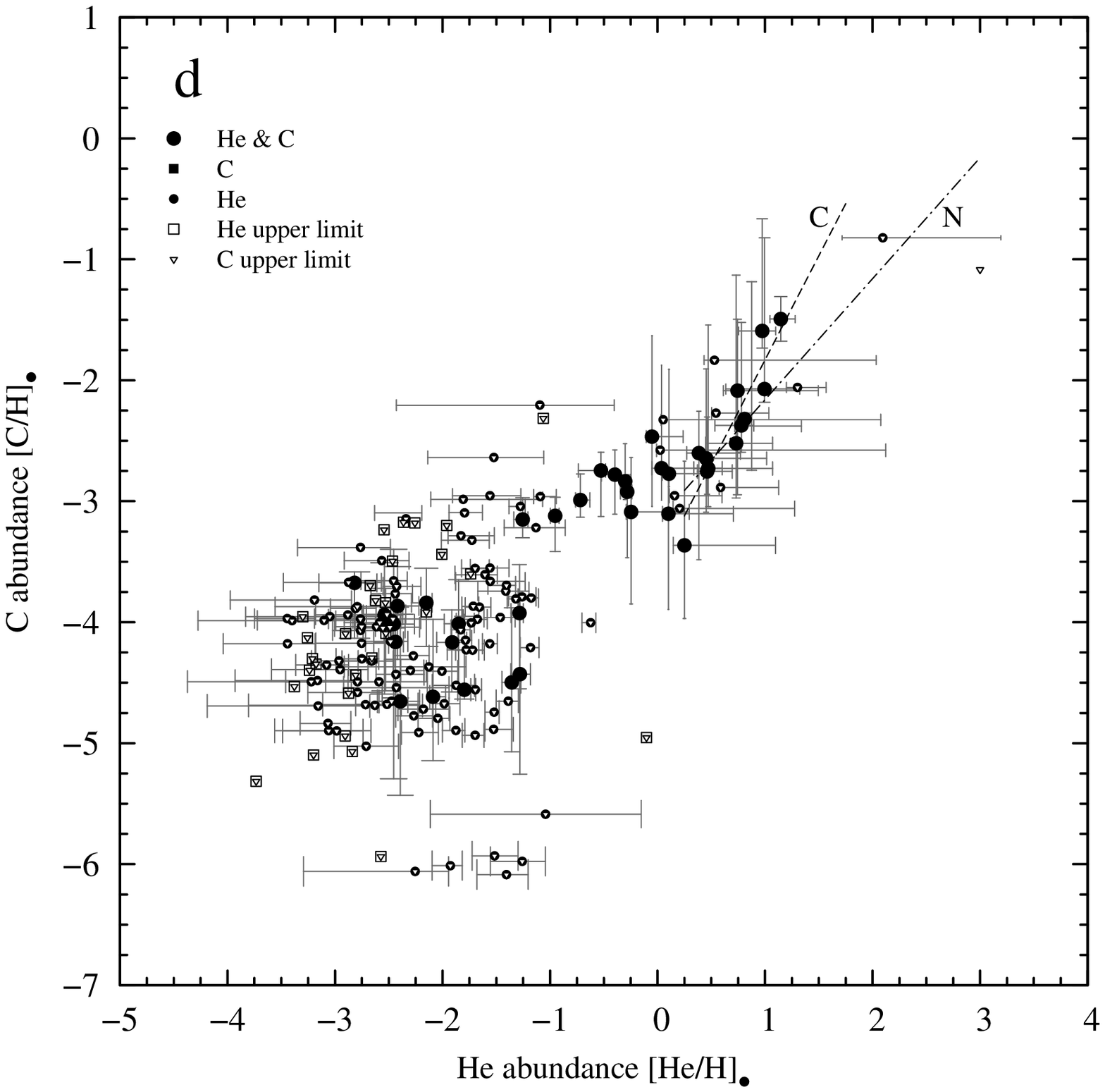}
\caption{Panels {\it a} and {\it d} of Figure \ref{Fig:cnoabn_nl} complemented with error bars and
 upper limit measurements. {\it Left:} Observed C abundances with respect to effective temperature.
 The vertical line at 38\,000 K separates sdB and sdO stars and the horizontal
 line shows the solar C abundance. {\it Right:} C and He abundance
 correlations. }
\label{Fig:a1}
\end{figure*}
\begin{figure*}
\includegraphics[width=0.42\linewidth,bb=67 154 551 700,clip=,angle=-90]{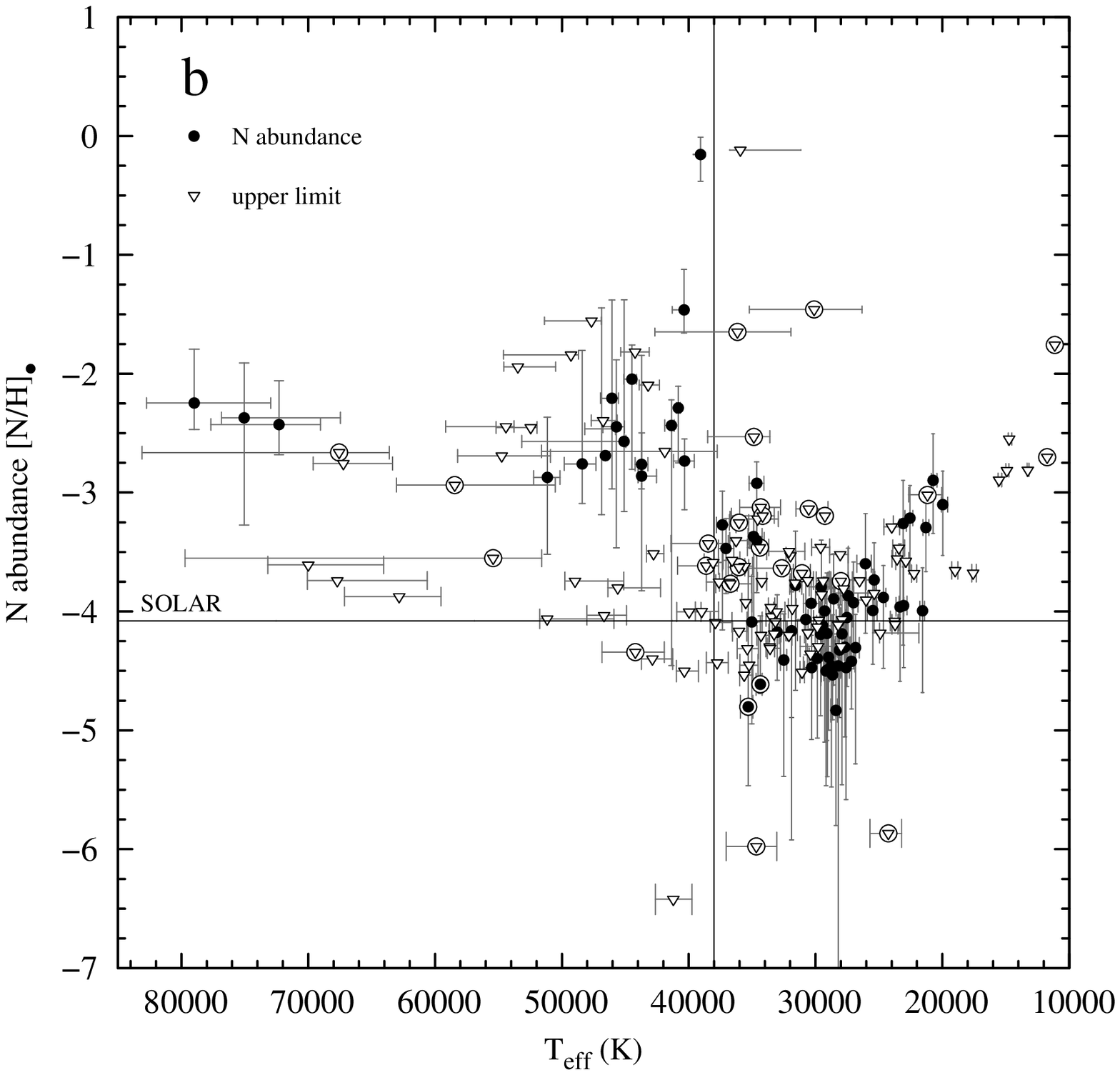}
\includegraphics[width=0.42\linewidth,bb=67 154 551 700,clip=,angle=-90]{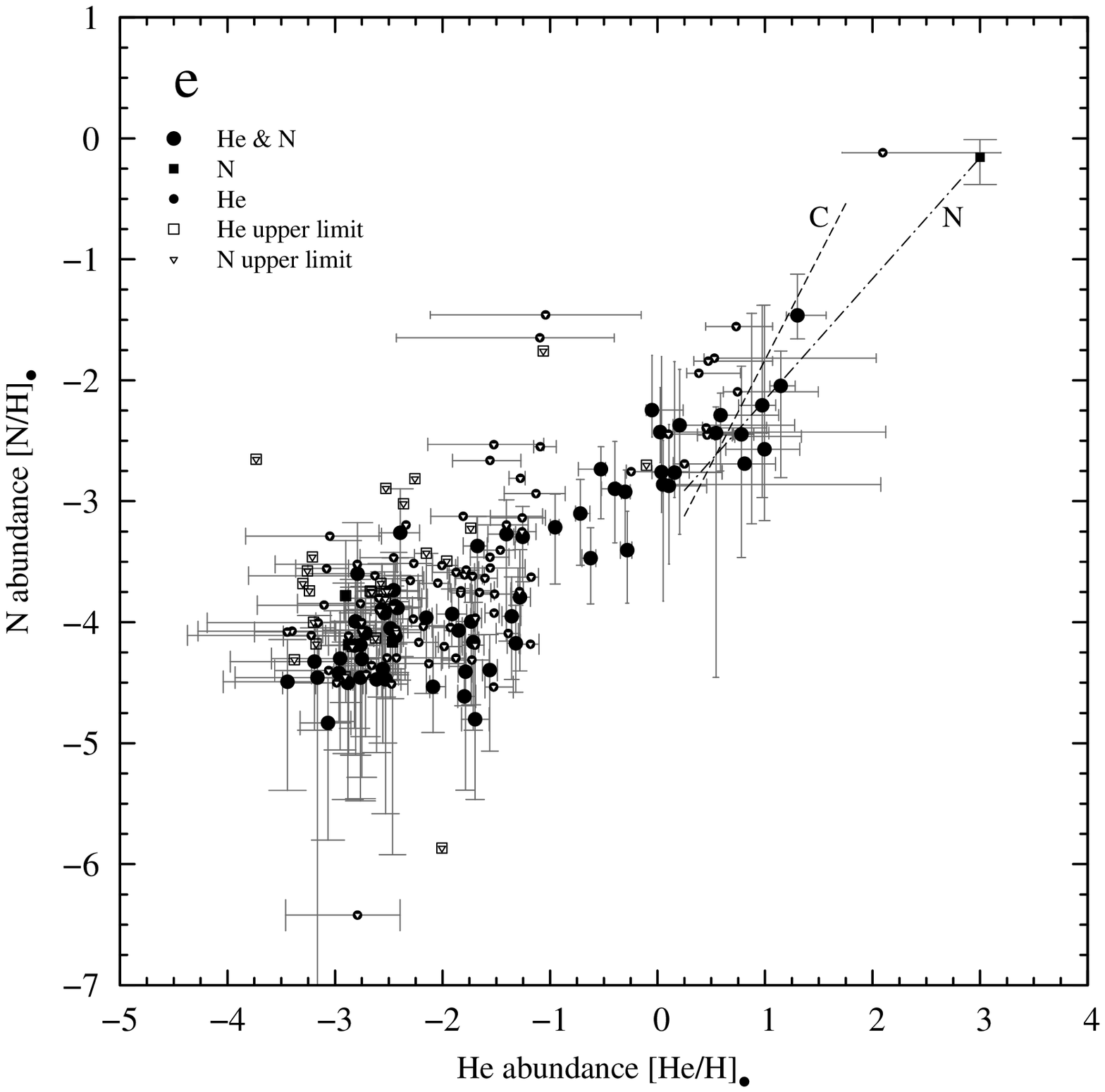}
\caption{Panels {\it b} and {\it e} of Figure \ref{Fig:cnoabn_nl} complemented with error bars and
 upper limit measurements. {\it Left:} Observed N abundances with respect to
 effective temperature.
 The vertical line at 38\,000 K separates sdB and sdO stars and the horizontal
 line shows the solar N abundance. {\it Right:} N and He abundance
 correlations.}
\label{Fig:a2}
\end{figure*}

\begin{figure*}
\includegraphics[width=0.55\linewidth,bb=67 154 551 700,clip=,angle=-90]{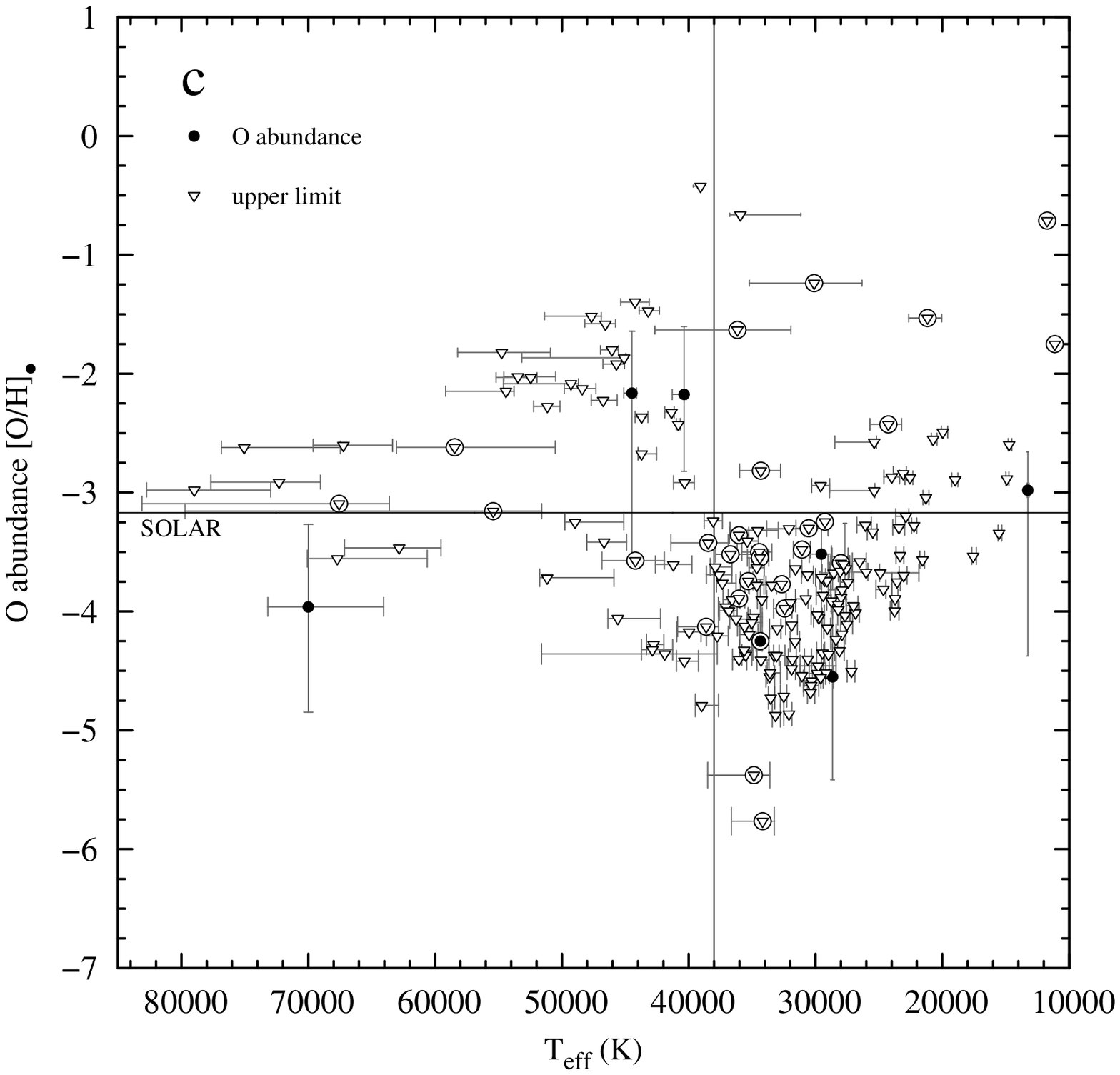}
\caption{Panel {\it c} of Figure \ref{Fig:cnoabn_nl} complemented with error bars and
 upper limit measurements shows our observed O abundances with respect to
 effective temperature.
 The vertical line at 38\,000 K separates sdB and sdO stars and the horizontal
 line shows the solar O abundance.}
\label{Fig:a3}
\end{figure*}
\begin{figure*}
\includegraphics[width=0.55\linewidth,bb=67 154 551 700,clip=,angle=-90]{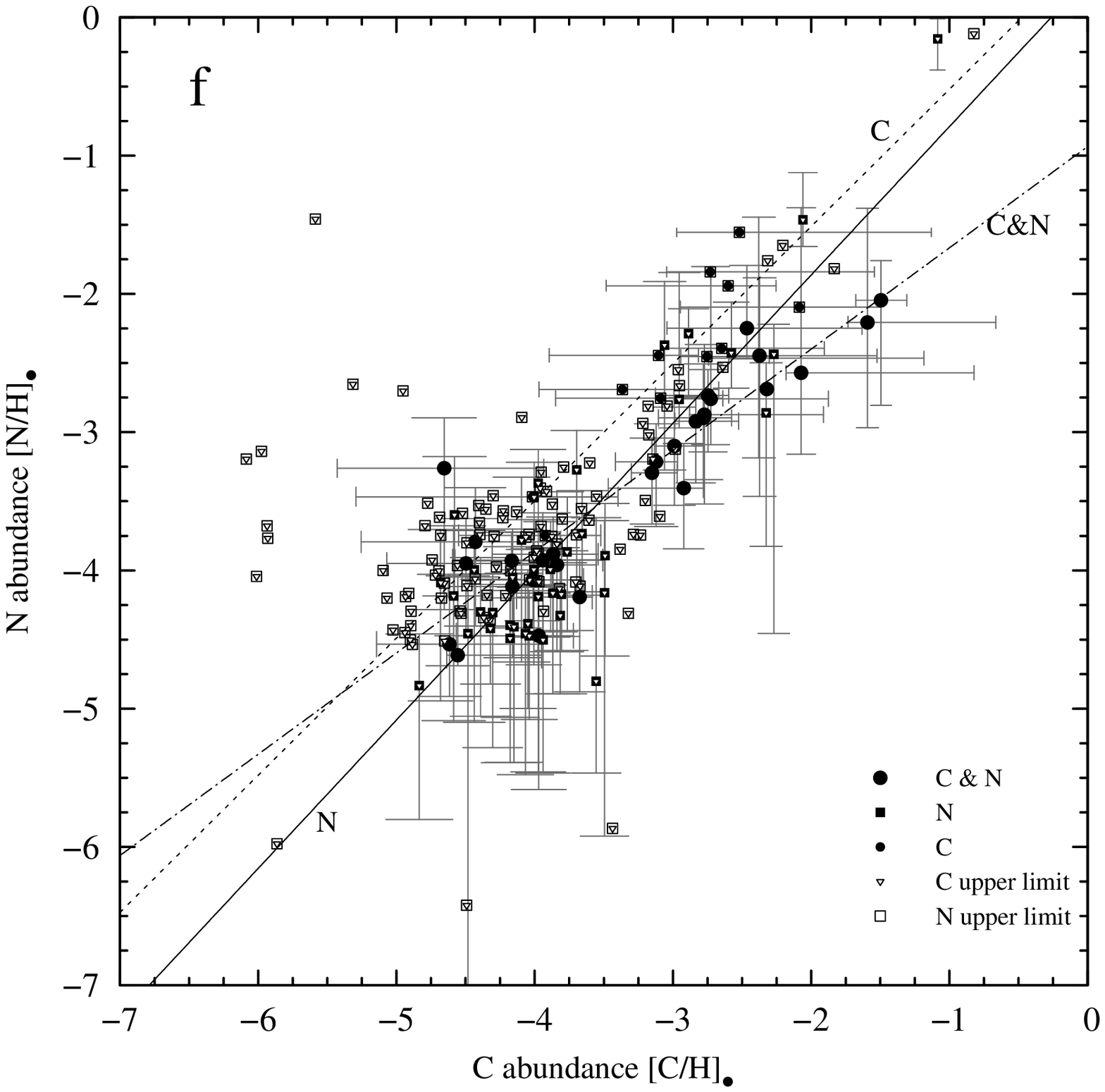}
\label{Fig:a4}
\caption{Panel {\it f} of Figure \ref{Fig:cnoabn_nl} complemented with
error bars and
 upper limit measurements showing the C and N abundance
 correlations found in the {\sl GALEX} sample.}
\end{figure*}

\end{appendix}

\end{document}

%% file: table_incl.tex
\setcounter{table}{1}
\setlength{\tabcolsep}{5pt}
\newpage
\begin{landscape}
\begin{table}
\begin{center}
\begin{minipage}[h]{250mm}
\renewcommand{\footnoterule}{\vspace*{-15pt}}
\setstretch{1.08}
\caption[]{Atmospheric parameters of hot subdwarf stars in the {\sl GALEX} survey.}
\begin{tabular}{l|cccllr@{.}lr@{.}lr@{.}lr@{.}lcrr|ll}
\hline
\multicolumn{1}{c|}{${\rm GALEXJ}$} & ${\rm Type}$   & $N_{\rm UV}$ & $V$ &\multicolumn{1}{c}{${\rm T}_{\rm eff}$}    & \multicolumn{1}{c}{$\log g$}
&\multicolumn{8}{c}{${\rm Abundances,\ }[{\rm X/H}]_\bullet
 {\rm\ or\ [Fe/H]}$}&\multicolumn{1}{c}{${\rm E}(B-V)$} &\multicolumn{1}{c}{$V-J$}&\multicolumn{1}{c|}{$J-H$}&${\rm Comments}$&${\rm Other\ name}$\\
\cline{7-14}                &    & ${\rm (mag)}$ & ${\rm (mag)}$&\multicolumn{1}{c}{${\rm (K)}$}
&\multicolumn{1}{c}{${\rm (cm\ s}^{-2}{\rm )}$} &\multicolumn{2}{c}{${\rm He}$}  &
\multicolumn{2}{c}{${\rm C}$} & \multicolumn{2}{c}{${\rm N}$}
&\multicolumn{2}{c}{${\rm O}$}& ${\rm (mag)}$ &${\rm (mag)}$&${\rm (mag)}$& &\\
\hline
{\bf 003531.98$+$245917.22}  &{\bf sdB}    &$13.61$&$14.16$&$35230^{+\ 740}_{-\ 700}$&$5.66^{+0.07}_{-0.15}$&$-2$&$91>$               &$-4$&$94>$               &$-4$&$45>$               &$-4$&$19>$          & 0.02 &$-0.474$&$-0.152$& & PG 0032+247         \\
{\bf 003639.99$+$375553.37}  &{\bf sdB}    &$12.75$&$14.07$&$38970^{+\ 500}_{-1330}$ &$5.89^{+0.10}_{-0.09}$&$-3$&$20>$               &$-5$&$10>$               &$-4$&$00>$               &$-4$&$79>$          & 0.01 &$-0.709$&$-0.097$& & FBS 0033+376        \\
{\bf 004729.22$+$095855.69}  &{\bf sdB}    &$10.44$&$10.24$&$24650^{+\ 590}_{-\ 200}$&$5.38^{+0.03}_{-0.05}$&$-2$&$42^{+0.20}_{-0.07}$&$-3$&$87^{+0.33}_{-0.44}$&$-3$&$88^{+0.27}_{-0.59}$&$-3$&$81>$          & 0.01 &$-0.576$&$-0.123$&\#7,t & HD 4539        \\
     004729.22$+$095855.69   &$\dots$      &$\dots$&$\dots$&$24000^{+\ 180}_{-\ 380}$&$5.21^{+0.03}_{-0.04}$&$-2$&$25^{+0.07}_{-0.09}$&$-3$&$89^{+0.13}_{-0.22}$&$-3$&$67^{+0.14}_{-0.26}$&$-3$&$92>$          & 0.00 &$\dots$ &$\dots$ &M,t & HD 4539          \\
\hline
\end{tabular}
\label{Tab:1}
\end{minipage}
\end{center}
\end{table}
\end{landscape}

\setcounter{table}{1}
\setlength{\tabcolsep}{5pt}
\newpage
\begin{landscape}
\begin{table}
\begin{center}
\begin{minipage}[h]{250mm}
\renewcommand{\footnoterule}{\vspace*{-15pt}}
\setstretch{1.08}
\caption[]{Atmospheric parameters of hot subdwarf stars in the {\sl GALEX} survey. Continued.}
\begin{tabular}{l|cccllr@{.}lr@{.}lr@{.}lr@{.}lcrr|ll}
\hline
\multicolumn{1}{c|}{${\rm GALEXJ}$} & ${\rm Type}$   & $N_{\rm UV}$ & $V$ &\multicolumn{1}{c}{${\rm T}_{\rm eff}$}    & \multicolumn{1}{c}{$\log g$}
&\multicolumn{8}{c}{${\rm Abundances,\ }[{\rm X/H}]_\bullet
 {\rm\ or\ [Fe/H]}$}&\multicolumn{1}{c}{${\rm E}(B-V)$} &\multicolumn{1}{c}{$V-J$}&\multicolumn{1}{c|}{$J-H$}&${\rm Comments}$&${\rm Other\ name}$\\
\cline{7-14}                &    & ${\rm (mag)}$ & ${\rm (mag)}$&\multicolumn{1}{c}{${\rm (K)}$}
&\multicolumn{1}{c}{${\rm (cm\ s}^{-2}{\rm )}$} &\multicolumn{2}{c}{${\rm He}$}  &
\multicolumn{2}{c}{${\rm C}$} & \multicolumn{2}{c}{${\rm N}$}
&\multicolumn{2}{c}{${\rm O}$}& ${\rm (mag)}$ &${\rm (mag)}$&${\rm (mag)}$& &\\
\hline
{\bf 050735.80$+$034815.00}  &{\bf sdB}    &$13.82$&$14.42$&$23990^{+\ 630}_{-\ 610}$&$5.42^{+0.08}_{-0.11}$&$-3$&$05^{+0.48}_{-0.78}$&$-3$&$95>$               &$-3$&$29>$               &$-2$&$87>$               & 0.10 &$-0.205$&$0.005$ &\#7,rv &                  \\
{\bf 055542.07$+$640812.60}  &{\bf sdO}    &$13.07$&$13.33$&$45600^{+\ 770}_{-3380}$ &$5.73^{+0.04}_{-0.39}$&$-2$&$59^{+0.45}_{-0.29}$&$-4$&$49>$               &$-3$&$80>$               &$-4$&$06>$               &\multicolumn{1}{r}{$<$0.22$\quad$}&$-0.381$&$-0.097$&f,b,na & FBS 0550+641     \\
{\bf 061325.26$+$342053.09}  &{\bf sdB}    &$13.60$&$13.63$&$34250^{+\ 330}_{-\ 390}$&$5.75^{+0.10}_{-0.06}$&$-1$&$28^{+0.04}_{-0.08}$&$-3$&$92^{+0.40}_{-0.63}$&$-3$&$75>$               &$-3$&$90>$               & 0.17 &$-0.408$&$0.027$ &na,r &                  \\
{\bf 063952.00$+$515658.00}  &{\bf sdB}    &$12.47$&$11.97$&$30400^{+\ 250}_{-\ 340}$&$5.58^{+0.13}_{-0.04}$&$-2$&$66^{+0.20}_{-0.28}$&$-4$&$32>$               &$-4$&$36>$               &$-4$&$68>$               & 0.19 &$-0.554$&$-0.142$&na,b,rv,r\footnote[2]{\citet{vuckovic12}},t & \\
\hline
\end{tabular}
\end{minipage}
\end{center}
\end{table}
\end{landscape}

\setcounter{table}{1}
\setlength{\tabcolsep}{5pt}
\newpage
\begin{landscape}
\begin{table}
\begin{center}
\begin{minipage}[h]{250mm}
\renewcommand{\footnoterule}{\vspace*{-15pt}}
\setstretch{1.08}
\caption[]{Atmospheric parameters of hot subdwarf stars in the {\sl GALEX} survey. Continued.}
\begin{tabular}{l|cccllr@{.}lr@{.}lr@{.}lr@{.}lcrr|ll}
\hline
\multicolumn{1}{c|}{${\rm GALEXJ}$} & ${\rm Type}$   & $N_{\rm UV}$ & $V$ &\multicolumn{1}{c}{${\rm T}_{\rm eff}$}    & \multicolumn{1}{c}{$\log g$}
&\multicolumn{8}{c}{${\rm Abundances,\ }[{\rm X/H}]_\bullet
 {\rm\ or\ [Fe/H]}$}&\multicolumn{1}{c}{${\rm E}(B-V)$} &\multicolumn{1}{c}{$V-J$}&\multicolumn{1}{c|}{$J-H$}&${\rm Comments}$&${\rm Other\ name}$\\
\cline{7-14}                &    & ${\rm (mag)}$ & ${\rm (mag)}$&\multicolumn{1}{c}{${\rm (K)}$}
&\multicolumn{1}{c}{${\rm (cm\ s}^{-2}{\rm )}$} &\multicolumn{2}{c}{${\rm He}$}  &
\multicolumn{2}{c}{${\rm C}$} & \multicolumn{2}{c}{${\rm N}$}
&\multicolumn{2}{c}{${\rm O}$}& ${\rm (mag)}$ &${\rm (mag)}$&${\rm (mag)}$& &\\
\hline
{\bf 084024.05$+$113911.53}   &{\bf   B}    &$13.89$&$13.19$&$13270^{+\  40}_{-\  90}$&$4.14^{+0.02}_{-0.02}$&$-1$&$27^{+0.04}_{-0.11}$&$-3$&$04>$               &$-2$&$81>$               &$-2$&$98^{+0.32}_{-1.40}$& 0.00 &$0.000$ &$-0.041$&na & NVSS J084025+113959\\
{\bf 084143.80$+$130431.00}   &{\bf He-sdO} &$12.91$&$13.66$&$47660^{+3710}_{-\ 770}$ &$5.68^{+0.29}_{-0.24}$&$0$&$73^{+0.34}_{-0.28}$ &$-2$&$52^{+1.39}_{-0.45}$&$-1$&$55>$               &$-1$&$52>$               & 0.12 &$-0.643$&$-0.154$&\#7,C\&N & PG 0838+133\\
{\bf 084535.54$+$194150.18}   &     BHB     &$13.33$&$13.34$&$21570^{+\ 210}_{-\ 170}$&$4.87^{+0.03}_{-0.04}$&$-1$&$73^{+0.03}_{-0.14}$&$-4$&$01>$               &$-3$&$99^{+0.36}_{-0.69}$&$-3$&$56>$               & 0.04 &$-0.328$&$-0.085$& & LB   400 ?       \\
{\bf 085137.75$+$244151.69}   &{\bf sdO}    &$12.68$&$14.35$&$70000^{+3190}_{-5940}$  &$5.56^{+0.07}_{-0.10}$&$-1$&$79^{+0.17}_{-0.14}$&$-3$&$09>$               &$-3$&$61>$               &$-3$&$96^{+0.69}_{-0.89}$& 0.04 &$0.035$ &$0.120$ &na & PG 0848+249      \\
\vspace{-2mm}\\
\hline
\end{tabular}
\end{minipage}
\end{center}
\end{table}
\end{landscape}

\setcounter{table}{1}
\setlength{\tabcolsep}{5pt}
\newpage
\begin{landscape}
\begin{table}
\begin{center}
\begin{minipage}[h]{250mm}
\renewcommand{\footnoterule}{\vspace*{-15pt}}
\setstretch{1.08}
\caption[]{Atmospheric parameters of hot subdwarf stars in the {\sl GALEX} survey. Continued.}
\begin{tabular}{l|cccllr@{.}lr@{.}lr@{.}lr@{.}lcrr|ll}
\hline
\multicolumn{1}{c|}{${\rm GALEXJ}$} & ${\rm Type}$   & $N_{\rm UV}$ & $V$ &\multicolumn{1}{c}{${\rm T}_{\rm eff}$}    & \multicolumn{1}{c}{$\log g$}
&\multicolumn{8}{c}{${\rm Abundances,\ }[{\rm X/H}]_\bullet
 {\rm\ or\ [Fe/H]}$}&\multicolumn{1}{c}{${\rm E}(B-V)$} &\multicolumn{1}{c}{$V-J$}&\multicolumn{1}{c|}{$J-H$}&${\rm Comments}$&${\rm Other\ name}$\\
\cline{7-14}                &    & ${\rm (mag)}$ & ${\rm (mag)}$&\multicolumn{1}{c}{${\rm (K)}$}
&\multicolumn{1}{c}{${\rm (cm\ s}^{-2}{\rm )}$} &\multicolumn{2}{c}{${\rm He}$}  &
\multicolumn{2}{c}{${\rm C}$} & \multicolumn{2}{c}{${\rm N}$}
&\multicolumn{2}{c}{${\rm O}$}& ${\rm (mag)}$ &${\rm (mag)}$&${\rm (mag)}$& &\\
\hline
{\bf 141115.90$-$305307.00}  &{\bf sdB}    &$11.63$&$11.90$&$30150^{+\ 290}_{-\ 310}$&$5.72^{+0.04}_{-0.07}$&$-1$&$58^{+0.03}_{-0.05}$&$-4$&$18>$               &$-4$&$40^{+0.29}_{-0.67}$&$-4$&$52>$               & 0.04 &$-0.986$&$-0.046$&\#7,ca,rv  & CD-30 11223      \\
{\bf 141115.90$-$305307.00}  &$\dots$      &$\dots$&$\dots$&$25080^{+1260}_{-\ 730}$ &$5.22^{+0.19}_{-0.11}$&$-1$&$39^{+0.10}_{-0.14}$&$-3$&$59>$               &$-3$&$16>$               &$-2$&$76>$               & 0.02 &$\dots$ &$\dots$ &\#11,ca,rv & CD-30 11223      \\
{\bf 141133.45$+$703736.95A} &{\bf sdB}    &$12.71$&$13.17$&$21170^{+1500}_{-1110}$  &$5.55^{+0.31}_{-0.23}$&$-2$&$36>$               &$-3$&$17>$               &$-3$&$02>$               &$-1$&$53>$               & 0.00 &$1.057$ &$0.109$ & &                  \\
     141133.45$+$703736.95B  &    F0.5V    &       &       &$7160$                   &$4.15$                 &\multicolumn{8}{c}{${\rm [Fe/H]}=0.25$}                                                               &       &$(0.333)$&$(0.069)$ &M &                  \\
\hline
\end{tabular}
\end{minipage}
\end{center}
\end{table}
\end{landscape}

\setcounter{table}{1}
\setlength{\tabcolsep}{5pt}
\newpage
\begin{landscape}
\begin{table}
\begin{center}
\begin{minipage}[h]{250mm}
\renewcommand{\footnoterule}{\vspace*{-15pt}}
\setstretch{1.08}
\caption[]{Atmospheric parameters of hot subdwarf stars in the {\sl GALEX} survey. Continued.}
\begin{tabular}{l|cccllr@{.}lr@{.}lr@{.}lr@{.}lcrr|ll}
\hline
\multicolumn{1}{c|}{${\rm GALEXJ}$} & ${\rm Type}$   & $N_{\rm UV}$ & $V$ &\multicolumn{1}{c}{${\rm T}_{\rm eff}$}    & \multicolumn{1}{c}{$\log g$}
&\multicolumn{8}{c}{${\rm Abundances,\ }[{\rm X/H}]_\bullet
 {\rm\ or\ [Fe/H]}$}&\multicolumn{1}{c}{${\rm E}(B-V)$} &\multicolumn{1}{c}{$V-J$}&\multicolumn{1}{c|}{$J-H$}&${\rm Comments}$&${\rm Other\ name}$\\
\cline{7-14}                &    & ${\rm (mag)}$ & ${\rm (mag)}$&\multicolumn{1}{c}{${\rm (K)}$}
&\multicolumn{1}{c}{${\rm (cm\ s}^{-2}{\rm )}$} &\multicolumn{2}{c}{${\rm He}$}  &
\multicolumn{2}{c}{${\rm C}$} & \multicolumn{2}{c}{${\rm N}$}
&\multicolumn{2}{c}{${\rm O}$}& ${\rm (mag)}$ &${\rm (mag)}$&${\rm (mag)}$& &\\
\hline
{\bf 173153.70$+$064706.00}  &{\bf sdB}    &$13.52$&$13.74$&$27780^{+1030}_{-\ 470}$ &$5.35^{+0.18}_{-0.07}$&$-2$&$53>$               &$-3$&$84>$               &$-3$&$81>$               &$-3$&$60>$               & 0.16 &$-0.710$&$-0.062$&\#7,he,s&                  \\
{\bf 173630.26$+$073601.04}  &{\bf sdB}    &$13.68$&$13.91$&$28100^{+\ 390}_{-\ 360}$&$5.43^{+0.07}_{-0.06}$&$-3$&$19^{+0.34}_{-0.78}$&$-3$&$82>$               &$-4$&$33^{+0.38}_{-0.57}$&$-4$&$33>$               & 0.08 &$-0.125$&$-0.053$&\#7,ca,s&                  \\
{\bf 173651.20$+$280635.00A} &{\bf sdB}    &$11.53$&$11.44$&$36160^{+6500}_{-4220}$  &$5.24^{+0.84}_{-0.84}$&$-1$&$09^{+0.69}_{-1.34}$&$-2$&$21>$               &$-1$&$65>$               &$-1$&$63>$               & 0.11 &$0.591$ &$0.214$ &\#11 & TYC 2084-448-1   \\
     173651.20$+$280635.00B  &     F7V     &       &       &$6240$                   &$4.00$                 &\multicolumn{8}{c}{${\rm [Fe/H]}=0.17$}                                                               &       &$(0.640)$&$(0.176)$ &M &                  \\
\hline
\end{tabular}
\end{minipage}
\end{center}
\end{table}
\end{landscape}

\setcounter{table}{1}
\setlength{\tabcolsep}{5pt}
\newpage
\begin{landscape}
\begin{table}
\begin{center}
\begin{minipage}[h]{250mm}
\renewcommand{\footnoterule}{\vspace*{-15pt}}
\setstretch{1.08}
\caption[]{Atmospheric parameters of hot subdwarf stars in the {\sl GALEX} survey. Continued.}
\begin{tabular}{l|cccllr@{.}lr@{.}lr@{.}lr@{.}lcrr|ll}
\hline
\multicolumn{1}{c|}{${\rm GALEXJ}$} & ${\rm Type}$   & $N_{\rm UV}$ & $V$ &\multicolumn{1}{c}{${\rm T}_{\rm eff}$}    & \multicolumn{1}{c}{$\log g$}
&\multicolumn{8}{c}{${\rm Abundances,\ }[{\rm X/H}]_\bullet
 {\rm\ or\ [Fe/H]}$}&\multicolumn{1}{c}{${\rm E}(B-V)$} &\multicolumn{1}{c}{$V-J$}&\multicolumn{1}{c|}{$J-H$}&${\rm Comments}$&${\rm Other\ name}$\\
\cline{7-14}                &    & ${\rm (mag)}$ & ${\rm (mag)}$&\multicolumn{1}{c}{${\rm (K)}$}
&\multicolumn{1}{c}{${\rm (cm\ s}^{-2}{\rm )}$} &\multicolumn{2}{c}{${\rm He}$}  &
\multicolumn{2}{c}{${\rm C}$} & \multicolumn{2}{c}{${\rm N}$}
&\multicolumn{2}{c}{${\rm O}$}& ${\rm (mag)}$ &${\rm (mag)}$&${\rm (mag)}$& &\\
\hline
{\bf 205956.79$-$423203.58}  &     BHB     &$11.54$&$11.43$&$20730^{+\ 150}_{-\ 310}$&$4.52^{+0.09}_{-0.04}$&$-0$&$40^{+0.01}_{-0.13}$&$-2$&$78^{+0.20}_{-0.33}$&$-2$&$90^{+0.39}_{-0.45}$&$-2$&$55>$               & 0.02 &$-0.499$&$-0.145$&\#7,ca& CD-43 14300      \\
{\bf 210031.71$+$145213.83A} &{\bf sdB}    &$13.51$&$13.09$&$34170^{+2470}_{-\ 920}$ &$5.79^{+0.20}_{-0.24}$&$-1$&$41^{+0.20}_{-0.27}$&$-3$&$61>$               &$-3$&$20>$               &$-3$&$76>$               & 0.03 &$0.611$ &$0.210$ &\#7,r&                  \\
     210031.71$+$145213.83B  & F5.5IV-V    &       &       &$6550$                   &$4.06$                 &\multicolumn{8}{c}{${\rm [Fe/H]}=0.31$}                                                               &       &$(0.622)$&$(0.173)$ &M &                  \\
{\bf 211041.43$+$142559.45}  &{\bf He-sdO} &$13.16$&$14.03$&$44470^{+\ 650}_{-\ 350}$&$6.23^{+0.08}_{-0.18}$&$1$&$15^{+0.13}_{-0.10}$ &$-1$&$49^{+0.19}_{-0.18}$&$-2$&$05^{+0.29}_{-0.76}$&$-2$&$16^{+0.52}_{-1.40}$& 0.03 &$-0.711$&$-0.164$&\#7,C\&N&                  \\
\hline
\end{tabular}
\end{minipage}
\end{center}
\end{table}
\end{landscape}

\begin{sidewaystable}
\vspace{2cm}
\setlength{\tabcolsep}{5.5pt}
\begin{minipage}[h]{150mm}
\renewcommand\thefootnote{\alph{footnote}}
\renewcommand{\footnoterule}{\vspace{-20pt}}
\setstretch{1.08}
{\small
\begin{tabular}{l|cccllr@{.}lr@{.}lr@{.}lr@{.}lcrr|ll}
\hline
\multicolumn{1}{c|}{${\rm GALEXJ}$} & ${\rm Type}$   & $N_{\rm UV}$ & $V$ &\multicolumn{1}{c}{${\rm T}_{\rm eff}$}    & \multicolumn{1}{c}{$\log g$}
&\multicolumn{8}{c}{${\rm Abundances,\ }[{\rm X/H}]_\bullet
 {\rm\ or\ [Fe/H]}$}&\multicolumn{1}{c}{${\rm E}(B-V)$} &\multicolumn{1}{c}{$V-J$}&\multicolumn{1}{c|}{$J-H$}&${\rm Comments}$&${\rm Other\ name}$\\
\cline{7-14}                &    & ${\rm (mag)}$ & ${\rm (mag)}$&\multicolumn{1}{c}{${\rm (K)}$}
&\multicolumn{1}{c}{${\rm (cm\ s}^{-2}{\rm )}$} &\multicolumn{2}{c}{${\rm He}$}  &
\multicolumn{2}{c}{${\rm C}$} & \multicolumn{2}{c}{${\rm N}$}
&\multicolumn{2}{c}{${\rm O}$}& ${\rm (mag)}$ &${\rm (mag)}$&${\rm (mag)}$& &\\
\hline
{\bf 071646.93$+$231930.16A} &{\bf   B}    &$13.32$&$11.63$&$11140\pm180$            &$4.39^{+0.06}_{-0.22}$&$-1$&$06>$               &$-2$&$31>$               &$-1$&$76>$               &$-1$&$75>$               & 0.00 &$0.105$ &$0.016$ &na &                  \\
     071646.93$+$231930.16B  &      A1V    &       &       &$9310$                   &$3.67$                 &\multicolumn{8}{c}{${\rm [Fe/H]}=0.11$}                                                               &             &$(0.012)$&$(0.001)$ &M &                  \\
{\bf 085158.89$-$171238.86}\footnote[1]{Target is taken from \citet{jimenez11}.} &{\bf He-sdO} &$12.16$&$11.40$&$39060^{+\ 580}_{-\ 150}$&$5.79^{+0.10}_{-0.04}$&=$3$&$00$                &$-1$&$08>$               &$-0$&$15^{+0.14}_{-0.23}$&$-0$&$42>$               & 0.04 &$-1.013$&$-0.139$&\#7,t & TYC 6017$-$419$-$1\\
{\bf 160011.80$-$643330.35}\Large{$^a$} &{\bf He-sdB} &$     $&$11.86$&$34640^{+\ 590}_{-\ 580}$&$6.02^{+0.08}_{-0.11}$&$-0$&$30^{+0.05}_{-0.04}$&$-2$&$83^{+0.31}_{-0.15}$&$-2$&$92^{+0.18}_{-0.45}$&$-3$&$78>$               & 0.03 &$-0.708$&$-0.167$&\#7 & TYC 9044$-$1653$-$1 \\
\hline
\end{tabular}
}
\end{minipage}
\vspace{-52mm}
\flushleft
{{\bf Table 3.} Other targets that are not part ot the {\sl GALEX} UV selection. \customlabel{Tab:6}{3} } 
\vspace{ 52mm}
\end{sidewaystable}